\pdfoutput=1
\documentclass[%
 aip,pof,
 amsmath,amssymb,
preprint,%
]{revtex4-1}
\usepackage{color}
\usepackage{graphicx}% Include figure files
\usepackage{dcolumn}% Align table columns on decimal point
\usepackage{bm}% bold math
\usepackage[mathlines]{lineno}% Enable numbering of text and display math
\usepackage{mathrsfs}
\usepackage[utf8]{inputenc}
\usepackage[T1]{fontenc}
\usepackage{mathptmx}
\usepackage{subfigure}
\usepackage{tabularx}%控制表格宽度
\usepackage{booktabs}%控制表格线宽
\usepackage{bm}
\usepackage{CJKutf8}
\usepackage{amsthm}
\usepackage{amsfonts}
\usepackage{float}
\usepackage{extarrows}
\begin{document}

\preprint{AIP/POF}
%\draft
\begin {CJK*} {UTF8} {gbsn} 
\title{Boundary vorticity dynamics of two-phase viscous flow}
% Force line breaks with \\
\author{Tao Chen~(陈涛)}
%\email{1601111553@pku.edu.cn}
\affiliation{
Department of Mechanics and Aerospace Engineering, Southern University of Science and Technology, Shenzhen 518055, China
}
%\affiliation{State Key Laboratory for Turbulence and Complex Systems, College of Engineering, Peking University, Beijing 100871, China}
\author{Tianshu Liu~(刘天舒)}
\email{tianshu.liu@wmich.edu}
\affiliation{Department of Mechanical and Aerospace Engineering, Western Michigan University, Kalamazoo, Michigan, 49008, USA}
%\\This line break forced with \textbackslash\textbackslash
%\author{Lian-Ping Wang~(王连平)}
%%\email{wanglp@sustech.edu.cn}
%\affiliation{
%	Guangdong-Hong Kong-Macao Joint Laboratory for Data-Driven Fluid Mechanics and Engineering Applications, Southern University of Science and Technology, Shenzhen 518055, China
%}
%\affiliation{
%	Southern Marine Science and Engineering Guangdong Laboratory (Guangzhou), 1119 Haibin Road, Nansha District, Guangzhou,
%	511458, China
%}
%\affiliation{
%	Guangdong Provincial Key Laboratory of Turbulence Research and Applications, Center for Complex Flows and Soft Matter Research and Department of Mechanics and Aerospace Engineering, Southern University of Science and Technology, Shenzhen 518055, Guangdong, China
%}

\date{\today}% It is always \today, today,

\begin{abstract}
\centerline{\bf Abstract}
From the Navier-Stokes-Korteweg (NSK) equations, the exact relations between the fundamental surface physical quantities for two-phase viscous flow with diffuse interface are derived, including density gradient, shear stress, vorticity, pressure, enstrophy flux and surface curvature. These theoretical results provide a solid foundation of 
the boundary/interfacial vorticity dynamics and a new tool for analysis of complex interfacial phenomena in two-phase viscous flows.
To demonstrate the application of the developed results, simulation of a droplet impacting and spreading on a solid wall is conducted by using a recently developed well-balanced discrete unified gas kinetic scheme (WB-DUGKS), focusing on spreading process when the separation bubbles form inside the droplet. The distributions of shear stress, pressure and enstrophy flux at the interface and wall are analyzed, particularly near the moving contact points and other characteristic points. This example gives an unique perspective to the physics of droplet impingement on a wall.
%First, we show that skin friction, surface pressure and BEF are effective indicators for local flow separation and attachment at the bottom wall. It is found that the regions near the moving contact points are the main vorticity sources for generating the negative BEF.
%Secondly, we show that the IEF is negative on the liquid-vapor interface, which has a highly negative peak in a small vicinity of each contact point and two negative peaks with lower magnitudes around the central region. Interestingly, these peak regions are found to be caused by different mechanisms revealed by the decomposition. Therefore, the total enstrophy flux across the closed surface of the droplet volume is negative, which implies the increase of the enstrophy inside the droplet. 
%By virtue of the conservation law, it is reasonable to conjecture that the increased enstrophy will make up the viscous dissipation and resist the surface tension work, thereby facilitating the maintenance of separation bubbles during spreading process. 
\end{abstract}

\maketitle
\end{CJK*}

%\begin{quotation}
%The ``lead paragraph'' is encapsulated with the \LaTeX\ 
%\verb+quotation+ environment and is formatted as a single paragraph before the first section heading. 
%(The \verb+quotation+ environment reverts to its usual meaning after the first sectioning command.) 
%Note that numbered references are allowed in the lead paragraph.
%%
%The lead paragraph will only be found in an article being prepared for the journal \textit{Chaos}.
%\end{quotation}
\section{Introduction}
Investigation on the interaction between vorticity and boundary (solid wall or interface) is of crucial importance to understand fundamental physics of viscous flows. Exploration along this direction leads to a general theory referred to as the {\it boundary (interfacial) vorticity dynamics}.~\cite{WuJZ1995,WuJZ2015book} The boundary could be a solid wall (either rigid or flexible) or an interface separating different phases. Numerous simulations and experiments have demonstrated complex vortical structures created from the boundary.~\cite{Tryggvason2006,Elghobashi2019,HeChengming2020}  
The vorticity-based theoretical methods offer an effective way to give technically accurate interpretations to these observed results.
However, existing studies (particularly theoretical results) along this direction are relatively limited.
Here, the relevant topics will be reviewed, including the boundary and interfacial vorticity dynamics constituting the basic aspects of the present theoretical development.

From the perspective of the boundary vorticity dynamics, a wall is the essential source of complexity of near-wall viscous flows. The vorticity is first created at the wall by virtue of the viscosity and the no-slip condition, and then it diffuses into the interior of the fluid.
Study of the boundary vorticity dynamics could be traced back to the pioneering work of Lighthill.~\cite{Lighthill1963}
For incompressible viscous flow past a two-dimensional (2D) stationary flat wall, he first introduced the concept of the boundary vorticity flux (BVF) defined as the wall-normal vorticity flux across the wall per unit area and per unit time. The BVF is physically interpreted as the strength of the vorticity source/sink distributed on the wall and is directly determined by the surface pressure gradient through a pair of coupled partial differential equations.
The definition of BVF was later generalized by Panton~\cite{Panton1984} to three-dimensional (3D) case. Further, a general intrinsic theory on the BVF corresponding to an arbitrarily moving and deforming wall was proposed by Wu and Wu~\cite{WuJZ1996,WuJZWuJM1998} for 3D viscous flows. They showed that the boundary vorticity at a given time is determined by the temporal-spatial accumulated effect of the BVF and the diffusion, convection and dissipation of the whole vorticity field inside the flow. They also proved that the total force and moment acted on a rigid body were expressed solely in terms of proper vectorial moments of relevant fluxes (including the BVF). Therefore, the action and reaction between the vorticity and the solid wall were rationally depicted in an unified framework.~\cite{WuJZWuJM1998,WuJZ2015book} Lyman~\cite{Lyman1990} also proposed an alternative definition of the BVF by absorbing the viscous contribution into Lighthill-Panton-Wu's definition at the expense of the physically intuitive concept of viscous diffusion flux in 3D case.
When integrated over a closed surface, both the integrals of these two definitions of the BVF are equal to the volume integral of the vorticity diffusion term in the transport equation. 

In addition to the vectorial BVF, the boundary enstrophy flux (BEF) $f_{\Omega}$ is introduced as an important scalar quantity to measure the enstrophy diffusion rate across the wall.~\cite{WuJZ1995,Liu2016MST,Liu2018AIA,ChenTao2021WEF} For incompressible viscous flow past a stationary flat wall, skin friction $\bm{\tau}$ (namely, wall shear stress) and surface pressure $p_{\partial B}$ have been identified as the footprints of near-wall coherent structures. By applying the Taylor-series expansion to the Navier-Stokes (NS) equations on a stationary flat wall, Bewley and Protas~\cite{Bewley2004} found that the near-wall flow variables in a small vicinity of the wall were uniquely determined by $\bm{\tau}$,~$p_{\partial B}$ and their relevant temporal-spatial derivatives at the wall.
Interestingly, Liu {\it et al.}~\cite{Liu2016MST,Liu2018AIA} found that skin friction $\bm{\tau}$ and surface pressure $p_{\partial B}$ were not independent but were intrinsically coupled through the BEF. A concise and exact $\bm{\tau}$~--~$p_{\partial B}$ relation was derived: $\bm{\tau}\bm{\cdot}\bm{\nabla}_{\partial B}p_{\partial B}=\mu f_{\Omega}$ ($\bm{\nabla}_{\partial B}$ is the surface tangential gradient operator), indicating that BEF was generated through the viscous coupling between the skin friction and surface pressure gradient. 
Relations between skin friction and other surface physical quantities were discussed and generalized by Liu and Woodiga,~\cite{LiuWoodiga2011} Chen {\it et al.}~\cite{ChenTao2019POF} and Miozzi {\it et al.}~\cite{Miozzi2016,Miozzi2019EXF}

Interestingly, by modeling the BEF properly, different on-wall footprints can be mutually inferred based on the $\bm{\tau}$~--~$p_{\partial B}$ relation, which is related to global measurements of skin friction and surface pressure.
On one hand, global skin friction field is difficult to measure while surface pressure field can be directly obtained using pressure-sensitive paint (PSP).~\cite{Liu2021PSPBook,Liu2019PAS}
A global skin friction field $\bm{\tau}$ can be extracted from the known surface pressure field $p_{\partial B}$ by solving the Euler-Lagrangian equation derived from the variational weak-form of the $\bm{\tau}$~--~$p_{\partial B}$ relation.~\cite{Liu2018AIA,ChenTao2019POF} One the other hand, using the $\bm{\tau}$~--~$p_{\partial B}$ relation, Cai {\it et al.}~\cite{CaiZM2022} proposed an approximate but efficient variational method to extract surface pressure field $p_{\partial B}$ from skin friction field $\bm{\tau}$ obtained by global luminescent oil-film (GLOF) measurements. Surprisingly, although the BEF cannot be known as an a priori, the constant BEF approximation still yields a satisfactory approximate solution of surface pressure compared to the ground truth, particularly for near-wall flow structures dominated by the skin friction divergence $\bm{\nabla}_{\partial B}\bm{\cdot}\bm{\tau}$. In fact, the skin friction divergence was the first invariant of the no-slip tensor proposed by Chong {\it et al.}~\cite{Chong2012} and was shown to be a critical quantity in characterizing sweep and ejection events in wall-bounded turbulence by Chen {\it et al.}~\cite{ChenTao2021POF} and Chen and Liu.~\cite{ChenTao2022AIPb} The significance of $\bm{\nabla}_{\partial B}\bm{\cdot}\bm{\tau}$ was also demonstrated by Liu {\it et al.}~\cite{Liu2021EXP} in identifying turbulent wedges in the boundary-layer transition front. Besides, Chen and Liu~\cite{ChenTao2022AIPa,ChenTao2022AIPb} also showed that the structure of near-wall Lamb vector was directly related to $\bm{\nabla}_{\partial B}\bm{\cdot}\bm{\tau}$, the surface vorticity divergence $\bm{\nabla}_{\partial B}\bm{\cdot}\bm{\omega}_{\partial B}$ and the BEF $f_{\Omega}$.

Furthermore, by applying the $\bm{\tau}$~--~$p_{\partial B}$ relation to an airfoil, Liu {\it et al.}~\cite{Liu2017AIAA} proposed a new aerodynamic force formula, explicitly revealing the critical role of viscosity in generating lift. The main consequence is that lift cannot be generated without the cost of generating the viscous drag at the same time. Without the fluid viscosity, both lift and drag are zero (D' Alembert paradox).~\cite{LiuTianshu2021} 
Chen {\it et al.}~\cite{ChenTao2021POF} simulated a turbulent channel flow at the frictional Reynolds number at $Re_{\tau}=180$ and found that the strong wall-normal velocity events (SWNVEs) induced by the near-wall quasi-streamwise vortices were strongly correlated with high-magnitude BEF regions, which accounted for the high intermittent feature of the viscous sublayer. Later, Chen {\it et al.}~\cite{ChenTao2021WEF} derived the exact relation between the temporal-spatial evolution rate of the wall-normal enstrophy flux (WNEF) at the wall and fundamental surface physical quantities (skin friction $\bm{\tau}$ and surface pressure $p_{\partial B}$). Note that the BEF field is the restriction of the WNEF field on the wall. Near the SWNVEs, it was shown that this evolution rate was  dominated by the wall-normal variation of the vortex stretching term.
%Closely associated with this research, Guerrero {\it et al.}~\cite{Guerrero2022JFM} performed a detailed statistical study based on the direct numerical simulation (DNS) of a turbulent pipe flow. They found that the rare back flow events were the signatures of near-wall self-sustaining processes, which were closely related to the enstrophy stretching and intensification mechanisms.

The second topic relevant to the present theoretical development is the interfacial vorticity dynamics, which focuses on surface vorticity and its viscous flux across the interface separating two fluids. Previous studies were performed by considering two-phase flow with a sharp interface with some physical properties (such as surface tension), where the interfacial thickness was zero and a physical quantity was allowed to have a jump across the interface.
Longuet-Higgins~\cite{LH1953} explained that the vorticity appeared on a 2D steady free surface (a special interface) as a direct consequence of the continuity of the tangential shear stress across the surface. He proved that the surface vorticity is just twice the product of the surface tangential velocity and the surface curvature.
Later, the concept of the BVF and some vorticity-based description were generalized to a free surface by Lugt,~\cite{Lugt1987} Rood~\cite{Rood1994} and Herrera~\cite{Herrera2010}. Wu~\cite{WuJZ1995} developed a general theory of interfacial vorticity dynamics, including the vorticity creation from the interface and the integrated reaction of the created vorticity to a closed interface. Three points were further claimed by Wu.~\cite{WuJZ1995} 
First, the need for balancing the tangential components of the surface deformation stress leads to the appearance of the surface vorticity, further deepening the interpretation of Longuet-Higgins.~\cite{LH1953} 
Secondly, generalized formulas of both the one-sided and net BVF are given, followed by the discussion on the generalized BEF as a complement. Third, the most remarkable aspect in 3D case is the appearance of a normal vorticity field, which has an extra contribution to the boundary-layer behaviour and to interfacial vortex-sheet velocity. Recently, vortex dynamical approaches were adopted by He {\it et al.}~\cite{HeChengming2020} to interpret the nonmonotonic viscous dissipation of off-center droplet collision. A half-domain helicity analysis identified a strong interaction between the ring-shaped vortices in the droplet interior and the line-shaped shear layer in the droplet interaction region during the stretching separation.
In addition, they derived a general relation between the total enstrophy and the total viscous dissipation rate for unsteady free-surface flows, and found that their differences were analytically attributed to the unbalanced flow, vorticity and velocity related to the phase interface. Moreover, Brøns {\it et al.}~\cite{Brons2014} and Terrington {\it et al.}~\cite{Terrington2020JFM} investigated the generation and conservation of vorticity for generalized fluid-fluid interfaces, where the total circulation rather than the vorticity was actually used as a tool to understand the associated physics. 

Instead of using the sharp interface model, this paper discusses the decomposition of the boundary and interfacial enstrophy fluxes for two-phase viscous flow from the perspective of diffuse interface, which has not been considered in the existing literature. Compared to the sharp interface model, the interfacial mixing layer has a finite thickness across which the physical properties of fluid vary smoothly from one phase to the other. The diffuse interface model is particularly suitable for the problems of near-critical liquid-vapor flows, complex topological changes and large deformations of the interface, interface breakup and coalescence, and so on, where the traditional sharp interface model may fail to give accurate prediction of the interfacial changes.~\cite{Anderson1998}
The vorticity dynamics of diffusive interface is the main theme to be explored here.  

This paper is organized as follows. In Section~\ref{Modeling of two-phase flow}, starting from the free energy description of two-phase flow system, the Navier-Stokes-Korteweg (NSK) equations are introduced along with the discussions on notation convention and separation of hydrodynamic pressure. In Section~\ref{Enstrophy fluxes for two-phase flow}, we derive explicit decomposition of the boundary and interfacial enstrophy fluxes for the diffuse-interface two-phase flow model. Then, numerical model and method are briefly described in Section~\ref{Numerical model and method}. In Section~\ref{simulation}, we simulate and analyze a droplet impact on a hydrophilic solid wall to demonstrate the application of these theoretical results. Finally, conclusions are drawn in Section~\ref{Conclusions and discussions} along with some discussions. Appendix~\ref{ProofLeeLin} gives the detailed proof of Eq.~\eqref{xx2}. Interface mean curvature and its physical interpretation are briefly discussed in Appendix~\ref{curvature}. 
Appendix~\ref{Code validation} shows standard numerical tests for steady droplet with different prescribed contact angles, demonstrating the effectiveness of geometric boundary condition combined with the WB-DUGKS.
Appendix~\ref{ESPQ} shows the reliability of our post-processing code to extract the surface physical quantities on a curved surface.
\newpage
\section{Modeling of two-phase flow}\label{Modeling of two-phase flow}
\subsection{Free energy description of two-phase flow system}
The thermodynamics of an isothermal two-phase flow system at equilibrium can be described through the second-gradient theory, where the corresponding Ginzburg-Landau free energy functional is~\cite{CahnHilliard1958,Anderson1998,Jamet2000NED}
\begin{eqnarray}
\mathcal{F}(\rho,\bm{\nabla}\rho)=\int_{V}\left(\psi(\rho)+\frac{1}{2}\kappa\lVert\bm{\nabla}\rho\rVert^2\right)dV,
\end{eqnarray}
where $\rho$ is the fluid density, $\kappa$ represents the interfacial free energy coefficient associated with the surface tension $\sigma_{s}$, and $V$ is the integral volume occupied by the fluids. The first term $\psi\left(\rho\right)$ is the bulk free energy density, while the second term $\kappa\lVert\bm{\nabla}\rho\rVert^2/2$ is the interfacial free energy density caused by the non-local molecular interaction.

\begin{figure}[h!]
	\centering
	\subfigure[]{
		\begin{minipage}[t]{0.5\linewidth}
			\centering
			\includegraphics[width=1.0\columnwidth,trim={5cm 3.0cm 5cm 3.2cm},clip]{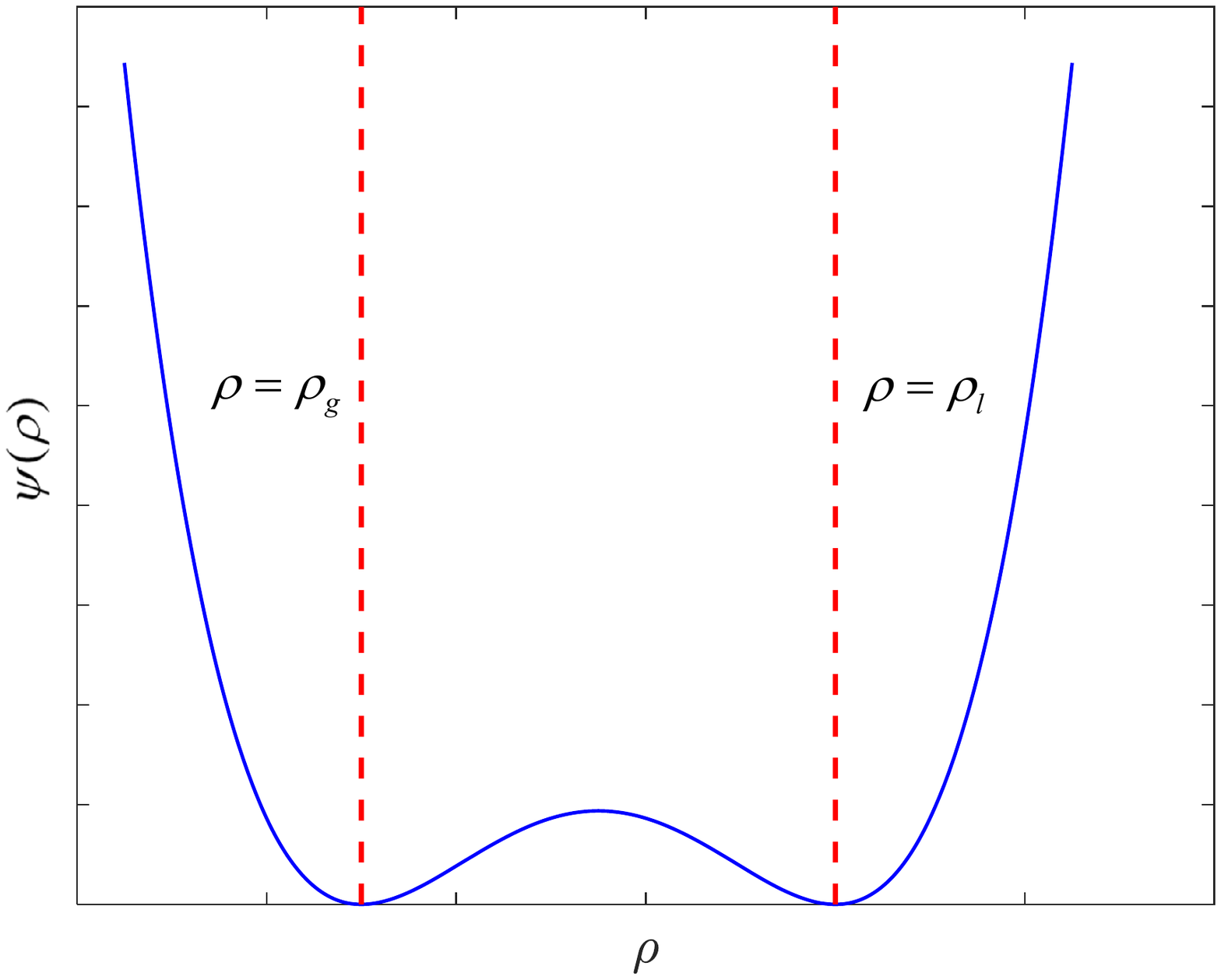}
			\label{psi_rho_profile}
		\end{minipage}%
	}%
	\subfigure[]{
		\begin{minipage}[t]{0.5\linewidth}
			\centering
			\includegraphics[width=1.0\columnwidth,trim={5cm 3.2cm 5cm 3.2cm},clip]{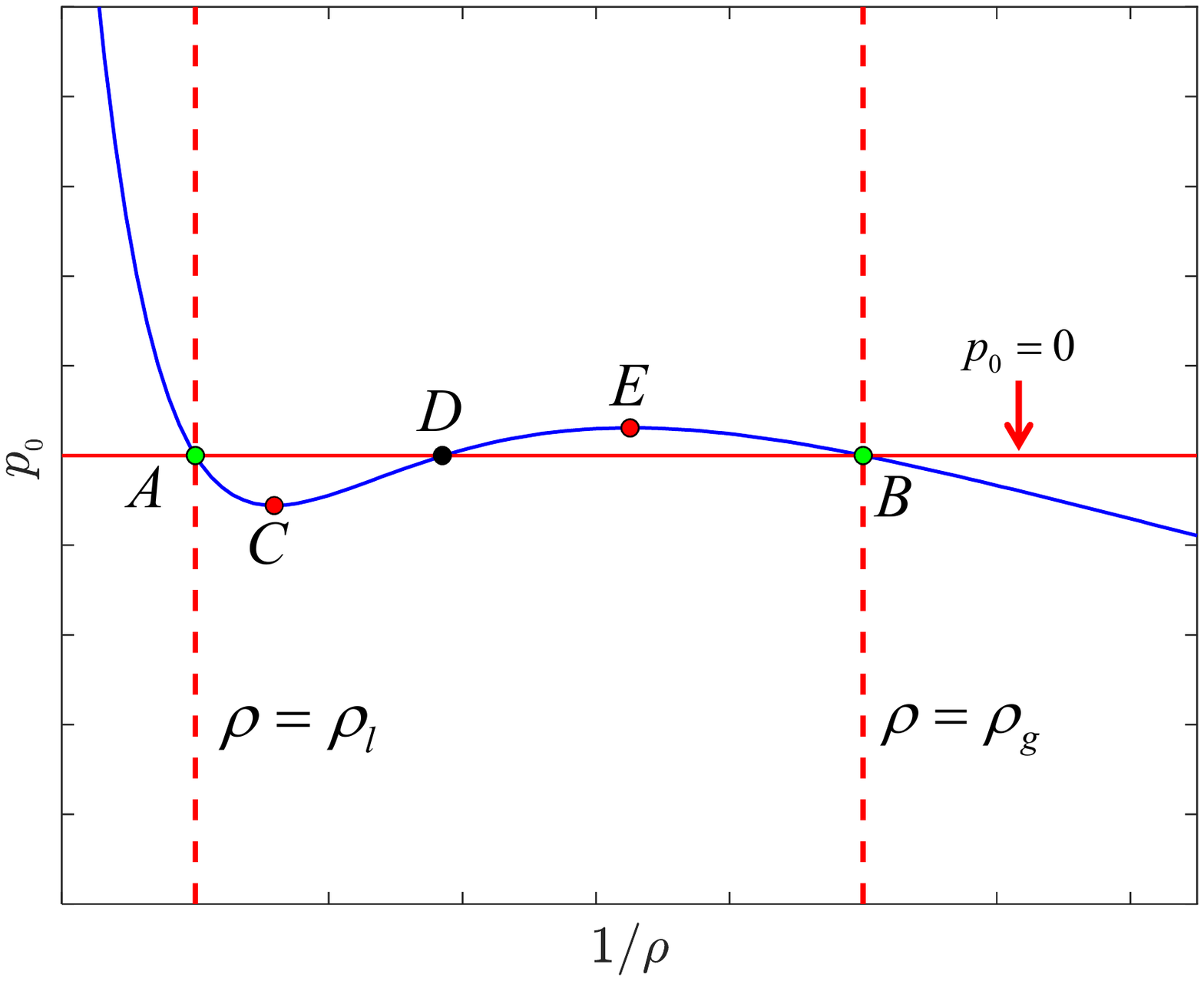}
			\label{Thermodynamic_pressure}
		\end{minipage}%
	}%
	\caption{(a) Double-well bulk free energy density $\psi(\rho)$. (b) Thermodynamic pressure $p_0$ as a function of $1/\rho$. In both the figures, the two vertical lines denote the saturated liquid and vapor densities, respectively.}
	\label{rrrr}
\end{figure}

When the bulk free energy density $\psi(\rho)$ is given, the thermodynamic pressure (namely, the equation of state) is evaluated by
\begin{eqnarray}\label{p0}
p_{0}=\rho\frac{\partial\psi}{\partial\rho}-\psi.
\end{eqnarray}
In the vicinity of the critical point of the equation of state (namely, for near-critical fluids), as illustrated in Fig.~\ref{psi_rho_profile}, the double-well approximation of the bulk free energy density can be employed, namely,~\cite{Jacqmin1999,Jamet2000NED}
\begin{eqnarray}\label{doublewell}
\psi(\rho)=\beta(\rho-\rho_{l})^{2}(\rho-\rho_{g})^2,
\end{eqnarray}
where $\beta$ is a positive constant, $\rho_{l}$ and $\rho_{g}$ represent the densities of separated phases at saturation, respectively (note that the real densities can have small variations with respect to the saturation values).

Substituting Eq.~\eqref{doublewell} into Eq.~\eqref{p0}, the corresponding thermodynamic pressure is 
\begin{eqnarray}\label{p0a}
p_{0}=\beta(\rho-\rho_{l})(\rho-\rho_{g})\left[3\rho^2-\rho_l\rho_g-\rho(\rho_{l}+\rho_{g})\right].
\end{eqnarray}
Fig.~\ref{Thermodynamic_pressure} shows the variation of $p_{0}$ with respect to the reciprocal of the density $1/\rho$ near the critical point (below the critical temperature). According to the Maxwell's equal-area rule, for a flat surface at equilibrium, there are three possible densities (points $A$, $D$ and $B$) corresponding to the pressure $p_{0}=0$.
The two points $A$ and $B$ represent the saturated liquid and gas densities $\rho_{l}$ and $\rho_{g}$, respectively.
The point $D$ lies in a mechanically unstable region $CE$ with $dp_0/d\rho<0$ which forces the fluids to separate into two stable phases.~\cite{LeeLin2005,HeXiaoyi1999}
It is noted that if $\psi(\rho)$ is properly selected, other kinds of equation of state can also be recovered, such as those of van der Waals,~\cite{Rowlinson1982} Carnahan-Starling,~\cite{CS1969} and so on. For the present discussion, we focuses on Eq.~\eqref{doublewell}.

The first-order variation of the free energy functional with respect to the density gives the chemical potential $\mu_{\rho}$, namely,
\begin{eqnarray}\label{murho}
\mu_{\rho}\equiv\frac{\delta\mathcal{F}}{\delta\rho}=\frac{\partial\psi}{\partial\rho}-\kappa\nabla^2\rho,
\end{eqnarray}
where the first term defines the bulk chemical potential
\begin{eqnarray}\label{mu0}
\mu_0\equiv\frac{\partial\psi}{\partial\rho}.
\end{eqnarray}
Using Eq.~\eqref{doublewell}, $\mu_0$ is expressed as
\begin{eqnarray}
\mu_{0}=4\beta(\rho-\rho_{l})(\rho-\rho_{g})\left(\rho-\frac{\rho_{l}+\rho_{g}}{2}\right).
\end{eqnarray}

\begin{figure}[h!]
	\centering
	\includegraphics[width=0.7\columnwidth,trim={0.1cm 3.3cm 0.1cm 3.7cm},clip]{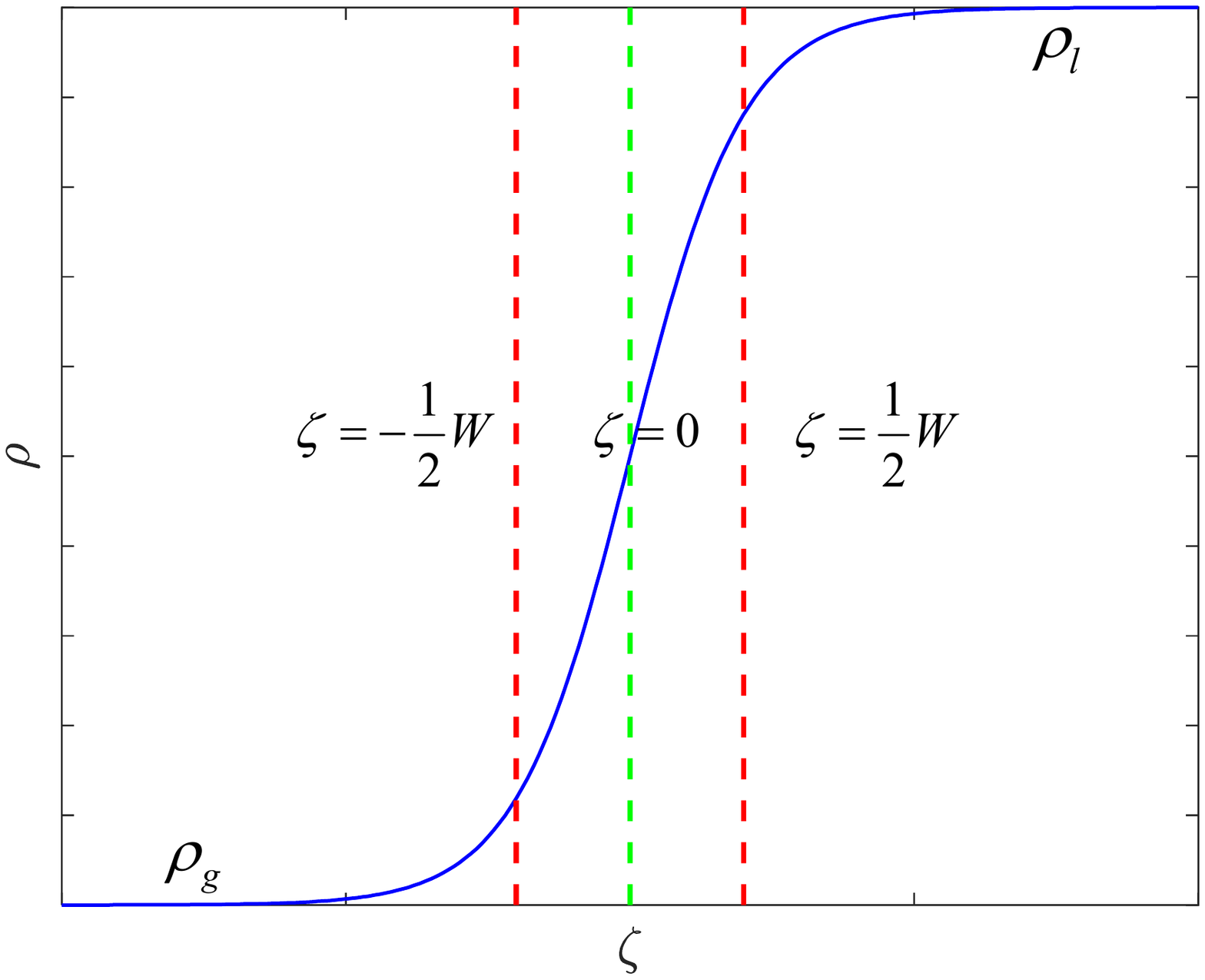}
	\caption{Sketch of the hyperbolic tangent density profile in Eq.~\eqref{density_profile}. Two vertical red lines indicate the coordinates $\zeta=\pm{W}/2$, and the green line indicates $\zeta=0$.} 
	\label{density_profile1}
\end{figure}

Considering a flat surface at equilibrium, one can solve the energy minimum equation  $\mu_{\rho}=0$ with the asymptotic boundary conditions $[d\rho/d\zeta]_{\zeta\rightarrow\pm\infty}=0$, $[\rho]_{\zeta\rightarrow+\infty}=\rho_{l}$ and $[\rho]_{\zeta\rightarrow-\infty}=\rho_{g}$, which results in a hyperbolic tangent density profile (see Fig.~\ref{density_profile1}),
\begin{eqnarray}\label{density_profile}
\rho(\zeta)=\frac{\rho_{l}+\rho_{g}}{2}+\frac{\rho_{l}-\rho_{g}}{2}\tanh\left(\frac{2\zeta}{W}\right),
\end{eqnarray}
where $\zeta$ is the surface normal coordinate measured from the point satisfying $\rho=(\rho_{l}+\rho_{g})/2$, and $W=(\rho_{l}-\rho_{g})^{-1}\sqrt{8\kappa/\beta}$ is a length parameter with the same order of magnitude as the interfacial thickness.
Though derived from the flat surface at equilibrium, Eq.~\eqref{density_profile} still provides a leading-order approximation in the presence of a curved surface, when the ratio of the interfacial thickness to the characteristic length scale is sufficiently small.

The surface tension coefficient $\sigma_{s}$ is equal to the integral of the free energy density across the interface per unit area.~\cite{Jacqmin1999} Using Eq.~\eqref{density_profile}, $\sigma_{s}$ is further evaluated as
\begin{eqnarray}
\sigma_{s}=\kappa\int_{-\infty}^{+\infty}\left(\frac{d\rho}{d\zeta}\right)^2d\zeta
=\frac{1}{6}(\rho_{l}-\rho_{g})^{3}\sqrt{2\kappa\beta}.
\end{eqnarray}
Conversely, $\beta$ and $\kappa$ can be determined through
\begin{eqnarray}
\beta=\frac{12\sigma_{s}}{W(\rho_{l}-\rho_{g})^4},~\kappa=\frac{3\sigma_{s}W}{2(\rho_{l}-\rho_{g})^2},
\end{eqnarray}
when $\sigma_{s}$, $W$ and two saturation densities ($\rho_{l}$ and $\rho_{g}$) are given a priori.

\subsection{Navier-Stokes-Korteweg equations}\label{NSK_equations}
The Navier-Stokes-Korteweg (NSK) equations for two-phase flow system can be expressed as
\begin{eqnarray}\label{c1}
\frac{\partial\rho}{\partial t}+\bm{\nabla}\bm{\cdot}(\rho\bm{u})=0,
\end{eqnarray}
\begin{eqnarray}\label{c2}
\frac{\partial(\rho\bm{u})}{\partial t}+\bm{\nabla}\bm\cdot\left(\rho\bm{uu}\right)=-\rho\bm{\nabla}\mu_{\rho}+\bm{\nabla}\bm{\cdot}\bm{\Pi},
\end{eqnarray}
where $\bm{u}$ is the macroscopic velocity, $\bm{\Pi}$ is the viscous stress tensor, namely,
\begin{eqnarray}\label{viscous_stress_tensor}
\bm{\Pi}\equiv2\mu\bm{S}+\lambda\vartheta\bm{I}.
\end{eqnarray}
In Eq.~\eqref{viscous_stress_tensor}, $\mu$ is the dynamic viscosity, $\vartheta\equiv\bm{\nabla}\bm{\cdot}\bm{u}$ is the dilatation, and $\bm{I}$ is the unit tensor. $\lambda\equiv\mu_{V}-(2/D)\mu$ is a viscosity coefficient, where $\mu_V$ is the bulk viscosity and $D$ is the spatial dimension. $\bm{S}$ (the strain rate tensor) and $\bm{A}$ (the rotation tensor) are respectively the symmetric and antisymmetric parts of the velocity gradient tensor $\bm{\nabla u}$:
\begin{eqnarray}
\bm{S}=\frac{1}{2}\left(\bm{\nabla}\bm{u}^{T}+\bm{\nabla}\bm{u}\right),~
\bm{A}=\frac{1}{2}\left(\bm{\nabla}\bm{u}^{T}-\bm{\nabla}\bm{u}\right).
\end{eqnarray}
For any given vector $\bm{\eta}$, the rotation tensor $\bm{A}$ satisfies the exact relation $2\bm{A}\bm{\cdot}\bm{\eta}=\bm{\omega}\times\bm{\eta}$ where $\bm{\omega}\equiv\bm{\nabla}\times\bm{u}$ is the vorticity. The enstrophy is defined as $\Omega=\omega^2/2$.

It is noted that both the pressure and surface tension force are incorporated in the chemical potential gradient force term $-\rho\bm{\nabla}\mu_{\rho}$ ({\it potential form}~\cite{GuoZhaoli2021,ZhangCH2021}). By using Eqs.~\eqref{p0} and~\eqref{mu0}, a thermodynamic identity can be obtained as
\begin{eqnarray}\label{identity}
\rho\bm{\nabla}\mu_{0}=\bm{\nabla}\left(\rho\mu_{0}\right)-\mu_{0}\bm{\nabla}\rho=\bm{\nabla}(p_{0}+\psi)-\bm{\nabla}\psi=\bm{\nabla}p_{0}.
\end{eqnarray}
Then, from Eqs.~\eqref{murho} and~\eqref{identity}, one can obtain an equivalent expression of the chemical potential gradient force ({\it pressure form}~\cite{LeeTaehun2006}):
\begin{eqnarray}\label{pressure_form}
-\rho\bm{\nabla}\mu_{\rho}=-\bm{\nabla}p_{0}+\kappa\rho\bm{\nabla}\nabla^2\rho.
\end{eqnarray}
In addition, by virtue of the following identity
\begin{eqnarray}\label{id1}
\rho\bm{\nabla}\nabla^2\rho=\bm{\nabla}\bm{\cdot}
\left[\left(\rho\nabla^{2}\rho+\frac{1}{2}\lVert\bm{\nabla}\rho\rVert^{2}\right)\bm{I}-\bm{\nabla}\rho\bm{\nabla}\rho\right],
\end{eqnarray}
Eq.~\eqref{pressure_form} can also be rewritten as ({\it divergence form})
\begin{eqnarray}
-\rho\bm{\nabla}\mu_{\rho}
=-\bm{\nabla}\bm{\cdot}\bm{P},
\end{eqnarray}
where $\bm{P}$ is referred to as the Korteweg pressure tensor:~\cite{HeDoolen2002}
\begin{eqnarray}\label{pressure_tensor}
\bm{P}=\left(p_{0}-\kappa\rho\nabla^2\rho-\frac{1}{2}\kappa\lVert\bm{\nabla}\rho\rVert^2\right)\bm{I}+\kappa\bm{\nabla}\rho\bm{\nabla}\rho\equiv p\bm{I}+\kappa\bm{\nabla}\rho\bm{\nabla}\rho.
\end{eqnarray}
In Eq.~\eqref{pressure_tensor}, $p$ is called the non-local total pressure including the thermodynamic pressure $p_{0}$, and two capillary contributions due to the density gradients. It is noted that Eq.~\eqref{pressure_tensor} is formally consistent with the pressure tensor derived from the kinetic Enskog-Vlasov equation which unifies the Enskog kinetic theory for short-range molecular interaction, and the mean field theory for long-range molecular interaction.~\cite{Chapman1970,HeDoolen2002}
\subsection{Notation convention}\label{Notation_convention}
For the diffuse interface model with finite interfacial thickness, the fluid density varies smoothly from one fluid to the other. The liquid-vapor interface $S$ is usually defined as the instantaneous isosurface of the density field where $\rho_S=(\rho_{l}+\rho_{g})/2$.
A physical quantity $\mathcal{F}$ with the subscript $S$ indicates its restriction on the interface $S$ to form a surface physical field $\mathcal{F}_{S}\equiv\left[\mathcal{F}\right]_{S}$. 
Correspondingly, the unit normal vector of $S$ is selected as $\bm{n}_{S}\equiv\left[\bm{\nabla}\rho\right]_S/\lVert\bm{\nabla}\rho\rVert_S$, which orients from the vapor to the liquid.
$\bm{\nabla}_{S}\equiv\bm{\nabla}-\bm{n}_{S}\partial/\partial{n}=(\bm{I}-\bm{n}_{S}\bm{n}_{S})\bm{\cdot}\bm{\nabla}$ denotes the surface gradient operator along the tangential direction of $S$, where $\bm{\nabla}$ is the 3D spatial gradient operator and $\partial/\partial{n}$ is the surface normal derivative. 

The surface curvature tensor can be written as $\bm{K}_{S}=-\bm{\nabla}_{S}\bm{n}_{S}$, which is solely determined by the  distribution of the unit normal vector field $\bm{n}_{S}$ along $S$. The trace of the surface curvature tensor is twice of the mean curvature $H\equiv(\kappa_{1}+\kappa_{2})/2$, namely, $tr(\bm{K})=-\bm{\nabla}_{S}\bm{\cdot}\bm{n}_{S}=\kappa_{1}+\kappa_{2}=2H$, where $\kappa_{1}$ and $\kappa_{2}$ are the two principal curvatures of $S$.
Physical interpretation and evaluation of the mean curvature can be found in Appendix~\ref{curvature}.
The surface Laplacian is denoted by ${\nabla}_{S}^{2}\equiv\bm{\nabla}_{S}\bm{\cdot}\bm{\nabla}_{S}$.

Similarly, we use $\partial B$ to represent a stationary solid boundary (either flat or curved), whose unit normal vector field $\bm{n}_{\partial B}$ directs from the wall to the interior of the fluid. 
The no-slip boundary condition is employed, namely, $\bm{u}_{\partial B}=\bm{0}$.
The notation convention used for the interface $S$ is also applicable for the solid wall $\partial B$, when the subscripts $S$ are changed to $\partial{B}$ relevant physical quantities.
\subsection{Separation of hydrodynamic pressure}\label{separationph}
We notice that Lee and Lin~\cite{LeeLin2005} proposed a new equivalent formulation of the pressure tensor ({\it stress form}), which reads as
\begin{eqnarray}\label{PLLa}
\bm{P}\equiv p_{h}^{(L-L)}\bm{I}-\bm{\Phi},
\end{eqnarray}
where $p_{h}^{(L-L)}$ and $\bm{\Phi}$ are respectively given by
\begin{eqnarray}\label{PLLb}
	p_{h}^{(L-L)}\equiv p_{0}-\kappa\rho\nabla^2\rho+\frac{1}{2}\kappa\lVert\bm{\nabla}\rho\rVert^2,~
\bm{\Phi}\equiv\kappa\left(\lVert\bm{\nabla}\rho\rVert^2\bm{I}-\bm{\nabla}\rho\bm{\nabla}\rho\right).
\end{eqnarray}
Lee and Lin~\cite{LeeLin2005} showed that $p_{h}^{(L-L)}$ was an effective definition of the hydrodynamic pressure, which varies smoothly across the interfacial region compared to that of the thermodynamic pressure $p_0$. Dimensional analysis shows that the hydrodynamic pressure is of the order $\mathcal{O}(Ma^2)$ if normalized by $\rho U_c^2$, where $U_c$ is the characteristic flow velocity and $Ma$ is the characteristic Mach number.
$\bm{\Phi}$ is the surface tension stress tensor, which has three real eigenvalues and three mutually perpendicular principal axes. One of the eigenvalues is zero, whose corresponding principal axis is parallel to the unit normal vector $\bm{n}_{S}$. The other two eigenvalues are equal to $\kappa\lVert\bm{\nabla}\rho\rVert^2$, whose integral across the interface gives the surface tension coefficient $\sigma_{s}$. Therefore, the corresponding principal axes can be chosen as any two perpendicular vectors lying in the tangent plane of $S$.
Obviously, the work done by the surface tension stress tensor $\bm{\Phi}$ will not change the total momentum in a close volume.

Alternatively, the chemical potential gradient force $-\rho\bm{\nabla}\mu_{\rho}$ can also be decomposed as
\begin{eqnarray}\label{phnew}
-\rho\bm{\nabla}\mu_{\rho}=-\bm{\nabla}(\rho\mu_{\rho})+\mu_{\rho}\bm{\nabla}\rho\equiv-\bm{\nabla}p_{h}+\rho\bm{F}_{\nabla}.
\end{eqnarray}
where $p_h$ is defined as the hydrodynamic pressure in the present paper and $\rho\bm{F}_{\nabla}$ is the interfacial force per unit volume locally proportional to the density gradient and the chemical potential.
In fact, by using the flat surface solution (i.e., Eq.~\eqref{density_profile}) as the leading-order approximation for a curved interface, it can be shown that
\begin{eqnarray}\label{xx2}
p_{h}\approx p_{h}^{(L-L)},~
\rho\bm{F}_{\nabla}\approx\bm{\nabla}\bm{\cdot}\bm{\Phi}.
\end{eqnarray}
Therefore, Eq.~\eqref{phnew} is essentially consistent with the decomposition of Lee and Lin (see Eqs.~\eqref{PLLa} and~\eqref{PLLb}), which indicates the rationality of hydrodynamic pressure separation using Eq.~\eqref{phnew}. Detailed proof of Eq.~\eqref{xx2} is given in Appendix~\ref{ProofLeeLin}. It is worth mentioning that different formulations of interfacial forces mentioned here and their numerical performances are also discussed by Zhang {\it et al.}~\cite{ZhangCH2021} and Liu {\it et al.}~\cite{LiuHaihu2014} under the framework of phase-field-based lattice Boltzmann method. 

\section{Enstrophy fluxes for two-phase flow}\label{Enstrophy fluxes for two-phase flow}
In this section, we derive explicit decomposition of the boundary and interfacial enstrophy fluxes for two-phase viscous flow under the framework of diffuse interface model, where all physical mechanisms causing these fluxes are clearly elucidated from theoretical perspective. These results can be considered as a natural extension of Wu's BVF theory~\cite{WuJZ1995} for the sharp interface model.
It is worth pointing out that the liquid-vapor interface has no inner structure in the sharp interface model and therefore only the jump of physical quantities across the interface can be studied. In contrast, for the diffuse interface model, the liquid and vapor phases are separated by a smooth transition layer, which allows for the investigation of enstrophy flux across the liquid-vapor interface $S$ inside this layer.
These theoretical results will be further demonstrated and discussed in Section~\ref{simulation} by using the simulation data.
\subsection{Interfacial enstrophy flux}\label{IEF}
The interfacial enstrophy flux (IEF) is defined as
\begin{eqnarray}
\varphi_{\Omega}\equiv\mu_{S}\left[\frac{\partial\Omega}{\partial{n}}\right]_{S}=\mu_{S}\bm{n}_{S}\bm{\cdot}\left[\bm{\nabla}\Omega\right]_{S},
\end{eqnarray}
where $\bm{n}_{S}$ is the unit normal vector of the liquid-vapor interface $S$ (see Section~\ref{Notation_convention}). Physically, the IEF $\varphi_{\Omega}$ measures the enstrophy diffusion rate across the interface $S$ per unit area per unit time. Positive IEF implies that the enstrophy diffuses from the liquid to the vapor, causing an increase (decrease) of the enstrophy in the vapor (liquid) side, and vice versa.

The NSK equations introduced in Section~\ref{NSK_equations} are formulated based on the primitive variables including density, thermodynamic pressure and velocity. The most primary derived fields that describe the local spatial variation of a velocity field are the vorticity $\bm{\omega}$ and the dilatation $\vartheta$. The former describes the shearing process while the latter that measures the isotropic expansion and compression processes is not significant in the bulk fluid phases under the low-speed isothermal assumption. Compared to the single-phase viscous flow, more physical mechanisms arise due to the presence of the phase interface and its interaction with the generated vorticity. On the one hand, the observed coherent vortical structures near the boundary are well described by the vorticity, which are considered as the sinews and muscles of fluid motions. On the other hand, the surface vorticity is one of the fundamental surface physical quantities, which is evaluated through the velocity derivative instead of velocity itself. Surface vorticity is directly related to surface shear stress (skin friction), surface pressure and surface curvature. The boundary vorticity dynamics to be explored here aims to provide physically direct and technically accurate interpretation to vorticity creation and interaction with the boundary, so that the vorticity-based formulation is more appropriate for our analysis.

Since the hydrodynamic pressure $p_{h}$ has been separated from the chemical potential force in Section~\ref{separationph}, Eq.~\eqref{c2} can be rewritten in the vorticity-based form:
\begin{eqnarray}\label{NSK2}
\rho\bm{a}=-\bm{\nabla}\bar{p}_{h}-\mu\bm{\nabla}\times\bm{\omega}+\rho\bm{F}_{\nabla}+\rho\bm{F}_{\mu},
\end{eqnarray}
where $\bar{p}_{h}\equiv p_h-(2\mu+\lambda)\vartheta$ is the modified pressure including the dilatation correction. $\rho\bm{F}_{\nabla}=\mu_{\rho}\bm{\nabla}\rho$
is the interfacial  force due to the density gradient (as introduced in Eq.~\eqref{phnew}). $\rho\bm{F}_{\mu}=-\bm{\nabla}\mu\times\bm{\omega}-2\bm{\nabla}\mu\bm{\cdot}\bm{B}$ is the force caused by the viscosity gradient, where $\bm{B}\equiv\vartheta\bm{I}-\bm{\nabla}\bm{u}^{T}$ is the surface deformation tensor previously studied by Batchelor,~\cite{Batchelor1967} Dishington~\cite{Dishington1965} and Wu.~\cite{WuJZ1995}

Physically, the material derivative of a surface element $\delta\bm{S}=\delta{S}\bm{n}_{S}$ is $D\delta\bm{S}/Dt=\delta{\bm{S}}\bm{\cdot}\bm{B}_{S}$. Wu~\cite{WuJZ1995,WuJZ2005JFM} further proved that $\bm{n}_{S}\bm{\cdot}\bm{B}_{S}=\left(\bm{\nabla}_{S}\bm{\cdot}\bm{u}_{S}\right)\bm{n}_{S}+\bm{W}\times\bm{n}_{S}$. The tangential velocity divergence $\bm{\nabla}_{S}\bm{\cdot}\bm{u}_{S}$ is equal to the relative change rate of the area $\delta{S}$, namely, $(\delta S)^{-1}(D\delta{S}/Dt)=\bm{\nabla}_{S}\bm{\cdot}\bm{u}_{S}$. $\bm{W}=\bm{W}_{\pi}+W_{n}\bm{n}$ is the local angular velocity of $\bm{n}_{S}$ (namely, the material derivative of the unit normal vector is $D\bm{n}_{S}/Dt=\bm{W}\times\bm{n}_{S}$),
 whose tangential and normal components are respectively given by
\begin{eqnarray}\label{Wpi}
	\bm{W}_{\pi}=-\bm{n}_{S}\times\left(\bm{\nabla}_{S}[u_n]_{S}+\bm{K}_{S}\bm{\cdot}\bm{u}_{S}\right),
\end{eqnarray}
\begin{eqnarray}\label{Wn}
	W_{n}=\frac{1}{2}\bm{n}_{S}\bm{\cdot}\left(\bm{\nabla}_{S}\times\bm{u}_{S}\right)=\frac{1}{2}\left[\omega_{n}\right]_{S}.
\end{eqnarray}
It is clear that $\bm{W}$ is solely determined by the geometry and motion of the surface, independent of the flow off the surface. $\bm{W}_{\pi}$ is determined by nonuniform normal surface motion, and the coupling between surface curvature and surface velocity. $W_{n}$ is equal to half of the wall-normal vorticity component on $S$. 

By applying Eq.~\eqref{NSK2} to the liquid-vapor interface $S$, we obtain
\begin{eqnarray}\label{xx4}
\bm{n}_{S}\times\rho_{S}\bm{a}_{S}
+\bm{n}_{S}\times\bm{\nabla}_{S}\left[\bar{p}_{h}\right]_{S}
-\bm{n}_{S}\times\rho_{S}\left[\bm{F}_{\nabla}\right]_{S}
-\bm{n}_{S}\times\rho_{S}\left[\bm{F}_{\mu}\right]_{S}
=-\mu_{S}\bm{n}_{S}\times\left[\bm{\nabla}\times\bm{\omega}\right]_{S},
\end{eqnarray}
where the term in right hand side can be further evaluated as
\begin{eqnarray}\label{xx5}
-\mu_{S}\bm{n}_{S}\times\left[\bm{\nabla}\times\bm{\omega}\right]_{S}
=-\mu_{S}\bm{\nabla}_{S}\left[\omega_{n}\right]_{S}
-\mu_{S}\bm{K}_{S}\bm{\cdot}\bm{\omega}_{S}
+\mu_{S}\left(\bm{\nabla}_{S}\bm{\cdot}\bm{\omega}_{S}\right)\bm{n}_{S}
+\mu_{S}\left[\frac{\partial\bm{\omega}}{\partial n}\right]_{S}.
\end{eqnarray}

By applying the Caswell-Wu decomposition~\cite{WuJZ2005JFM} to the strain rate tensor $\bm{S}$, the surface shear stress $\bm{\tau}_{S}\equiv\bm{n}_{S}\bm{\cdot}\bm{\Pi}_{S}\bm{\cdot}\left(\bm{I}-\bm{n}_{S}\bm{n}_{S}\right)$ can be expressed as
\begin{eqnarray}\label{Skin_friction}
\bm{\tau}_{S}=\mu_{S}\bm{\omega}_{r}\times\bm{n}_{S}
\equiv\bm{\tau}_{\bm{\omega}}+\bm{\tau}_{_{\bm{W}}},
\end{eqnarray}
where $\bm{\omega}_{r}\equiv\bm{\omega}_{S}-2\bm{W}$ is the relative vorticity, and $-2\bm{W}$ is the additional vorticity in view of its contribution to $\bm{\tau}_{S}$. 
It follows from Eq.~\eqref{Wn} that $\bm{\omega}_{r}$ must be tangential to the interface $S$ and orthogonal to  $\bm{\tau}_{S}$. $\bm{\tau}_{\bm{\omega}}\equiv\mu_{S}\bm{\omega}_{S}\times\bm{n}_{S}$ is the vorticity-induced surface shear stress. $\bm{\tau}_{_{\bm{W}}}\equiv-2\mu_{S}\bm{W}\times\bm{n}_{S}=-2\mu_{S}(D\bm{n}_{S}/Dt)$ is caused by the surface angular velocity $\bm{W}$, which is called the surface deformation shear stress.~\cite{WuJZ1995}

By taking a dot product of both sides of Eq.~\eqref{xx4} with the surface vorticity $\bm{\omega}_{S}$ and using Eq.~\eqref{xx5}, we obtain the decomposition of $\varphi_{\Omega}$ as
\begin{eqnarray}\label{pp0}
\varphi_{\Omega}=\varphi_{\Omega}^{(1)}+\varphi_{\Omega}^{(2)}+\varphi_{\Omega}^{(3)}+\varphi_{\Omega}^{(4)}+\varphi_{\Omega}^{(5)},
\end{eqnarray}
where the five terms in the right hand side are respectively
\begin{subequations}\label{pp}
\begin{eqnarray}\label{phia}
\varphi_{\Omega}^{(1)}=\frac{1}{\mu_{S}}\bm{\tau}_{\bm{\omega}}\bm{\cdot}\bm{\nabla}_{S}\left[\bar{p}_{h}\right]_{S},
\end{eqnarray}
\begin{eqnarray}\label{phib}
\varphi_{\Omega}^{(2)}=\frac{1}{\mu_{S}}\bm{\tau}_{\bm{\omega}}\bm{\cdot}\rho_{S}\bm{a}_{S},
\end{eqnarray}
\begin{eqnarray}\label{phic}
\varphi_{\Omega}^{(3)}=-\frac{1}{\mu_{S}}\bm{\tau}_{\bm{\omega}}\bm{\cdot}\rho_{S}\left[\bm{F}_{\mu}\right]_{S},
\end{eqnarray}
\begin{eqnarray}\label{phid}
\varphi_{\Omega}^{(4)}=-\frac{1}{\mu_{S}}\bm{\tau}_{\bm{\omega}}\bm{\cdot}\rho_{S}\left[\bm{F}_{\nabla}\right]_{S},
\end{eqnarray}
\begin{eqnarray}\label{phie}
\varphi_{\Omega}^{(5)}=\mu_{S}\bm{\omega}_{S}\bm{\cdot}\bm{\nabla}_{S}\left[\omega_{n}\right]_{S}+\mu_{S}\bm{\omega}_{S}\bm{\cdot}\bm{K}_{S}\bm{\cdot}\bm{\omega}_{S}-\mu_{S}\left(\bm{\nabla}_{S}\bm{\cdot}\bm{\omega}_{S}\right)\left[\omega_{n}\right]_{S}.
\end{eqnarray}
\end{subequations}

In Eq.~\eqref{pp},
$\varphi_{\Omega}^{(1)}$, $\varphi_{\Omega}^{(2)}$ and $\varphi_{\Omega}^{(3)}$ represent different contributions to the total IEF $\varphi_{\Omega}$, which result from the surface pressure gradient, the surface acceleration and the force due to the viscosity gradient coupled with the vorticity-induced surface shear stress.
$\varphi_{\Omega}^{(4)}$ generally does not vanish for a given material surface whose unit normal vector is not parallel to the density gradient. Since the interface $S$ is usually defined as an isosurface of the density field at a given time instant, we have $\varphi_{\Omega}^{(4)}=0$ for this case.
The viscous term  $\varphi_{\Omega}^{(5)}$ only exists in the 3D viscous flow, which is contributed by the non-vanishing surface-normal vorticity distribution and vorticity-curvature interaction.

Supposing that the dynamic viscosity $\mu$ is a function of the density $\rho$, Eq.~\eqref{phic} can be evaluated as
\begin{eqnarray}\label{phic1}
\varphi_{\Omega}^{(3)}=-\left[\frac{d\mu}{d\rho}\right]_{S}\lVert\bm{\nabla}\rho\rVert_{S}\bm{\omega}_{S}\bm{\cdot}\bm{\omega}_{r},
\end{eqnarray}
or equivalently 
\begin{eqnarray}\label{phic2}
\varphi_{\Omega}^{(3)}
=-\frac{1}{\mu_{S}^2}\left[\frac{d\mu}{d\rho}\right]_{S}\lVert\bm{\nabla}\rho\rVert_{S}\bm{\tau}_{\bm{\omega}}\bm{\cdot}\bm{\tau}_{S}.
\end{eqnarray}
Eqs.~\eqref{phic1} and~\eqref{phic2} indicate that $\varphi_{\Omega}^{(3)}$ is influenced by the coupling among surface vorticity $\bm{\omega}_{S}$, surface angular velocity $\bm{W}$ (or vorticity-induced surface shear stress $\bm{\tau}_{\bm{\omega}}$ and surface deformation shear stress $\bm{\tau}_{\bm{W}}$), and the magnitude of density gradient $\lVert\bm{\nabla}\rho\rVert_{S}$ on $S$.

\subsection{Boundary enstrophy flux}\label{BEF2}
The boundary enstrophy flux (BEF) suggested by Wu~\cite{WuJZ1995,WuJZ2015book} and Liu {\it et al.}~\cite{Liu2016MST,Liu2018AIA} measures the vorticity diffusion rate across the solid wall, which is defined as
\begin{eqnarray}
f_{\Omega}\equiv\mu_{\partial B}\left[\frac{\partial\Omega}{\partial{n}}\right]_{\partial B}=\mu_{\partial B}\bm{n}_{\partial B}\bm{\cdot}\left[\bm{\nabla}\Omega\right]_{\partial B},
\end{eqnarray}
where $\bm{n}_{\partial B}$ is the unit normal vector of the wall, pointing from the wall to the fluid.
For single-phase incompressible viscous flow past a stationary wall,
BEF is shown as an intriguing quantity which is particularly related to the distinct topological features in complex viscous flows including separation and attachment lines, isolated critical points in a skin friction field.~\cite{Liu2018AIA,ChenTao2019POF,ChenTao2021WEF} For turbulent flows, the BEF is also related to the high intermittency of the viscous sublayer.~\cite{ChenTao2021WEF,ChenTao2021POF}

However, after careful literature review, we see that the concept of BEF is not yet generalized to two-phase viscous flow with diffuse interface.
Compared to single-phase flow, more physical contributions to the BEF could arise due to the presence of the diffuse interface, in particular to the region near the three-phase contact points.
Similar to the discussion in Section~\ref{IEF}, for a stationary flat wall, the BEF is decomposed as
\begin{eqnarray}\label{BEF}
	f_{\Omega}\equiv f_{\Omega}^{(1)}+f_{\Omega}^{(2)}+f_{\Omega}^{(3)}+f_{\Omega}^{(4)},
\end{eqnarray}
where the terms in the right hand side of Eq.~\eqref{BEF} are, respectively,
\begin{subequations}\label{BEF_decompose}
\begin{eqnarray}\label{fomega1}
f_{\Omega}^{(1)}=\frac{1}{\mu_{\partial B}}\bm{\tau}\bm{\cdot}\bm{\nabla}_{\partial B}\left[\bar{p}_{h}\right]_{\partial B},
\end{eqnarray}
	\begin{eqnarray}\label{fomega2}
	f_{\Omega}^{(2)}=-\frac{1}{\mu_{\partial B}}\bm{\tau}\bm{\cdot}\left[\rho\bm{F}_{\mu}\right]_{\partial B}=-2\Omega_{\partial B}\left[\frac{\partial\mu}{\partial n}\right]_{\partial B}+\frac{2}{\mu}\vartheta_{\partial B}\bm{\tau}\bm{\cdot}\bm{\nabla}_{\partial B}\mu_{\partial B},
	\end{eqnarray}
	\begin{eqnarray}\label{fomega3}
	f_{\Omega}^{(3)}=-\frac{1}{\mu_{\partial B}}\bm{\tau}\bm{\cdot}\left[\rho\bm{F}_{\nabla}\right]_{\partial B}=-\frac{[\mu_{\rho}]_{\partial B}}{\mu_{\partial B}}\bm{\tau}\bm{\cdot}\bm{\nabla}_{\partial B}\rho_{\partial B},
	\end{eqnarray}
	\begin{eqnarray}\label{fomega4}
f_{\Omega}^{(4)}=\mu_{\partial B}\bm{\omega}_{\partial B}\bm{\cdot}\bm{K}\bm{\cdot}\bm{\omega}_{\partial B}.
	\end{eqnarray}
\end{subequations}

In Eq.~\eqref{BEF_decompose}, $f_{\Omega}^{(1)}$ is caused by the viscous coupling between skin friction (namely, the wall shear stress $\bm{\tau}=\mu_{\partial B}\bm{\omega}_{\partial B}\times\bm{n}_{\partial B}$) and surface pressure gradient. $f_{\Omega}^{(2)}$ is caused by the coupling between surface vorticity, surface dilatation and viscosity gradient. $f_{\Omega}^{(3)}$ is due to the coupling between skin friction and surface density gradient. It is expected that $f_{\Omega}^{(2)}$ and $f_{\Omega}^{(3)}$ should concentrate in the region near the three-phase moving contact line (point). $f_{\Omega}^{(4)}$ is formally interpreted as the viscous interaction between surface vorticity and surface curvature.

\section{Numerical model and method}\label{Numerical model and method}
The above exact relations for BEF and IEF are applicable for any two-phase viscous flow.
In order to demonstrate its practical application, a well-balanced discrete unified gas kinetic scheme (WB-DUGKS) most recently proposed by Zeng {\it et al.}~\cite{Zengwei2022} is adopted for the present simulation, which is an extension of Guo's well-balanced lattice Boltzmann model (WB-LBM).~\cite{GuoZhaoli2021}
Their numerical tests show that the spurious velocity for a stationary droplet can be eliminated completely with this model, which also exhibits superior ability to capture moving surface.
Since the details of WB-DUGKS and its validation have been well presented in the published paper for both stationary and moving surfaces,~\cite{Zengwei2022} we briefly summarize the main contents of the model and discuss some new points as follows.

In the WB-DUGKS, the following discrete Boltzmann model is solved:
\begin{eqnarray}
\frac{\partial f_{\alpha}}{\partial t}+\bm{\xi}_{\alpha}\bm{\cdot}\bm{\nabla}f_{\alpha}=\frac{f_{\alpha}^{eq}-f_{\alpha}}{\tau}+F_{\alpha}, \alpha=0,\cdots,Q-1,
\end{eqnarray}
where $f_{\alpha}(\bm{x},\bm{\xi}_{\alpha},t)$ is the particle distribution function with the discrete particle velocity $\bm{\xi}_{\alpha}$ at the spatial location $\bm{x}$ and time $t$. The subscript $\alpha$ denotes the discrete particle velocity direction and $Q$ is the total number of discrete particle velocities used.
$\tau$ is the relaxation time, which is related to the kinematic viscosity $\nu$ through the relation $\nu=c_{s}^{2}\tau$. $c_{s}=\sqrt{RT}$ is the pseudo speed of sound, $R$ is the gas constant and $T$ is the temperature.

The equilibrium distribution function $f_{\alpha}^{eq}$ is designed as~\cite{GuoZhaoli2021,Zengwei2022}
\begin{equation}
f_{\alpha}^{eq} = \left\{
\begin{aligned}
&\rho-(1-\omega_{0})\rho_{0}+\omega_{0}\rho s_{0}(\bm{u}), &\alpha=0 \\[2pt]
&\omega_{\alpha}[\rho_{0}+\rho s_{\alpha}(\bm{u})],&\alpha\neq{0},
\end{aligned} \right.
\end{equation}
where
$\rho_{0}$ is a numerical constant, $\omega_{\alpha}$ ($\alpha=0,\cdots,Q-1$) are the weighting factors, $s_{\alpha}(\bm{u})$ is given by
\begin{eqnarray}
s_{\alpha}(\bm{u})=\frac{\bm{\xi}_{\alpha}\bm{\cdot}\bm{u}}{c_{s}^{2}}+\frac{(\bm{\xi}_{\alpha}\bm{\cdot}\bm{u})^2}{2c_{s}^{4}}-\frac{\bm{u}\bm{\cdot}\bm{u}}{2c_{s}^{2}}.
\end{eqnarray}
 The choice of $\rho_{0}$ may influence the stability of the numerical scheme by ensuring the positivity of the equilibrium, but does not influence the results. Following Zeng {\it et al.},~\cite{Zengwei2022} we set $\rho_{0}=0$ in the simulations. The well-known D2Q9 discrete particle velocity model is adopted ($c_s^2=1/3$, $\omega_{0}=4/9$, $\omega_{1,2,3,4}=1/9$ and $\omega_{5,6,7,8}=1/36$).
In addition, the mesoscopic forcing term $F_{\alpha}$ is designed as~\cite{GuoZhaoli2021,Zengwei2022}
\begin{eqnarray}
F_{\alpha}=\omega_{\alpha}\left[
\begin{aligned}
&\frac{\bm{\xi}_{\alpha}\bm{\cdot}(-\rho\bm{\nabla}\mu_{\rho})}{c_s^2}+\frac{\bm{u}(-\rho\bm{\nabla}\mu_{\rho}+c_s^2\bm{\nabla}\rho)\bm{:}(\bm{\xi}_{\alpha}\bm{\xi}_{\alpha}-c_s^2\bm{I})}{c_s^4}\\
&+\frac{1}{2}\left(\frac{\bm{\xi}_{\alpha}^{2}}{c_s^2}-D\right)\left(\bm{u}\bm{\cdot}\bm{\nabla}\rho\right)
\end{aligned}
\right],
\end{eqnarray}
Compared to the Maxwellian distribution function, the second-order moment of this new equilibrium distribution $\sum_{\alpha=0}^{Q-1}\bm{\xi}_{\alpha}\bm{\xi}_{\alpha}f_{\alpha}^{eq}$ is equal to $\rho\bm{uu}$ instead of $c_{s}^{2}\rho\bm{I}+\rho\bm{uu}$. Therefore, the use of force $\bm{\nabla}(c_s^2\rho)$ in the force term of the standard Lattice Boltzmann equation (LBE) is circumvented.

The fluid density and velocity are updated through the particle velocity moments of the discrete distribution functions, namely,
\begin{eqnarray}
\rho=\sum_{\alpha=0}^{Q-1}f_{\alpha},~\rho\bm{u}=\sum_{\alpha=0}^{Q-1}\bm{\xi}_{\alpha}f_{\alpha}.
\end{eqnarray}
Through the Chapman-Enskog analysis,~\cite{Chapman1970} Eqs.~\eqref{c1} and~\eqref{c2} can be recovered from the present model at the continuum limit, but the resulting viscous stress tensor has a fixed bulk-to-shear viscosity ratio $\mu_{V}/\mu=(D+2)/D$. Although the original WB-DUGKS is adopted in our simulation, it is worth mentioning that, by unifying the Chapman-Enskog expansion and the Hermite expansion, this restriction on the ratio of bulk to shear viscosity can be removed by adding another source term $\omega_{\alpha}\frac{1}{2}\left(\frac{D+2}{D}-\frac{\mu_V}{\mu}\right)\rho \bm{\nabla}\bm{\cdot}\bm{u}\left(\frac{\bm{\xi}_{\alpha}^{2}}{c_s^2}-D\right)$ to the mesoscopic forcing term $F_{\alpha}$. 

\section{Numerical simulation and analysis}\label{simulation}
\subsection{Description of physical problem}\label{Description of physical problem}
In this section, we simulate a droplet impact on a hydrophilic solid wall with prescribed contact angle $\theta=\pi/3$ to demonstrate the application of the derived results. The reason for such choice is that this contact-line problem involves not only the moving and deforming liquid-vapor interface $S$, but also the hydrodynamic interaction between the two fluids and the solid wall $\partial{B}$. Therefore, both BEF and IEF can be studied. In particular, physical features of BEF and IEF near the moving contact lines can be observed during droplet impact on the wall.

\begin{figure}[h!]
	\centering
	\includegraphics[width=0.9\columnwidth,trim={0.1cm 6.1cm 0.1cm 6.1cm},clip]{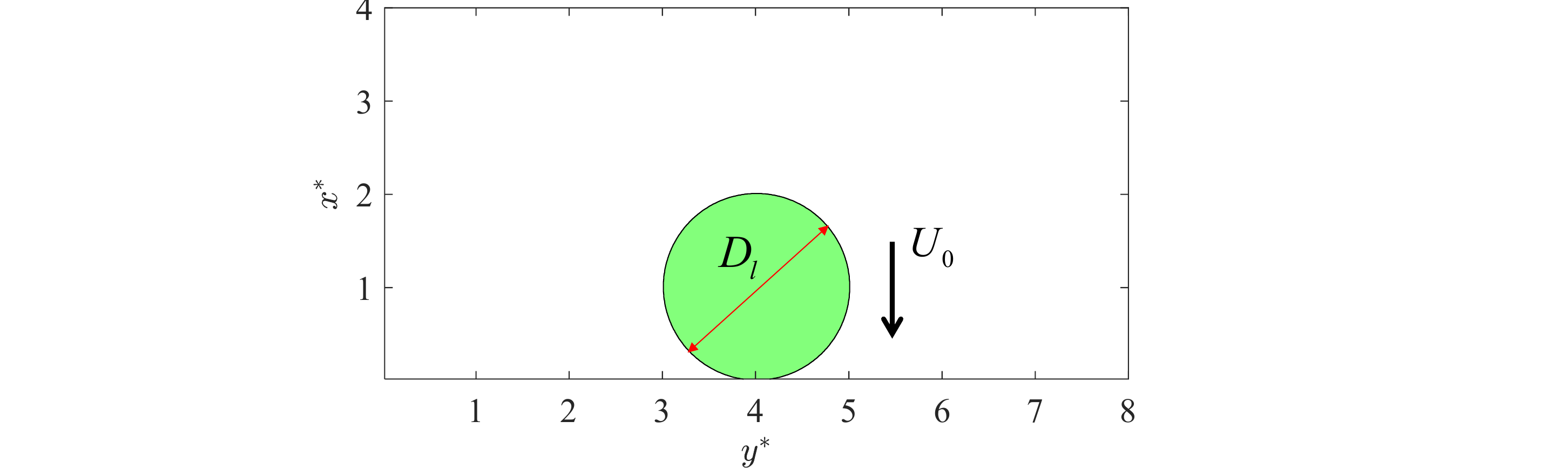}
	\caption{The initial state of droplet impact on the solid wall. $D_{l}$ represents the droplet diameter and $U_{0}$ is the initial speed of the droplet.} 
	\label{initial_state1}
\end{figure}

Initially, a liquid droplet with the diameter $D_{l}$ (the radius $R_{l}=D_{l}/2$) and the speed $U_{0}$ is surrounded by the ambient vapor, as shown in Fig.~\ref{initial_state1}. The whole computational domain is  $L_{x}\times L_{y}=200\Delta{x}\times400\Delta{x}$, and the initial droplet diameter $D_{l}$ is equal to 100 $\Delta{x}$, where $\Delta{x}$ is the grid spacing. The droplet center is initially located at $(x_{0},y_{0})=(D_{l}/2, 2D_{l})$.
Periodic boundary conditions are applied in the horizontal $y$-direction. For the vertical $x$-direction, the outflow boundary condition is applied for the top boundary, while 
the on-wall bounce back scheme is imposed at the solid bottom wall.
The geometric wetting boundary condition proposed by Ding and Spelt~\cite{Ding2007} is used to describe the wettability of ideal solid wall $\partial{B}$ with prescribed contact angle $\theta$, namely,
\begin{eqnarray}\label{geobc}
\bm{n}_{\partial B}\bm{\cdot}\bm{\nabla}\rho=-\tan\left(\frac{\pi}{2}-\theta\right)
\lVert\bm{\nabla}\rho-(\bm{n}_{\partial B}\bm{\cdot}\bm{\nabla}\rho)\bm{n}_{\partial B}\rVert.
\end{eqnarray}
Eq.~\eqref{geobc} relates the wall-normal derivative of density at the wall with its tangential derivative, which actually provides a Neumann boundary condition for WB-DUGKS. It is noted that combined with the Allen-Cahn (A-C) equation, the geometric wetting boundary condition has been successfully employed by Liang {\it et al.}~\cite{LiangHong2019} to simulate droplet impact dynamics with high density ratio. The code validation of the boundary condition in Eq.~\eqref{geobc} with the WB-DUGKS is shown in Appendix~\ref{Code validation} for different prescribed contact angles.

Since the double-well form of the bulk free energy density may cease to be valid away from the critical point of the equation of state,~\cite{LeeTaehun2006} the near-critical fluids with small density ratio $\rho_{l}/\rho_{g}=5$ is considered. Constant relaxation time $\tau$ is used, which implies that the kinematic viscosity ratio is $\nu_{l}/\nu_{g}=1$. Therefore, the dynamic viscosity ratio is $\mu_{l}/\mu_{g}=5$. In addition, the system is also depicted by the Weber number $We$ and the Reynolds number $Re$, namely,
\begin{eqnarray}
We=\frac{\rho_{l}U_{0}^{2}D_{l}}{\sigma_{s}},~Re=\frac{\rho_{l}U_{0}D_{l}}{\mu_{l}}.
\end{eqnarray}
In the present simulation, we set $We=50$ and $Re=85.7$.

The density and velocity fields are respectively initialized as
\begin{eqnarray}
\rho=\rho_{l}\phi+\rho_{g}(1-\phi),~u_{x}=-U_0\phi,~u_{y}=0,
\end{eqnarray}
where the function $\phi$ is given by
\begin{eqnarray}
\phi=\frac{1}{2}\left[1+\tanh\left(\frac{2\left(R_{l}-\sqrt{(x-x_{0})^2+(y-y_{0})^2}\right)}{W}\right)\right].
\end{eqnarray}
The ratio of the interfacial thickness parameter $W$ to the droplet diameter $D_{l}$ is $W/D_{l}=0.05$ (or $W=5\Delta{x}$). The ratio of the time step $\Delta{t}$ to the relaxation times $\tau$ is $dt/\tau\approx0.86$. In addition, the second-order isotropic central schemes are adopted for the evaluation of $\bm{\nabla}\phi$ and $\nabla^2\phi$ in the whole domain, where $\phi$ is an arbitrary scalar field.~\cite{LiangHong2019,Zengwei2022} 

We ran the WB-DUGKS code for 50000 steps at the resolution $N_{x}\times N_{y}=200\times400$ on the Taiyi cluster at the Southern University of Science and Technology. Our DUGKS code, parallelized using the one-dimensional domain decomposition strategy for two-dimensional problem, was tested with 80 cores. The total central processing unit (CPU) time is 21390.28 s and the wall clock time is 394 s.
 
\subsection{Analysis of boundary enstrophy flux}
\subsubsection{Evolution of droplet morphology}

\begin{figure}[h!]
	\centering
	\subfigure[$t^*=0$]{
		\begin{minipage}[t]{0.33\linewidth}
			\centering
			\includegraphics[width=1.0\columnwidth,trim={1cm 1.8cm 1cm 2.6cm},clip]{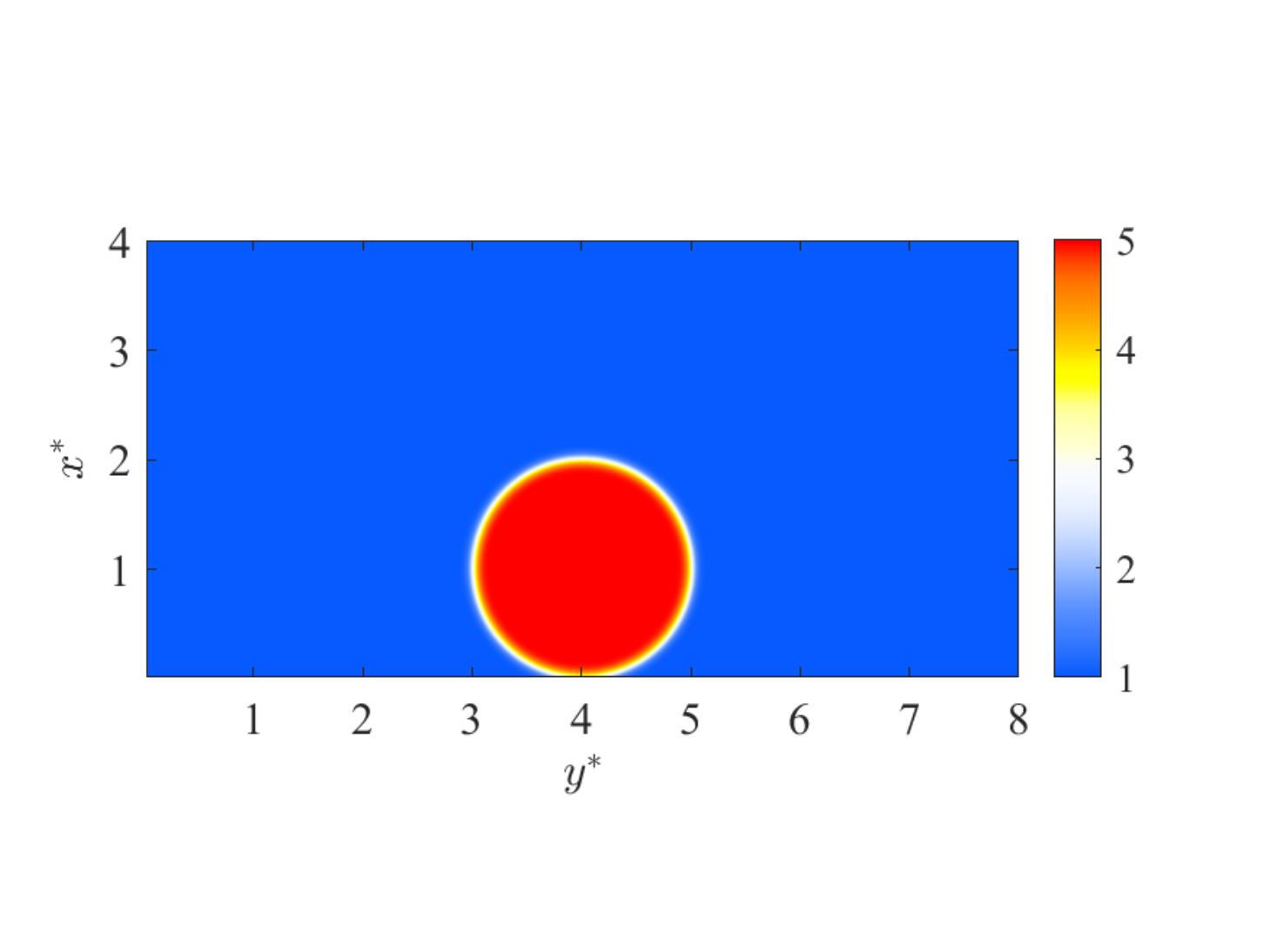}
			\label{Time_step_0}
		\end{minipage}%
	}%
	\subfigure[$t^*=0.3$]{
		\begin{minipage}[t]{0.33\linewidth}
			\centering
			\includegraphics[width=1.0\columnwidth,trim={1cm 1.8cm 1cm 2.0cm},clip]{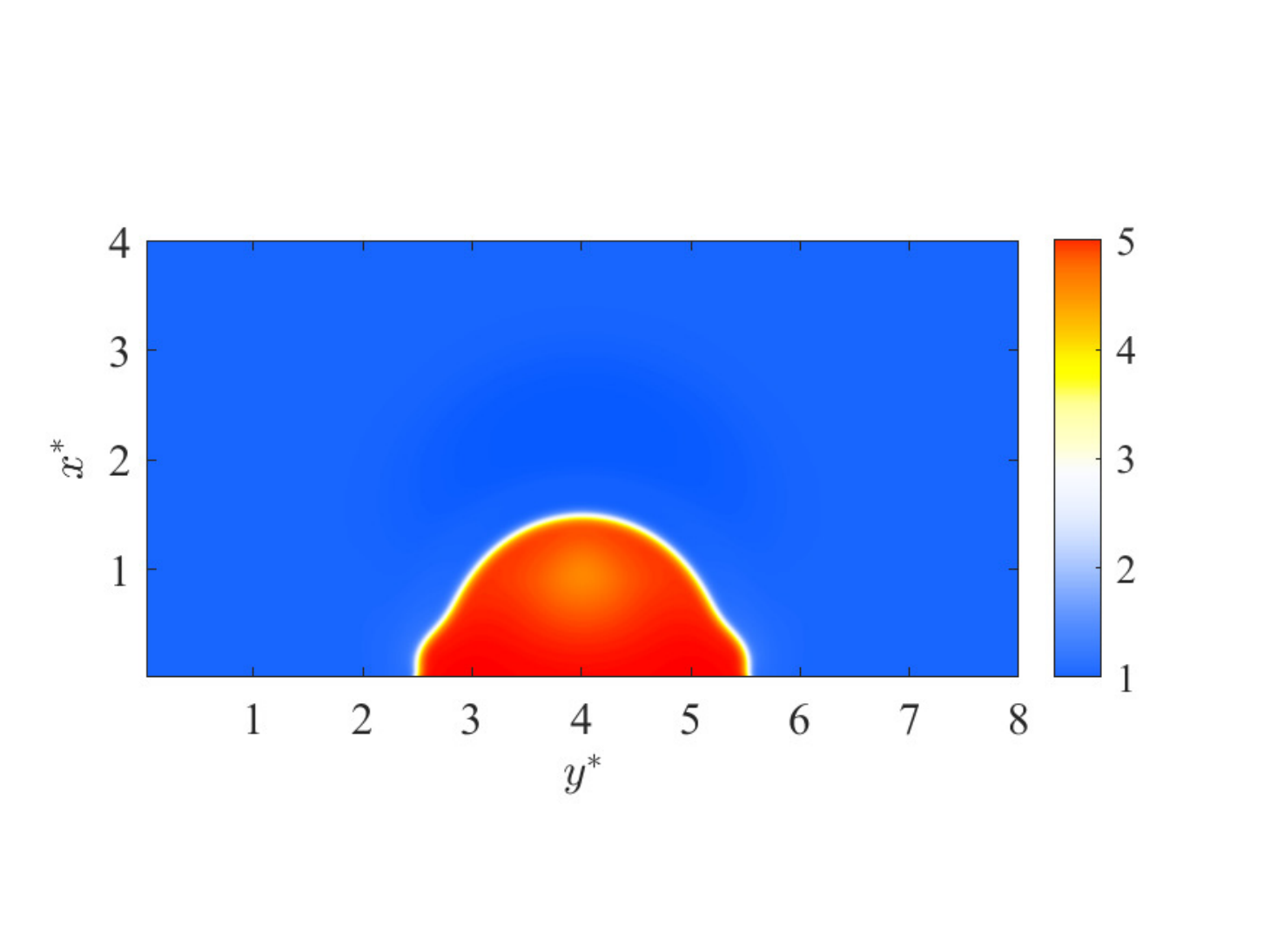}
			\label{Time_step_1000}
		\end{minipage}%
	}%
	\subfigure[$t^*=0.6$]{
		\begin{minipage}[t]{0.33\linewidth}
			\centering
			\includegraphics[width=1.0\columnwidth,trim={1cm 1.8cm 1cm 2.0cm},clip]{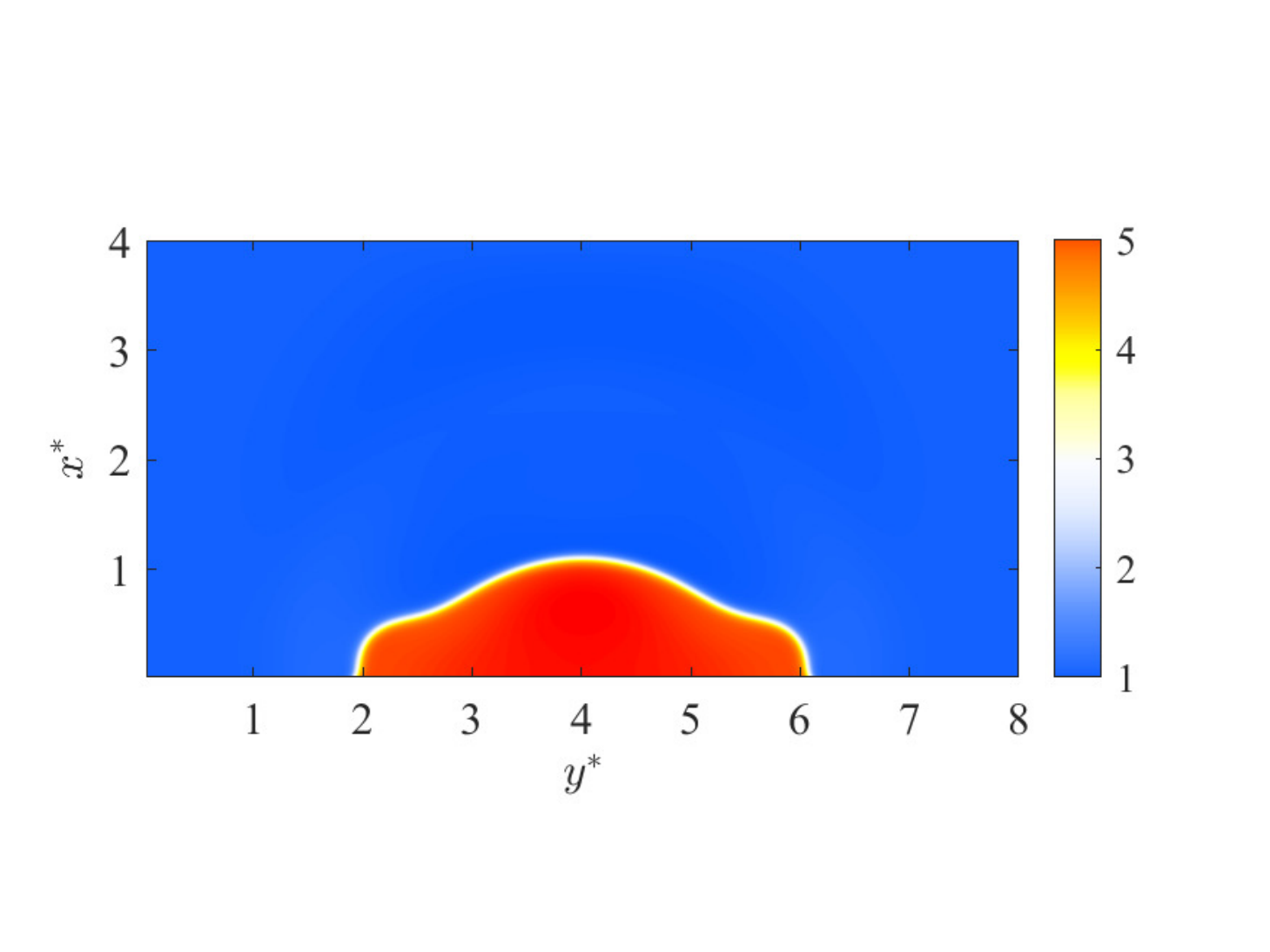}
			\label{Time_step_2000}
		\end{minipage}%
	}%

\subfigure[$t^*=1.2$]{
	\begin{minipage}[t]{0.33\linewidth}
		\centering
		\includegraphics[width=1.0\columnwidth,trim={1cm 1.8cm 1cm 2.0cm},clip]{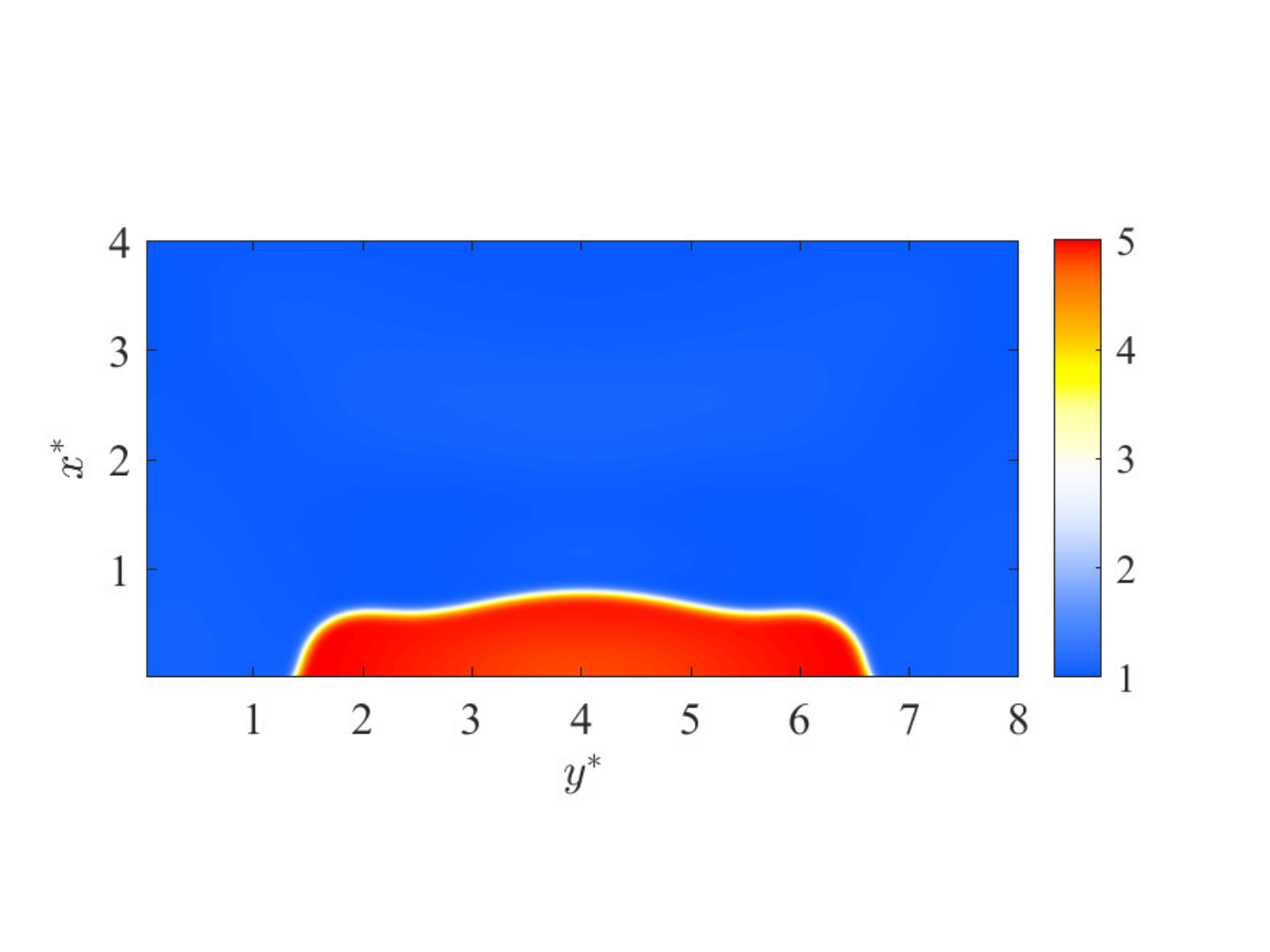}
		\label{Time_step_4000}
	\end{minipage}%
}%
\subfigure[$t^*=1.8$]{
	\begin{minipage}[t]{0.33\linewidth}
		\centering
		\includegraphics[width=1.0\columnwidth,trim={1cm 1.8cm 1cm 2.0cm},clip]{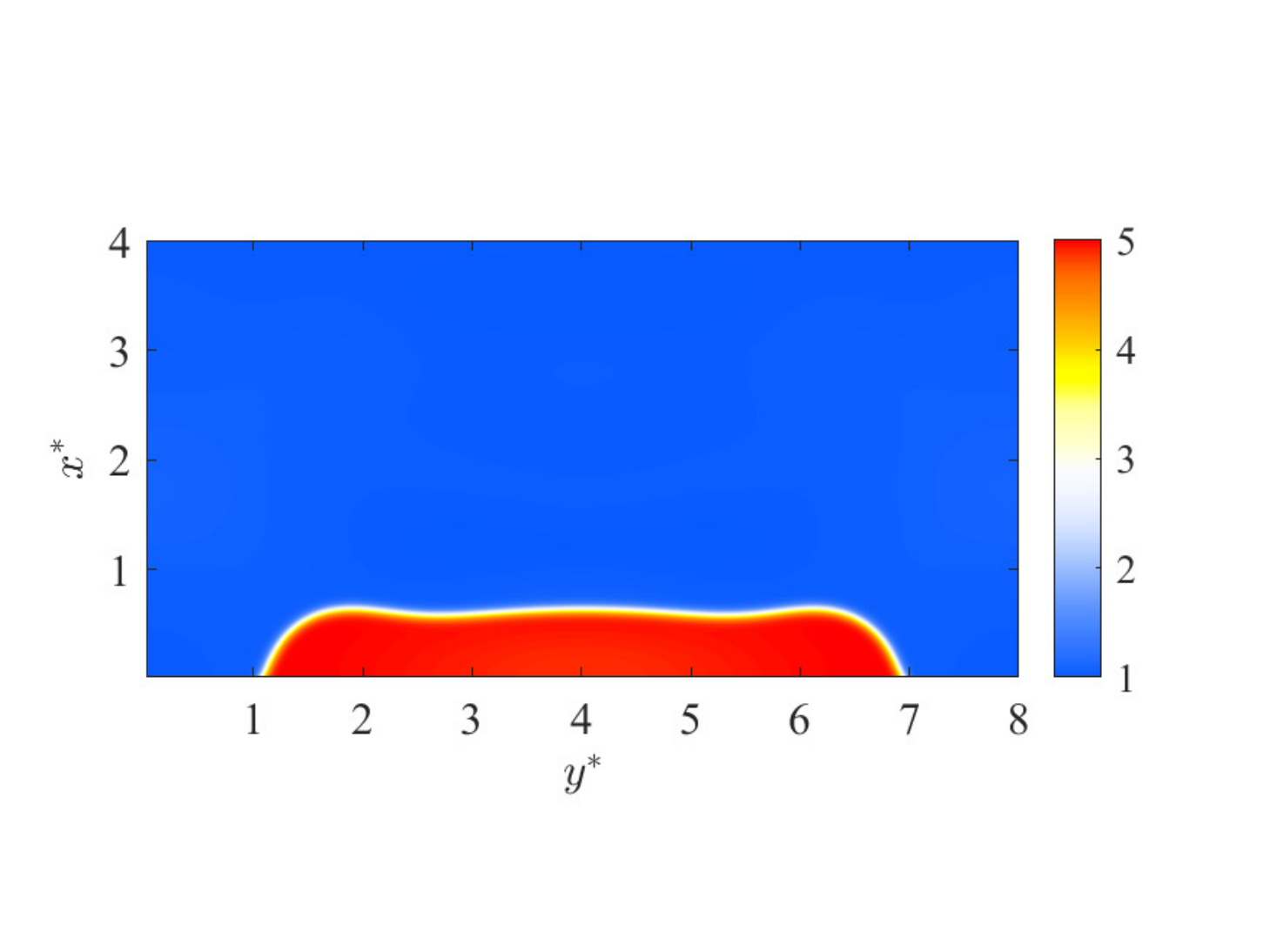}
		\label{Time_step_6000}
	\end{minipage}%
}%
	\subfigure[$t^*=4.8$]{
		\begin{minipage}[t]{0.33\linewidth}
			\centering
			\includegraphics[width=1.0\columnwidth,trim={1cm 1.8cm 1cm 2.0cm},clip]{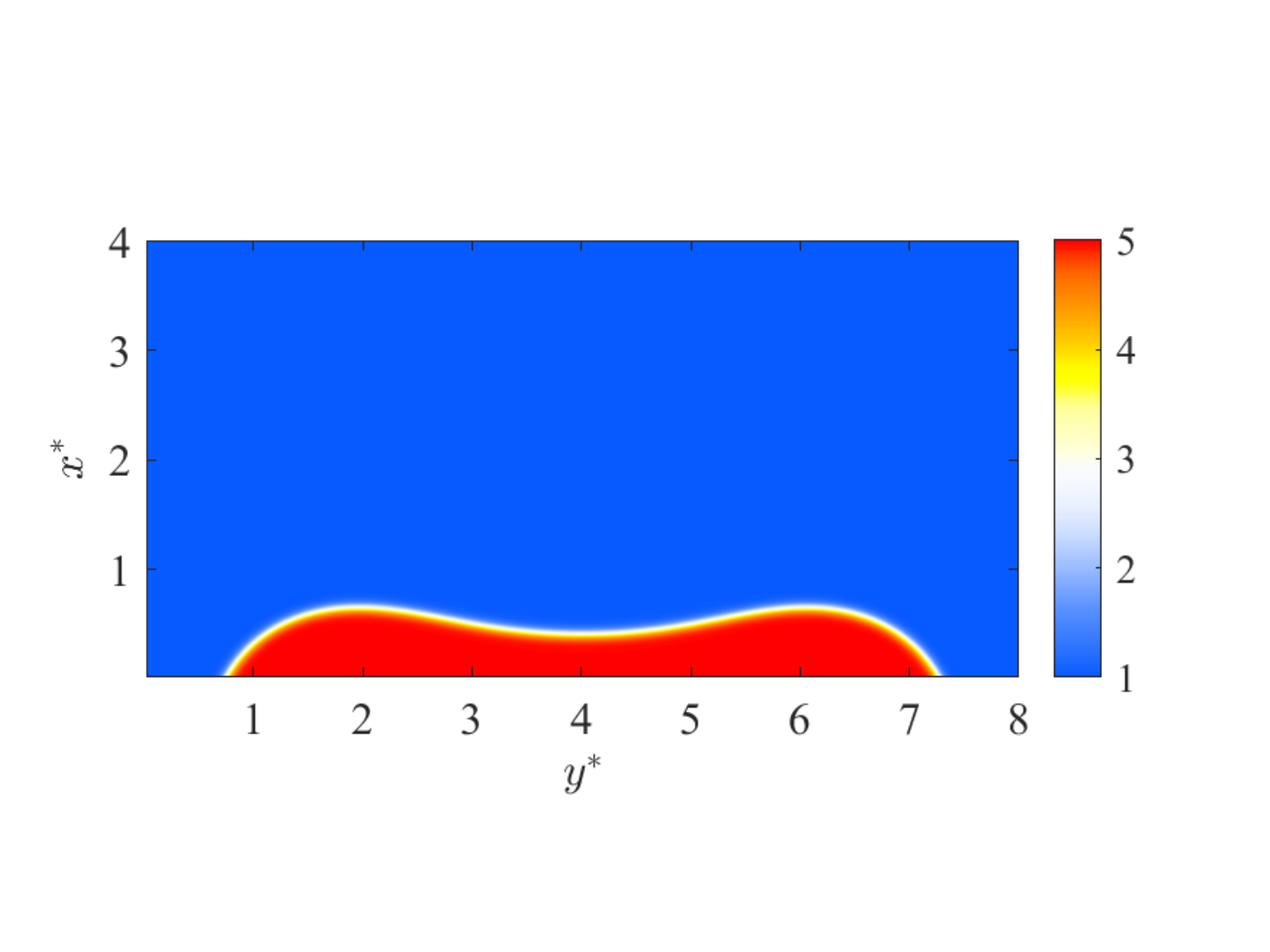}
			\label{Time_step_16000}
		\end{minipage}%
	}%

	\subfigure[$t^*=8.4$]{
	\begin{minipage}[t]{0.33\linewidth}
		\centering
		\includegraphics[width=1.0\columnwidth,trim={1cm 1.8cm 1cm 2.0cm},clip]{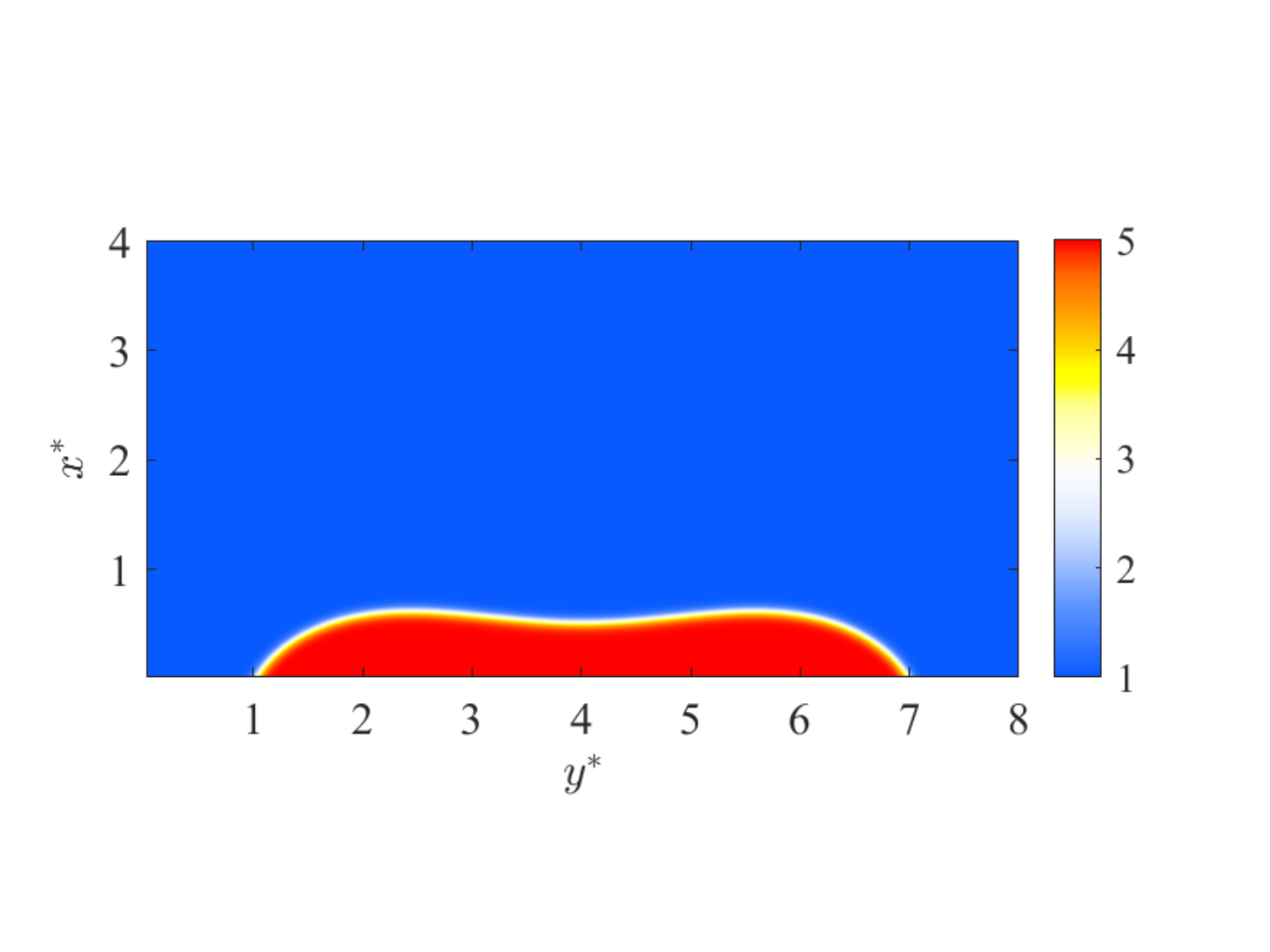}
		\label{Time_step_28000}
	\end{minipage}%
}%
	\subfigure[$t^*=11.4$]{
		\begin{minipage}[t]{0.33\linewidth}
			\centering
			\includegraphics[width=1.0\columnwidth,trim={1cm 1.8cm 1cm 2.0cm},clip]{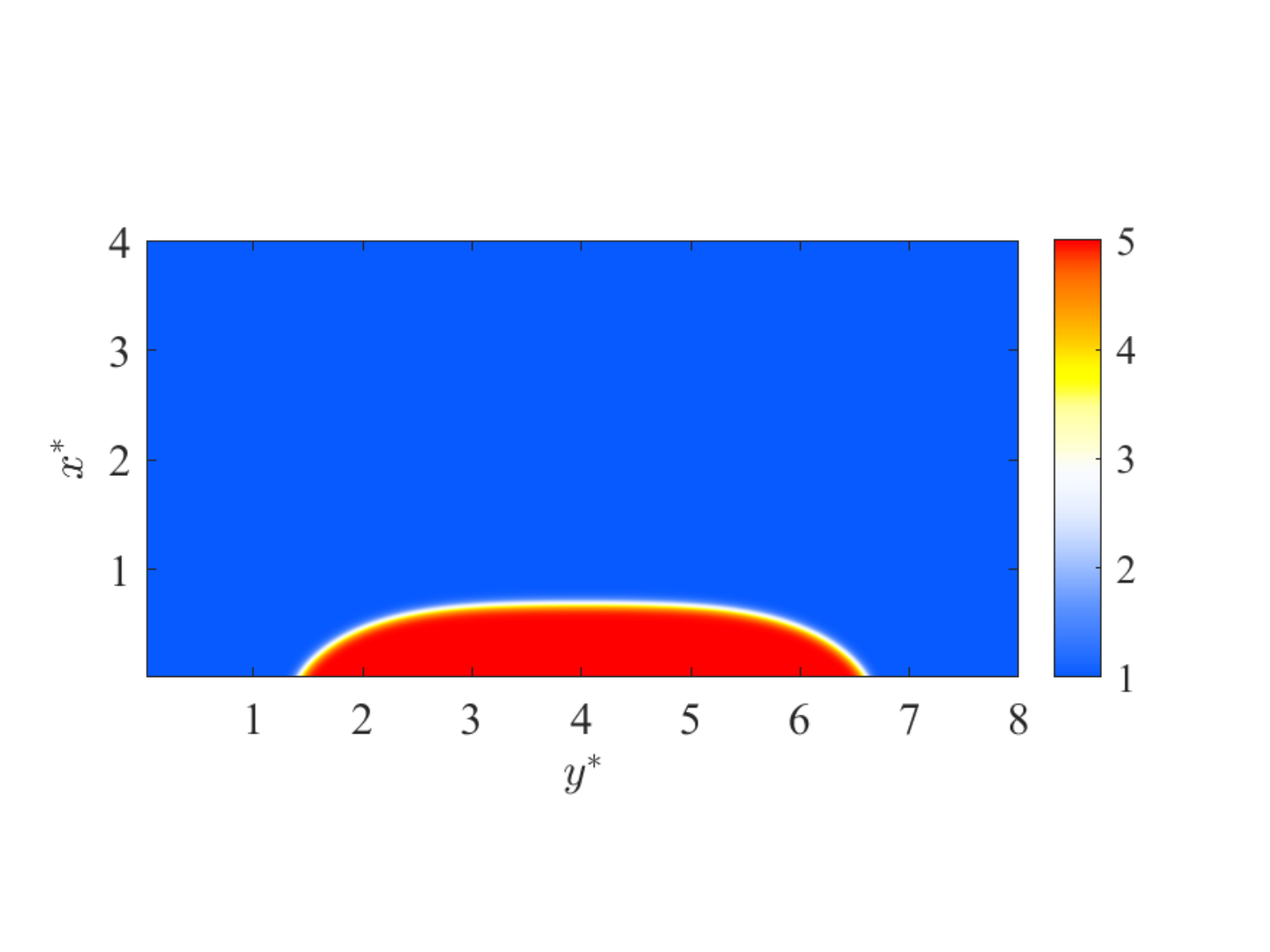}
			\label{Time_step_38000}
		\end{minipage}%
	}%
	\subfigure[$t^*=15$]{
	\begin{minipage}[t]{0.33\linewidth}
		\centering
		\includegraphics[width=1.0\columnwidth,trim={1cm 1.8cm 1cm 2.0cm},clip]{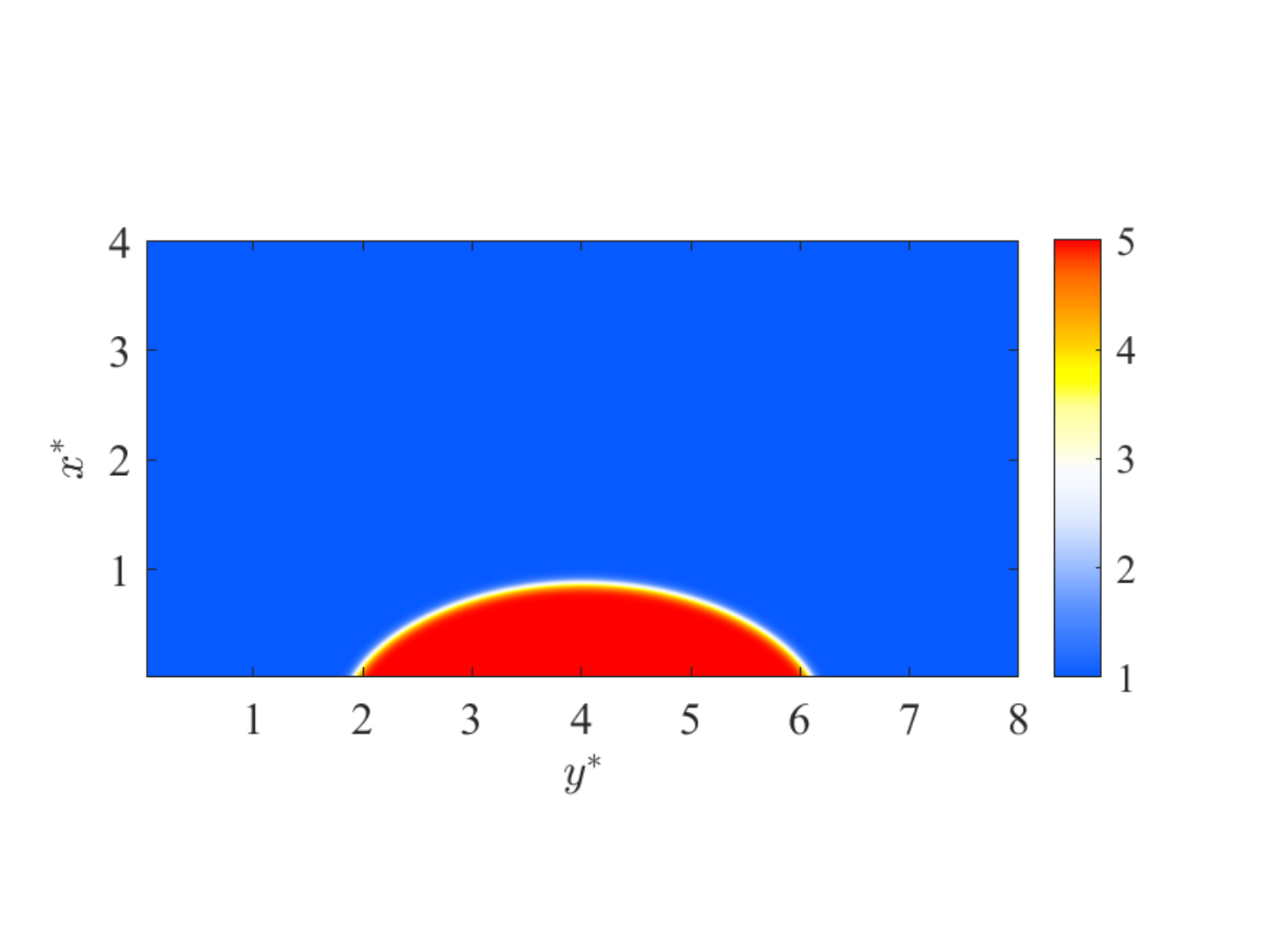}
		\label{Time_step_50000}
	\end{minipage}%
}%
	\caption{Snapshots of droplet morphology at different time instants. (a) $t^{*}=0.0$, (b) $t^{*}=0.3$, (c) $t^*=0.6$, (d) $t^*=1.2$, (e) $t^{*}=1.8$, (f) $t^{*}=4.8$, (g) $t^*=8.4$, (h) $t^{*}=11.4$ and (i) $t^{*}=15$. The colorbar denotes the value of density normalized by the gas density $\rho_{g}$.} 
	\label{shape_evolution}
\end{figure}
Fig.~\ref{shape_evolution} shows the time evolution of the droplet morphology after its impact on the hydrophilic solid wall, where the time is normalized by $D_{l}/U_{0}$. At the initial stage, the droplet has sufficient kinetic energy to overcome the surface tension work and the viscous dissipation. It is observed that the contact length increases with time due to the adhesive force between the droplet and the wall. After reaching the maximum spreading radius, the droplet undergoes a contracting process dominated by the surface tension force. Although the droplet morphology during its impact on the wall has been studied and reported in the existing literature, physical features of BEF and its relation with associated surface physical quantities are never considered and investigated previously. To demonstrate the theoretical results established in Section~\ref{Enstrophy fluxes for two-phase flow}, the typical snapshots at $t^{*}=4.8$ and $11.4$ will be analyzed in detail as follows.

\subsubsection{Spreading process}
\begin{figure}[h]
	\centering
	\subfigure[]{
		\begin{minipage}[t]{0.5\linewidth}
			\centering
			\includegraphics[width=1.0\columnwidth,trim={0.3cm 1.8cm 0.3cm 2.5cm},clip]{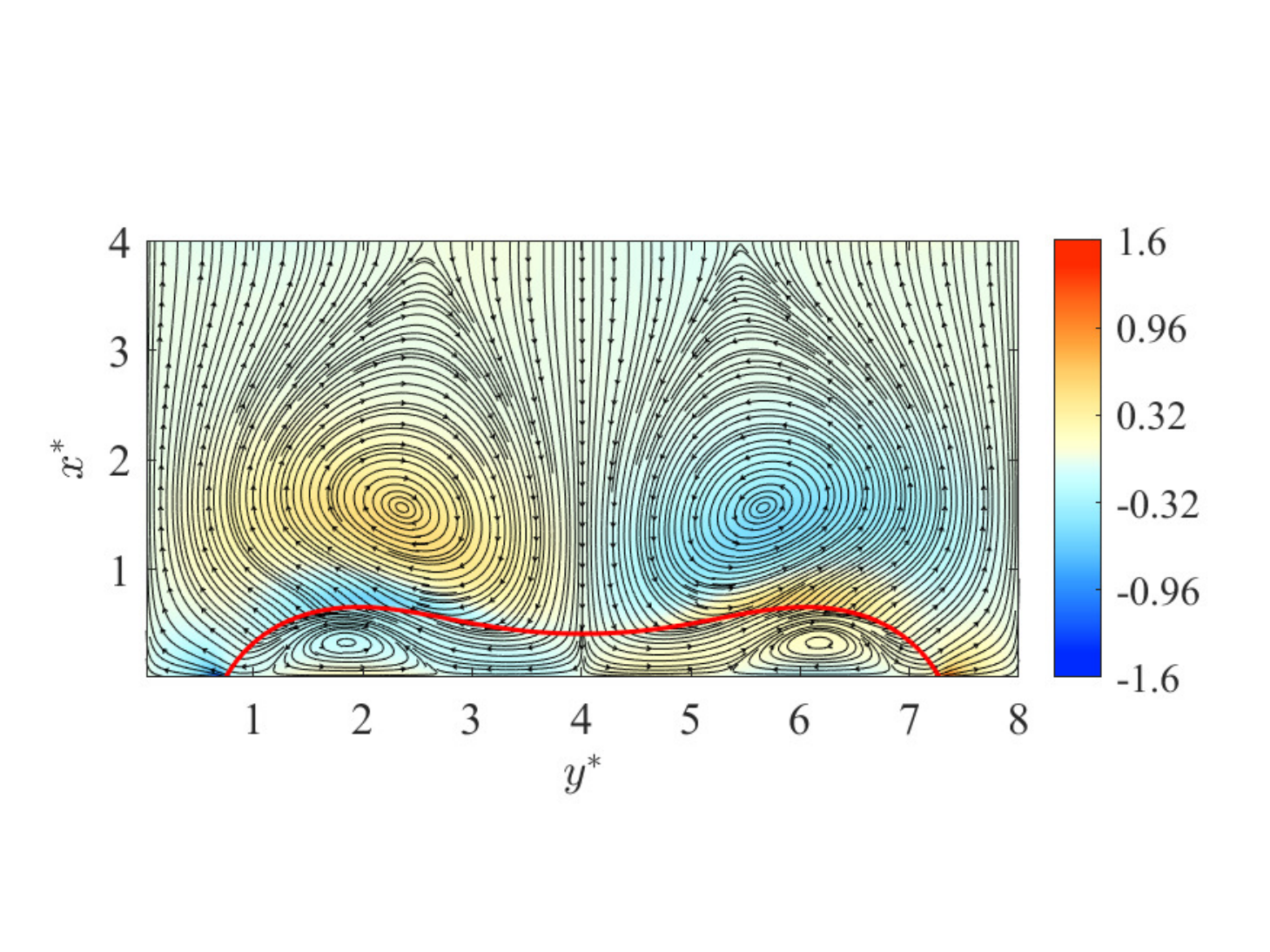}
			\label{dropimpact_V2_60dg_streamlines}
		\end{minipage}%
	}%
	\subfigure[]{
	\begin{minipage}[t]{0.5\linewidth}
		\centering
		\includegraphics[width=1.0\columnwidth,trim={0.2cm 2.6cm 0.2cm 3.4cm},clip]{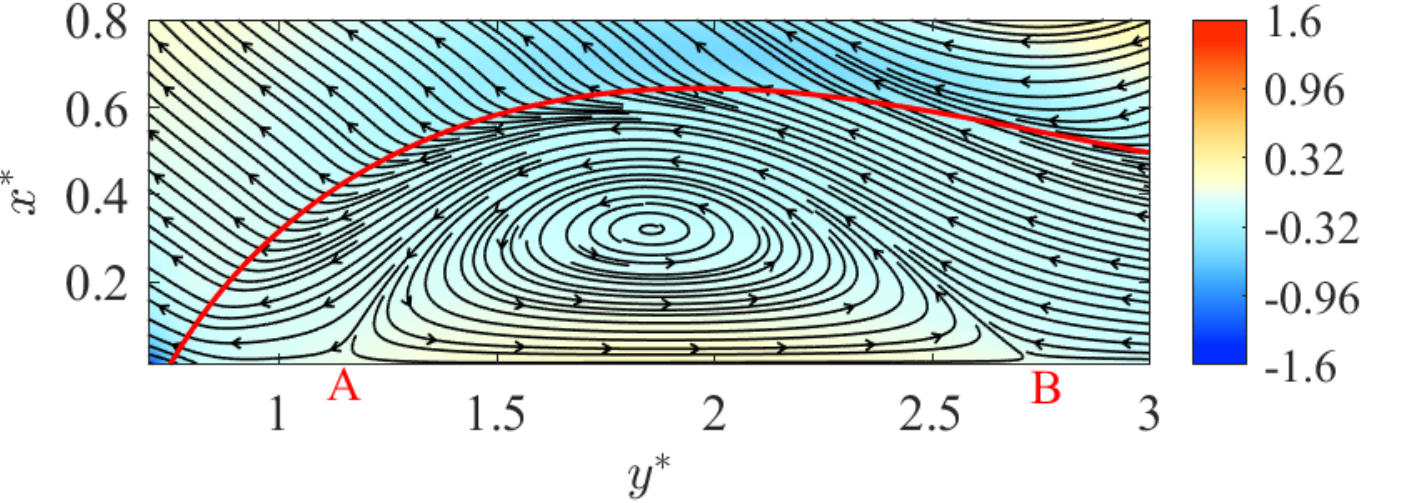}
		\label{local_1}
	\end{minipage}%
}%

	\subfigure[]{
		\begin{minipage}[t]{0.5\linewidth}
			\centering
			\includegraphics[width=1.0\columnwidth,trim={0.1cm 0.1cm 0.1cm 0.6cm},clip]{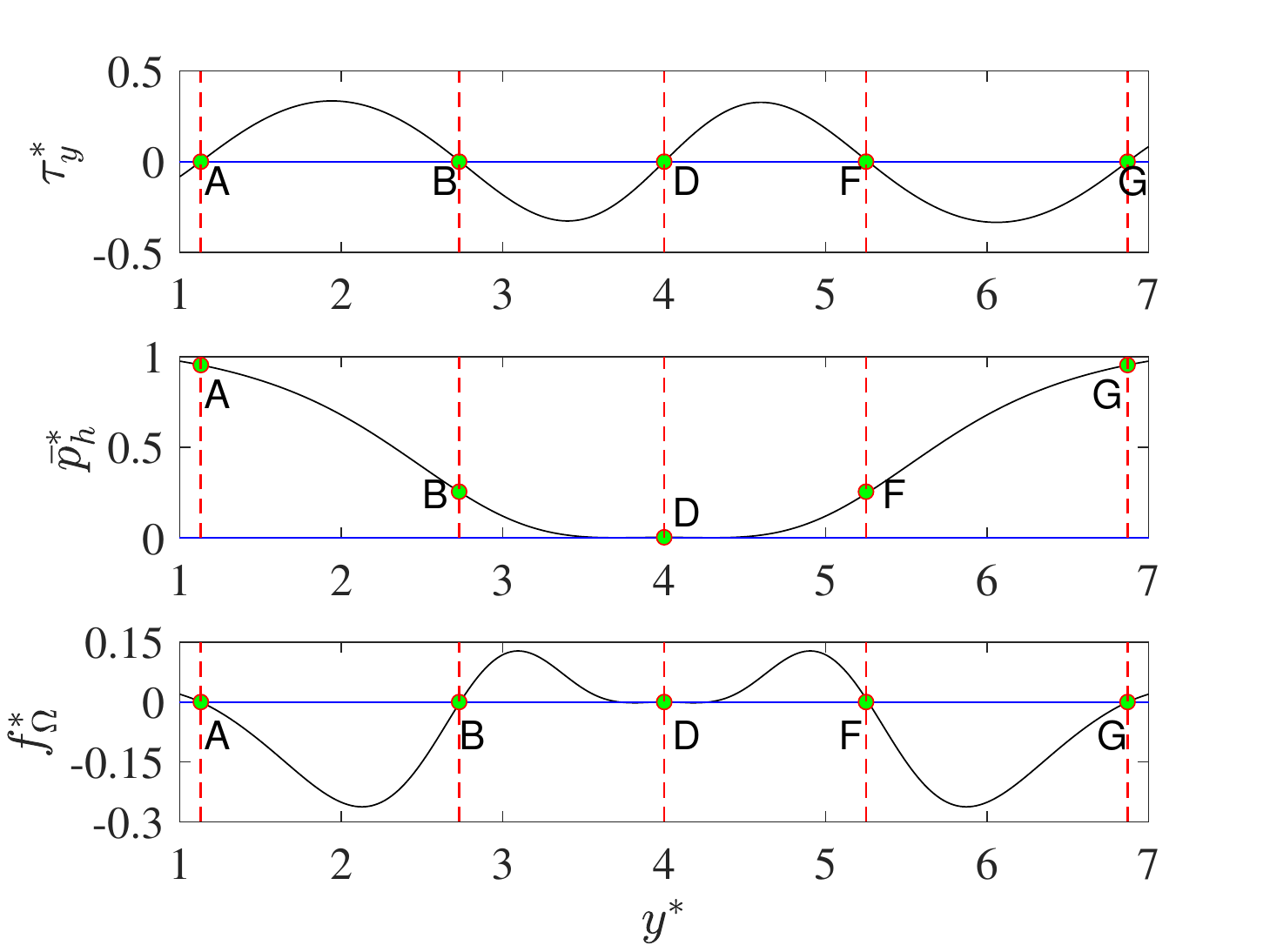}
			\label{dropimpact_V2_60dg_tau_p_BEF}
		\end{minipage}%
	}%
	\caption{(a) Instantaneous streamlines (black) superposed on the contour map of the normalized vorticity $\omega_{z}^{*}\equiv\omega_{z}/(U_{0}/D_{l})$ and the liquid-vapor interface $S$ defined by $\rho=(\rho_{l}+\rho_{g})/2$ (red). (b) A zoom-in view of the separation bubble in (a). (c) Normalized skin friction $\tau_y^*\equiv\tau_{y}/max\{\tau_{y}\}$, surface pressure $\bar{p}_{h}^{*}\equiv\bar{p}_{h}/max\{\bar{p}_{h}\}$, and BEF $f_{\Omega}^{*}\equiv f_{\Omega}/max\{f_{\Omega}\}$, where $max$ denotes the maximum value along the bottom wall. Time $t^*=4.8$.} 
	\label{BEF4p8}
\end{figure}
Fig.~\ref{dropimpact_V2_60dg_streamlines} shows a typical snapshot during the droplet spreading process along the bottom wetting wall $\partial B$ at $t^*=4.8$. The droplet interface $S$ (red line) with the constant density $\rho=(\rho_{l}+\rho_{g})/2$ is extracted. Two counter-rotating primary vortexes with concentrated vorticity magnitudes are identified above the interface $S$ due to its interaction with the downward viscous flow. It is clearly observed that the droplet spreads and elongates with two distinct separation bubbles formed near the ending contact points of the interface $S$. A zoom-in view of the separation bubble is provided in Fig.~\ref{local_1} for better demonstration of the flow details near the separation (B) and attachment (A) points.
Fig.~\ref{dropimpact_V2_60dg_tau_p_BEF} displays the distributions of normalized skin friction $\tau_{y}$, surface pressure $[\bar{p}_{h}]_{\partial B}$, and BEF $f_{\Omega}$ on $\partial B$ below the droplet. Five zero skin friction points are identified: $A$, $D$ and $G$ are three attachment points, while $B$ and $F$ are two separation points. From the distribution of surface pressure $[\bar{p}_{h}]_{\partial B}$, we see that the two separation bubbles are formed due to the adverse pressure gradient. Combining with Eqs.~\eqref{BEF} and~\eqref{BEF_decompose} , it is found that the total BEF $f_{\Omega}$ is dominated by $f_{\Omega}^{(1)}$, which is determined by the viscous coupling between the skin friction $\tau_y$ and surface pressure gradient $[\partial\bar{p}_{h}/\partial y]_{\partial B}$. Therefore, the five zero skin friction points should also be five zero-crossing points of the BEF. We observe that the BEF remains negative below the two separation bubbles, which indicates that the newly created boundary vorticity will enhance the existing near-wall vorticity. In contrast, positive BEF regions arise inside the regions $BD$ and $DF$, which indicates that the
newly created boundary vorticity will attenuate the original near-wall vorticity.

\begin{figure}[h]
	\centering
	\subfigure[]{
		\begin{minipage}[t]{0.5\linewidth}
			\centering
			\includegraphics[width=1.0\columnwidth,trim={0.1cm 0.1cm 0.1cm 1.2cm},clip]{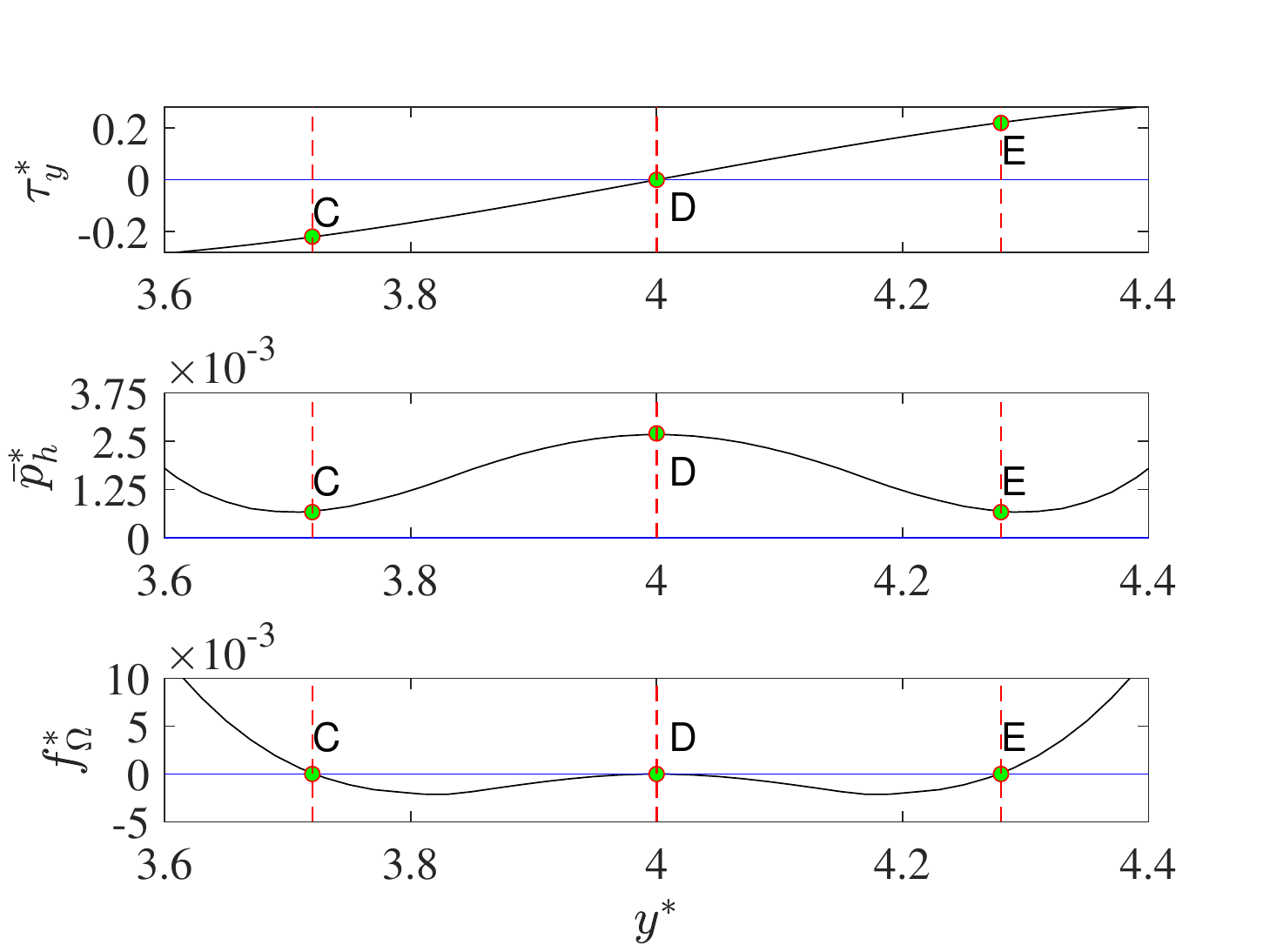}
			\label{fff1}
		\end{minipage}%
	}%
	\subfigure[]{
		\begin{minipage}[t]{0.5\linewidth}
			\centering
			\includegraphics[width=1.0\columnwidth,trim={0.1cm 0.09cm 0.1cm 0.6cm},clip]{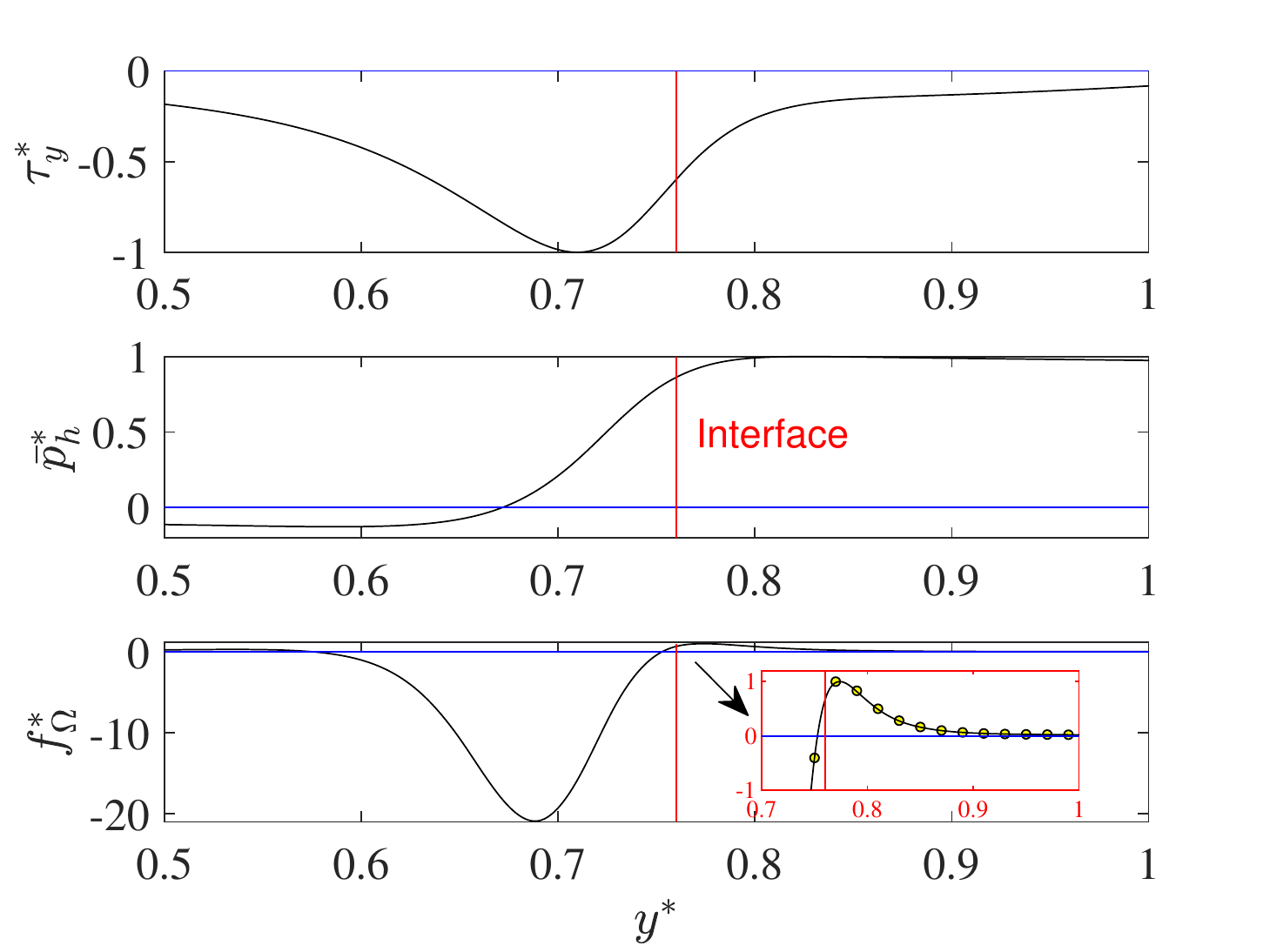}
			\label{fff2}
		\end{minipage}%
	}%
	\caption{Zoom-in views of the normalized skin friction $\tau_{y}^{*}$, surface pressure $\bar{p}_{h}^{*}$ and BEF $f_{\Omega}^{*}$ (a) below the droplet and (b) around the moving contact point (indicated by the red solid line $y^*=0.76$). Time $t^*=4.8$.} 
	\label{fff12}
\end{figure}

Intuitively, a local pressure maximum point should exist at the central point $D$ due to the wall-normal impinging motion, which is not clearly seen in Fig.~\ref{dropimpact_V2_60dg_tau_p_BEF}. Thus, in Fig.~\eqref{fff1}, we provide three zoom-in views of the same physical quantities around the point $D$. Indeed, $D$ is a local pressure maximum point, with two new neighbouring points $C$ and $E$ indicating local pressure minima (where $[\partial\bar{p}_{h}/\partial y]_{\partial B}=0$). Therefore, $C$ and $E$ become another two zero-crossing points of the BEF. These basic features demonstrate the complexity of droplet-wall interaction.
\begin{figure}[h!]
	\centering
	\includegraphics[width=0.6\columnwidth,trim={0.0cm 0.1cm 0.0cm 0.6cm},clip]{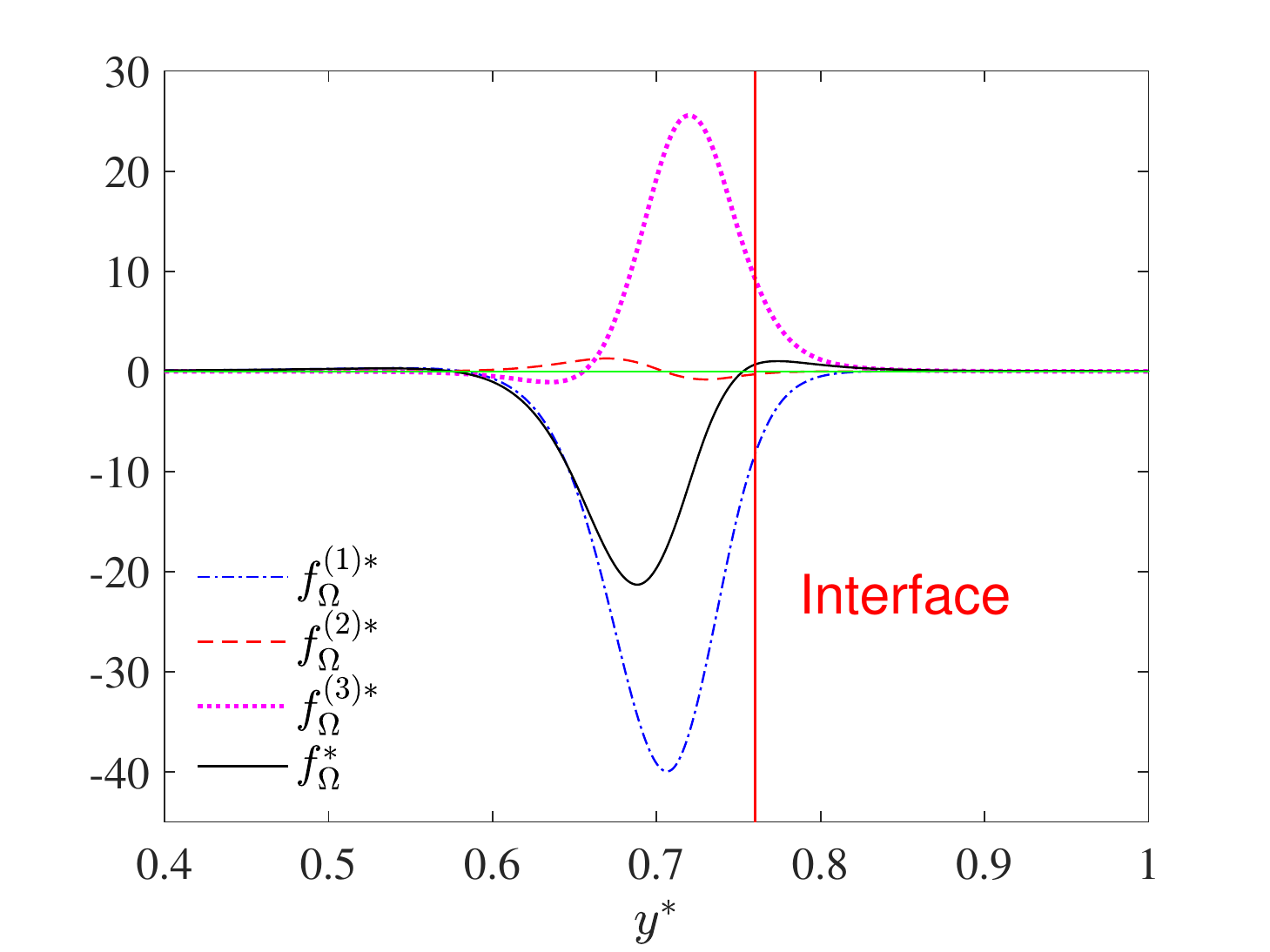}
	\caption{Comparison of different contributions to the normalized total BEF $f_{\Omega}^{*}$ at $t^*=4.8$. All the BEF terms are normalized by $max\{f_{\Omega}\}$ for better comparison. The red solid line indicates the location of the contact point $(x^*,y^*)=(0,0.76)$. Time $t^*=4.8$.} 
	\label{compare_BEF_terms}
\end{figure}

Compared to single-phase flow, features of surface quantities deserve more attention near the moving contact point. In Fig.~\ref{fff2}, we provide the distributions of normalized $\tau_{y}$, $[\bar{p}_{h}]_{\partial B}$, and $f_{\Omega}$ near the moving contact point. A highly negative peak of the skin friction $\tau_{y}$ is observed near the left hand side of the interface. The pressure $[\bar{p}_{h}]_{\partial B}$ varies smoothly across the interface with a relatively high magnitude of its positive gradient. Interestingly, the BEF also shows a highly negative peak at $y^*=0.7$ close to the interface. Therefore, compared to the region away from the contact point, the region corresponding to this negative BEF peak is the main vorticity source on the wall. In addition, a positive peak of the BEF with relatively smaller magnitude is also observed in the right hand side of the interface. 

The reason causing such distribution of the total BEF $f_{\Omega}$ near the contact point can be well explained using Eqs.~\eqref{BEF} and~\eqref{BEF_decompose}. In Fig.~\ref{compare_BEF_terms}, we compare different contributions to the total BEF $f_{\Omega}$. Since $f_{\Omega}^{(4)}$ vanishes in 2D viscous flow, $f_{\Omega}$ is determined by the sum of $f_{\Omega}^{(1)}$, $f_{\Omega}^{(2)}$ and $f_{\Omega}^{(3)}$. $f_{\Omega}^{(1)}$ is typically negative due to the negative peak of skin friction $\tau_{y}$ and the positive surface pressure gradient $[\partial\bar{p}_h/\partial y]_{\partial B}$. $f_{\Omega}^{(3)}$ is caused by the coupling between skin friction $\bm{\tau}$ and the interfacial force due to the density gradient $[\rho\bm{F}_{\bm{\nabla}}]_{\partial B}$, which also shows a positive peak near the interface.
Compared to $f_{\Omega}^{(1)}$ and $f_{\Omega}^{(3)}$, the contribution from $f_{\Omega}^{(2)}$ caused by the coupling between $\bm{\tau}$ and $[\rho\bm{F}_{\mu}]_{\partial B}$ is very small and therefore can be neglected. Therefore, the highly negative peak of the total BEF $f_{\Omega}$ is dominated by $f_{\Omega}^{(1)}$ whose magnitude is higher than that $f_{\Omega}^{(3)}$ in the corresponding region.
By comparison, the reason for the positive peak of the BEF in the right hand side of the interface is attributed to the fact that the magnitude of $f_{\Omega}^{(3)}$ slightly exceeds that of $f_{\Omega}^{(1)}$ in that region.

\subsubsection{Contracting process}
\begin{figure}[t]
	\centering
	\subfigure[]{
		\begin{minipage}[t]{0.5\linewidth}
			\centering
			\includegraphics[width=1.0\columnwidth,trim={0.3cm 1.8cm 0.3cm 2.5cm},clip]{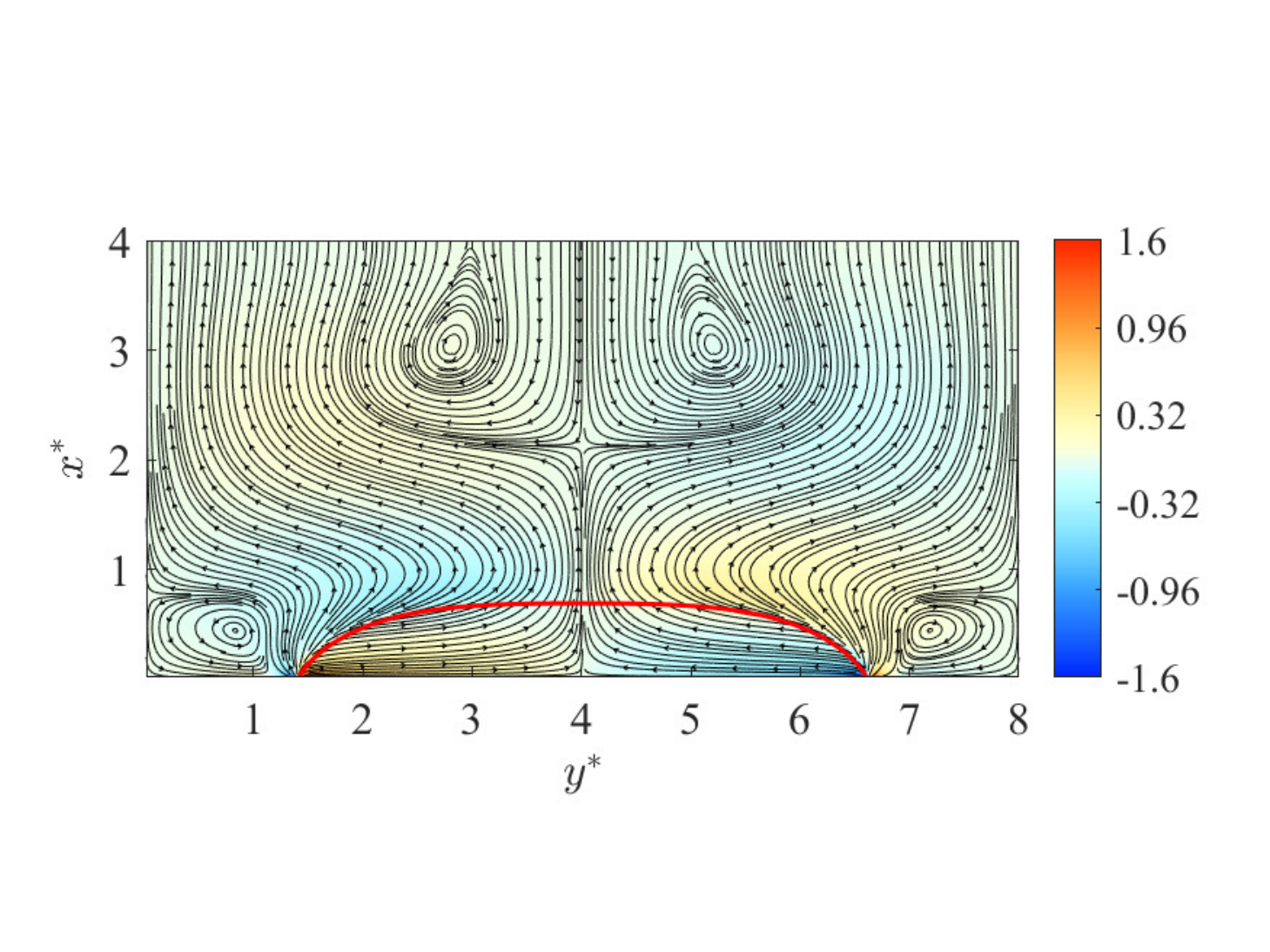}
			\label{dropimpact_V2_60dg_streamlines_b}
		\end{minipage}%
	}%
	\subfigure[]{
	\begin{minipage}[t]{0.5\linewidth}
		\centering
		\includegraphics[width=1.0\columnwidth,trim={0.2cm 2.6cm 0.2cm 3.4cm},clip]{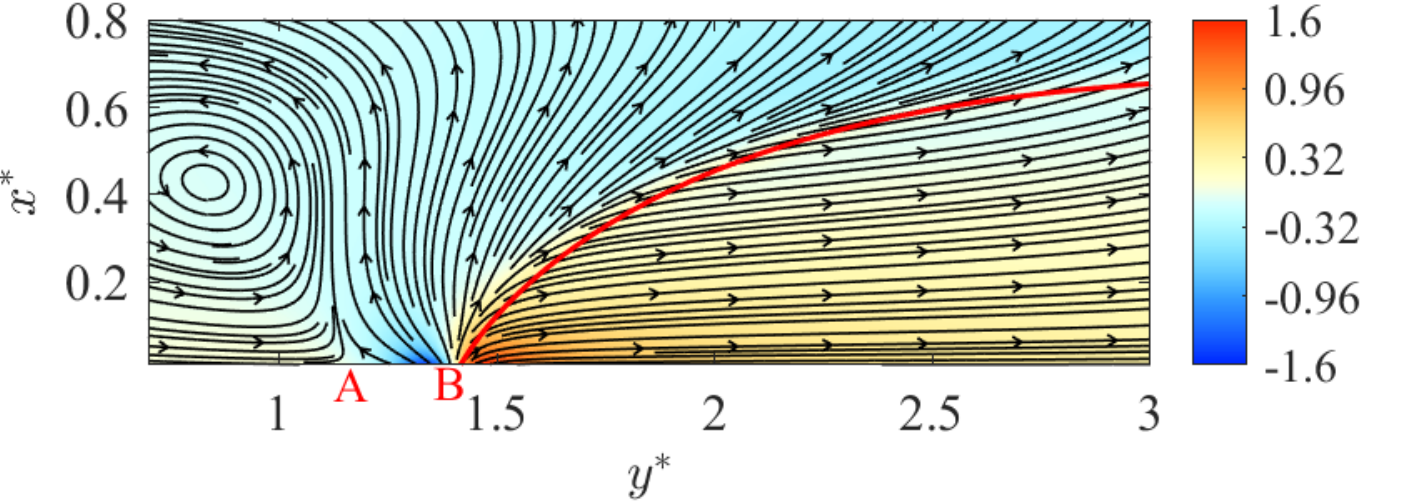}
		\label{local_2}
	\end{minipage}%
}%

	\subfigure[]{
	\begin{minipage}[t]{0.5\linewidth}
		\centering
		\includegraphics[width=1\columnwidth,trim={0cm 0cm 0cm 0cm},clip]{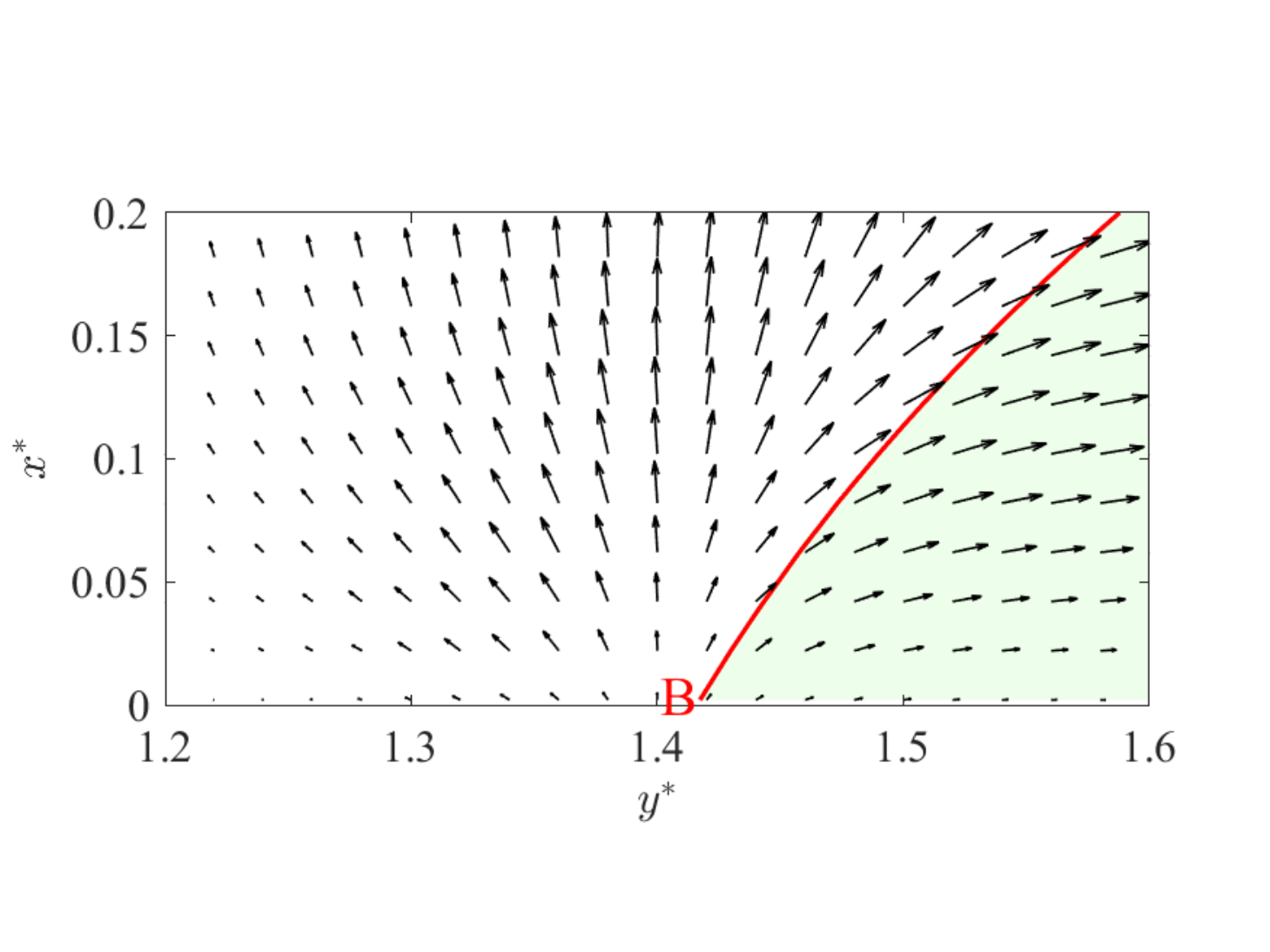}
		\label{local_3}
	\end{minipage}%
}%
	\subfigure[]{
		\begin{minipage}[t]{0.5\linewidth}
			\centering
			\includegraphics[width=1.0\columnwidth,trim={0.1cm 0.1cm 0.1cm 0.6cm},clip]{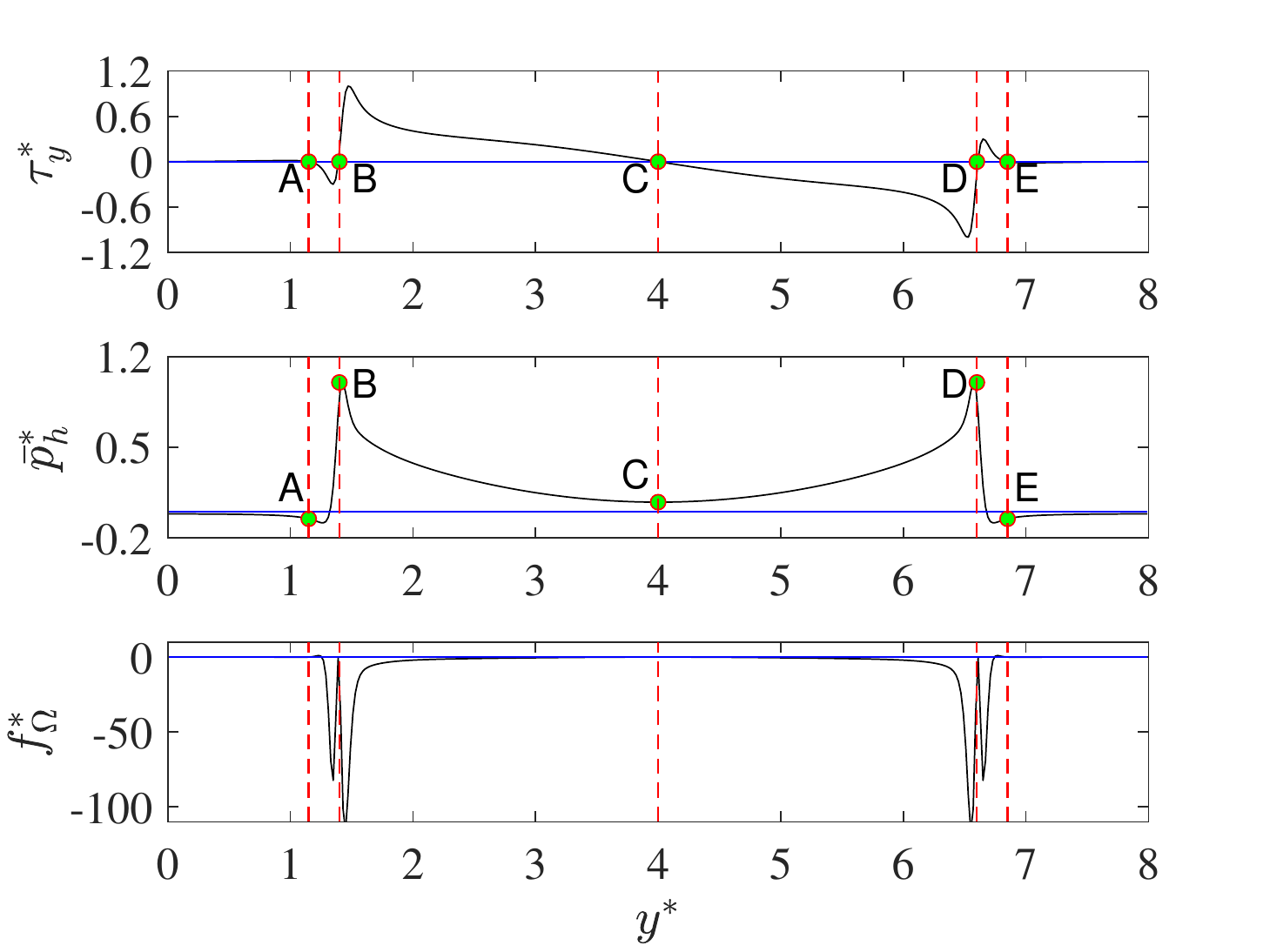}
			\label{dropimpact_V2_60dg_tau_p_BEF_b}
		\end{minipage}%
	}%
	\caption{(a) Instantaneous streamlines (black) superposed on the contour map of the normalized vorticity $\omega_{z}^{*}\equiv\omega_{z}/(U_{0}/D_{l})$ and the liquid-vapor interface $S$ defined by $\rho=(\rho_{l}+\rho_{g})/2$ (red). Two zoom-in views of (b) the streamlines and (c) the velocity vector field around the wall point $B$. (d) Normalized skin friction $\tau_y^*\equiv\tau_{y}/max\{\tau_{y}\}$, surface pressure $\bar{p}_{h}^{*}\equiv\bar{p}_{h}/max\{\bar{p}_{h}\}$, and BEF $f_{\Omega}^{*}\equiv f_{\Omega}/max\{f_{\Omega}\}$, where $max$ denotes the maximum value along the bottom wall. Time $t^*=11.4$.}
	\label{f1}
\end{figure}
It is interesting to examine the physical features of the surface quantities and the near-wall flow pattern during the short contracting process after the spreading. Figs.~\ref{dropimpact_V2_60dg_streamlines_b} and~\ref{local_2} show the streamlines around the droplet at $t^*=11.4$. 
In Fig.~\ref{local_2}, the streamline pattern in a small vicinity of the wall point $B$ indicates a mass source created in the contact line region, namely, $\left[\bm{\nabla}\bm{\cdot}\bm{u}\right]_{\partial B}>0$ holds at the point $B$. A zoom-in view of the vector arrows around the point $B$ is shown in Fig.~\ref{local_3} for better observation. Moreover, the streamlines depict a clearly divergent flow field everywhere within an area around the wall point $B$
inside a semi-circle of a small radius. We shall explain this apparently counter-intuitive divergent flow pattern around the point $B$ from both the mathematical
and physical viewpoints.

First, the mass conservation law in Eq.~\eqref{c1} implies that the velocity divergence (namely, the dilatation) is expressed as
\begin{eqnarray}\label{ch1}
\vartheta\equiv\bm{\nabla}\bm{\cdot}\bm{u}=-\frac{1}{\rho}\frac{\partial\rho}{\partial t}-\bm{u}\bm{\cdot}\frac{\bm{\nabla}\rho}{\rho}.
\end{eqnarray}
In most of the regions, the two terms in the right hand side of Eq.~\eqref{ch1} should almost balance each other in such a low-Mach-number flow.
By using the no-slip boundary condition $\bm{u}_{\partial B}=\bm{0}$, it follows that
$\left[\bm{\nabla}\bm{\cdot}\bm{u}\right]_{\partial B}(t,B)=-\frac{1}{\rho_{\partial B}}\left[\frac{\partial\rho}{\partial t}\right]_{\partial B}(t,B)$ holds at the wall point $B$.
The interfacial mixing layer has a finite thickness across which the density increases monotonically from the vapor phase to the liquid phase with a smooth transition.
The low-Mach-number isothermal assumption implies the weak compressibility inside the interfacial region so that the density distribution across the interface remains almost unchanged during the droplet retraction within a very short time interval $\Delta t$. However, the retraction will cause lower density value at the fixed wall point $B$ at the time instant $t+\Delta{t}$ so that 
$\rho(t+\Delta{t},B)<\rho(t,B)$ holds.
Therefore, we have $\left[\partial\rho/\partial{t}\right]_{\partial B}(t,B)<0$ and $\vartheta_{\partial B}\equiv\left[\bm{\nabla}\bm{\cdot}\bm{u}\right]_{\partial B}(t,B)>0$, which are consistent with the direct observation in Figs.~\ref{local_2} and~\ref{local_3}. 
It should be claimed that although the weak compressibility does not produce a significant effect on the density variation, it is important for the interpretation of the observed streamline pattern. The fluid near the phase interface and
the contact line is weakly compressible so that a non-vanishing velocity divergence in the near-wall region is physically reasonable.

Secondly, by performing the near-wall Taylor-series expansion in terms of the wall-normal coordinate in a small vicinity of the wall, it has been proved that the horizontal velocity component is dominated by the skin friction at the first order and is modified by the surface pressure gradient at the second order. The wall-normal velocity component is determined by the velocity divergence at the first order and is slightly modified by the second-order term contributed by the skin friction divergence and other physical effects.~\cite{ChenTao2022AIPa} When the first-order terms are retained, the near-wall flow pattern should be divergent in a small vicinity of the wall point $B$ due to the positive velocity divergence.

Thirdly, under the present diffuse-interface description, $\partial{p}_{0}/\partial{\rho}$ could become negative in the interfacial region, which may trigger isothermal phase change due to pressurization or depressurization.~\cite{LeeLin2003,Baroudi2020} The evaporation or condensation could happen in a small vicinity of the contact line, which is driven by the curvature of the diffuse interface. Under the isothermal approximation, the temperature difference caused by these phase changes will not give rise to a large temperature variation. However, the phase change can boost the contact line motion with the no-slip boundary condition, which naturally removes the contact line singularity.~\cite{Rednikov2019, Baroudi2020}

For our case, the evaporation induced by the convex surface shape near the contact line could further enhance the near-wall divergent streamline pattern and the formation of the counter-intuitive mass source created on the wall. Recently, Baroudi and Lee~\cite{Baroudi2020} simulated the spreading of a 2D liquid droplet in contact with an atmosphere of its pure vapor on a solid wall and found that a concave contact meniscus caused a lower equilibrium vapor pressure (namely, $p_{v}<p_{v}^{sat}$, $p_{v}$ is the vapor pressure and $p_{v}^{sat}$ is the saturation vapor pressure) near the contact line, which was well explained by the Kelvin equation.~\cite{Shanahan2001a,Shanahan2001b} As a result, the excess vapor pressure drives local condensation from the ambient saturated vapor, and thus the formation of the mass sink (namely, the slight supersaturation of the liquid vapor) near the contact line. The net condensation rate $J_{c}$ (namely, the condensation current) is shown to be proportional to the excess vapor pressure, that is, $J_{c}\propto\Delta{p}_{v}=p_{v}^{sat}-p_{v}$. Although their simulated case (without an initial impact velocity) is simpler than our case, the velocity field observed in their paper is very similar to our result in Fig.~\ref{local_3}. It should be noted that, in their work, the surface shape is concave and a mass sink is formed due to the condensation current. Similarly, for our case, according to the Kelvin equation,~\cite{Shanahan2001a} the convex surface shape causes a higher equilibrium vapor pressure (namely, $p_{v}>p_{v}^{sat}$) near the contact line and thus induces the evaporation current $J_{e}\propto\Delta{p}_{v}=p_{v}-p_{v}^{sat}$, which accounts for the divergent streamline pattern in a small vicinity of the wall and the mass source on the wall.

In addition, the divergent streamline pattern around the wall point $B$ inside a small radius can be understood as a comprehensive effect caused by the anti-clockwise vortex motion in the left hand side of $B$, the evaporation current driven by the surface curvature and the retraction motion of the droplet towards the positive direction.

\begin{figure}[t]
	\centering
	\subfigure[]{
		\begin{minipage}[t]{0.5\linewidth}
			\centering
			\includegraphics[width=1.0\columnwidth,trim={0.1cm 0.1cm 0.1cm 0.75cm},clip]{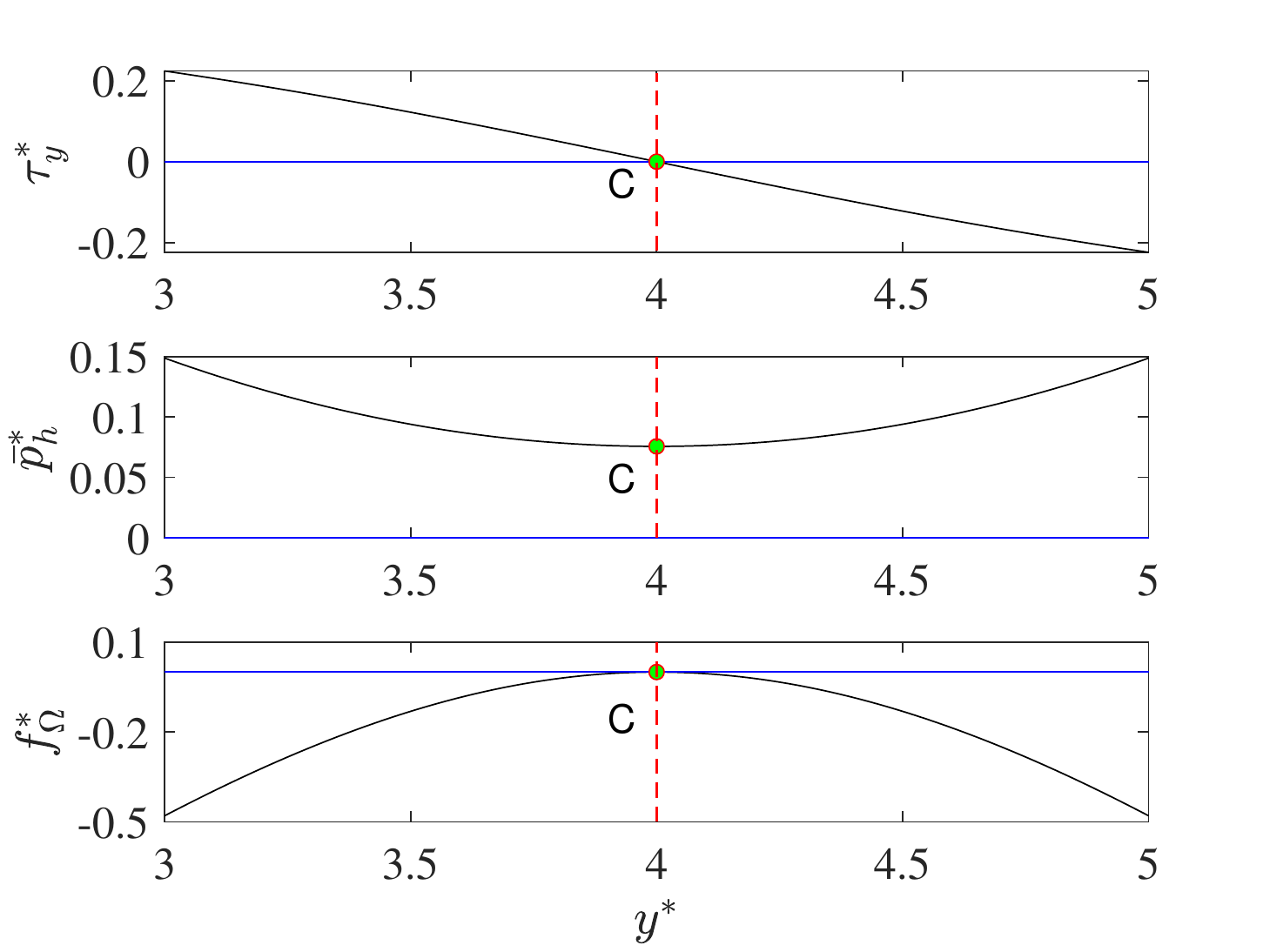}
			\label{dropimpact_V2_60dg_tau_p_BEF_local}
		\end{minipage}%
	}%
	\subfigure[]{
		\begin{minipage}[t]{0.5\linewidth}
			\centering
			\includegraphics[width=1.0\columnwidth,trim={0.1cm 0.09cm 0.1cm 0.6cm},clip]{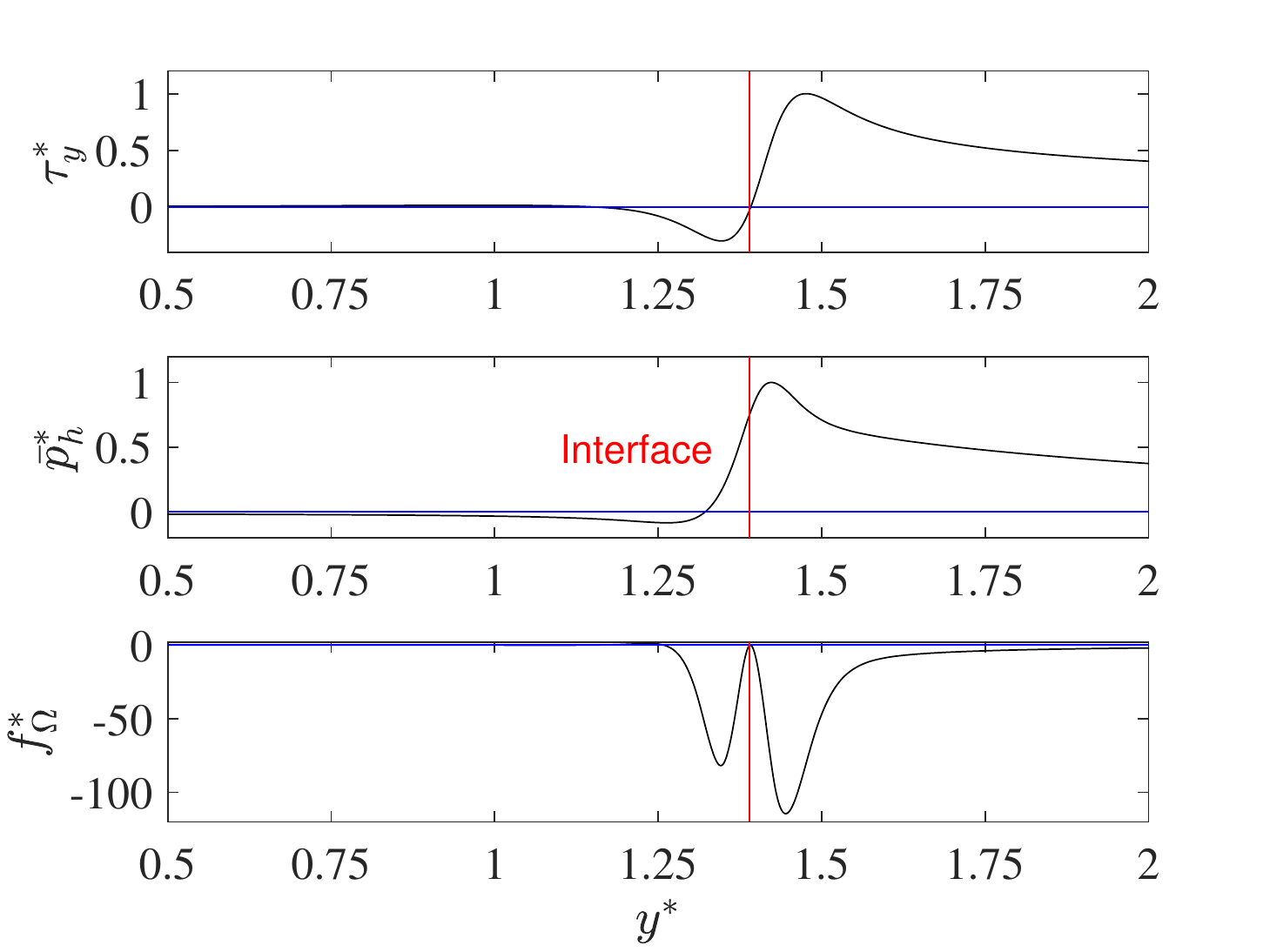}
			\label{dropimpact_V2_60dg_tau_p_BEF_mc_b}
		\end{minipage}%
	}%
	\caption{Zoom-in views of the normalized skin friction $\tau_{y}^{*}$, surface pressure $\bar{p}_{h}^{*}$ and BEF $f_{\Omega}^{*}$ (a) below the droplet and (b) around the moving contact point (indicated by the red solid line $y^*=0.76$). Time $t^*=11.4$.} 
	\label{xxx}
\end{figure}

The corresponding distributions of the normalized $\tau_{y}$, $[\bar{p}_{h}]_{\partial B}$ and $f_{\Omega}$ on the bottom wall are given in Fig.~\ref{dropimpact_V2_60dg_tau_p_BEF_b}. Five zero skin friction points can be found (namely, $A$~--~$E$). $A$ (or $E$) locates between a small region with reverse flow close to the contact point and a secondary vortex. $B$ and $D$ just coincide with the moving contact points where negative and positive skin friction lines intersect.
It is interesting to note that $B$ and $D$ also represent local pressure maxima, while two local pressure minimum points lie between $AB$ and $DE$.
Generally, the position of zero skin friction point and local pressure extreme point is not consistent, which is already observed in single-phase flow.~\cite{ChenTao2021WEF}
However, due to the presence of phase interface, $B$ and $D$ provide a counterexample for possible consistency of zero skin friction point and local pressure extreme point.
In addition, the central zero skin friction point $C$ below the upwelling flow also has local minimum surface pressure.

\begin{figure}[h!]
	\centering
	\includegraphics[width=0.6\columnwidth,trim={0.0cm 0.1cm 0.0cm 0.6cm},clip]{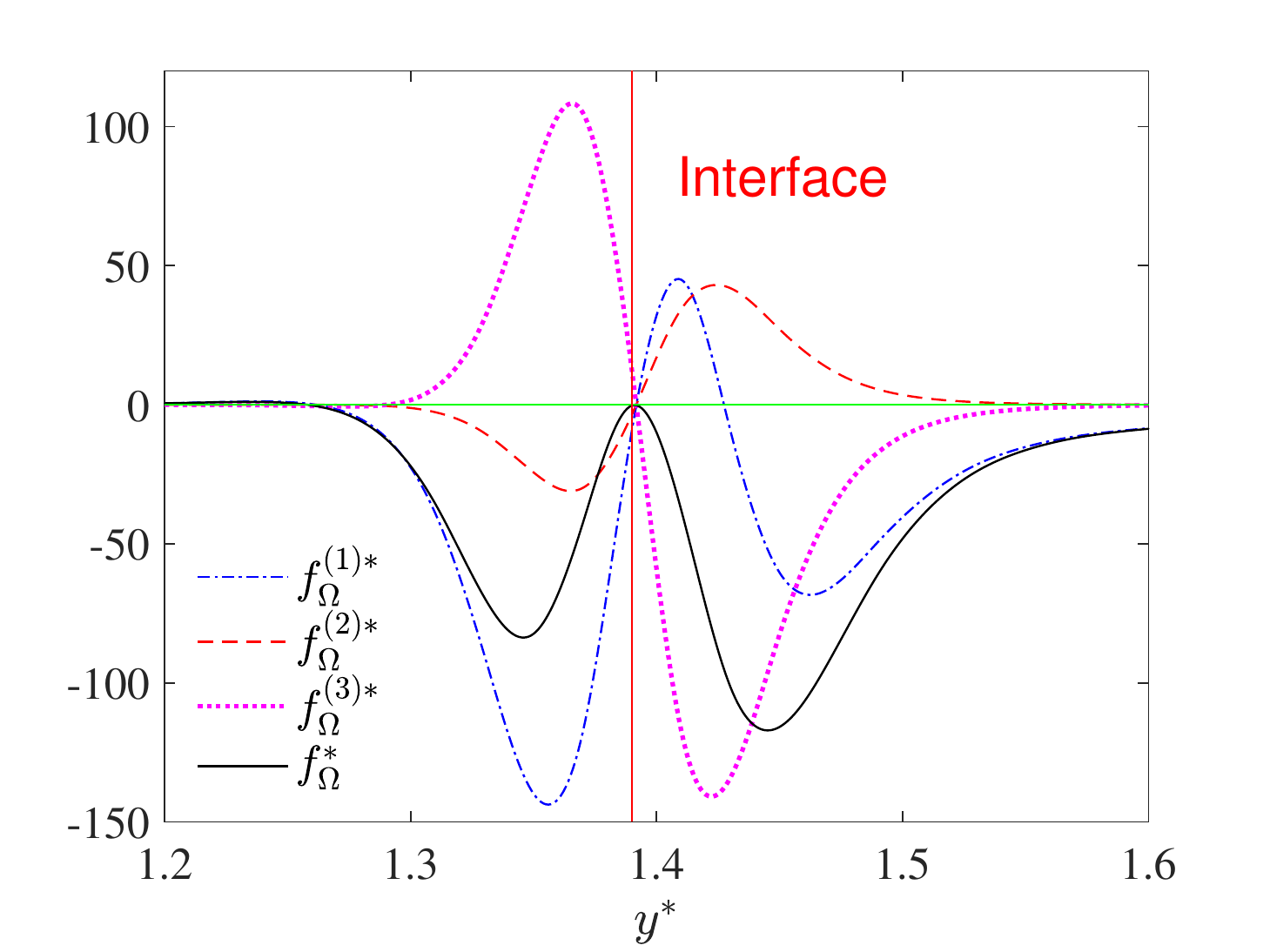}
	\caption{Comparison of different contributions to the normalized total BEF $f_{\Omega}^{*}$. All the BEF terms are normalized by $max\{f_{\Omega}\}$ for better comparison. The red solid line indicates the location of the moving contact line $y^*=1.39$. Time $t^*=11.4$.} 
	\label{dropimpact_V2_60dg_BEF123_mc}
\end{figure}

Since the signs of skin friction and surface pressure gradient are different on each side of the central point $C$, the BEF remains non-positive in the vicinity of $C$, as shown in Fig.~\ref{dropimpact_V2_60dg_tau_p_BEF_local}.
In fact, the BEF is typically negative for almost all the regions, particularly near the two contact points. Zoom-in views of $\tau_{y}$, $[\bar{p}_{h}]_{\partial B}$ and $f_{\Omega}$ are provided in Fig.~\ref{dropimpact_V2_60dg_tau_p_BEF_mc_b}. We observe that the sign of $f_{\Omega}$ coincides with the sign of ${\tau}_y\cdot[\partial{\bar{p}_{h}}/\partial y]_{\partial B}$ in most regions except for the small region near the right hand side of the contact point (interface). In this region, both the skin friction and surface pressure gradient are positive, which implies $f_{\Omega}>0$ in this region. However, only negative values ($f_{\Omega}<0$) can be seen in Fig.~\ref{dropimpact_V2_60dg_tau_p_BEF_mc_b}. Hence, there must be other physical mechanism to neutralize the positive $f_{\Omega}$, which directly changes the vorticity diffusion near the contact point.

In order to give a reasonable physical interpretation, by using Eqs.~\eqref{BEF} and~\eqref{BEF_decompose}, we compare different contributions to $f_{\Omega}$ in Fig.~\eqref{dropimpact_V2_60dg_BEF123_mc}. $f_{\Omega}^{(1)}$ comes from the dot product of skin friction $\bm{\tau}_y$ and surface pressure gradient $[\partial{\bar{p}_{h}}/\partial{y}]_{\partial B}$, which indeed shows a positive peak close to the right hand side of interface and remains negative in other regions. This positive $f_{\Omega}^{(1)}$-peak is enhanced by $f_{\Omega}^{(2)}$ due to the coupling between skin friction $\bm{\tau}$ and the force $[\rho\bm{F}_{\mu}]_{\partial B}$, and is attenuated by $f_{\Omega}^{(3)}$ as a result of that between $\bm{\tau}$ and $[\rho\bm{F}_{\bm{\nabla}}]_{\partial B}$. Due to higher magnitude of  $f_{\Omega}^{(3)}$ than that of the sum of $f_{\Omega}^{(1)}$ and $f_{\Omega}^{(2)}$, this positive peak eventually disappears. For the region near the left hand side of the moving contact line, we have $f_{\Omega}^{(1)}<0$, $f_{\Omega}^{(2)}<0$ and $f_{\Omega}^{(3)}>0$, whose comprehensive effect yields a negative peak of $f_{\Omega}$.

We also analyze the simulation results at other time instants (not shown here for simplicity). Overall, the analysis of the structure of the BEF shows that the regions near the moving contact points are the main vorticity source on the wall, compared to other common regions.
For single-phase incompressible viscous flow past a stationary flat wall, $f_{\Omega}$ is only generated through the coupling between skin friction and surface pressure gradient.~\cite{Liu2018AIA,ChenTao2019POF,ChenTao2021POF} In contrast, for two-phase viscous flow considered here, we see that the generation of $f_{\Omega}$ is a result of competition of different coupling terms shown in Eqs.~\eqref{BEF} and~\eqref{BEF_decompose}, which demonstrates its peculiarity due to the presence of the diffuse interface.
Although different contributions to the total BEF could be either positive or negative at different locations,
the total BEF $f_{\Omega}$ is typically negative, with extremely high magnitude near the contact points.

\subsection{Analysis of interfacial enstrophy flux}
\begin{figure}[h!]
	\centering
	\subfigure[]{
		\begin{minipage}[t]{0.5\linewidth}
			\centering
			\includegraphics[width=1.0\columnwidth,trim={0.1cm 0.1cm 0.1cm 0.3cm},clip]{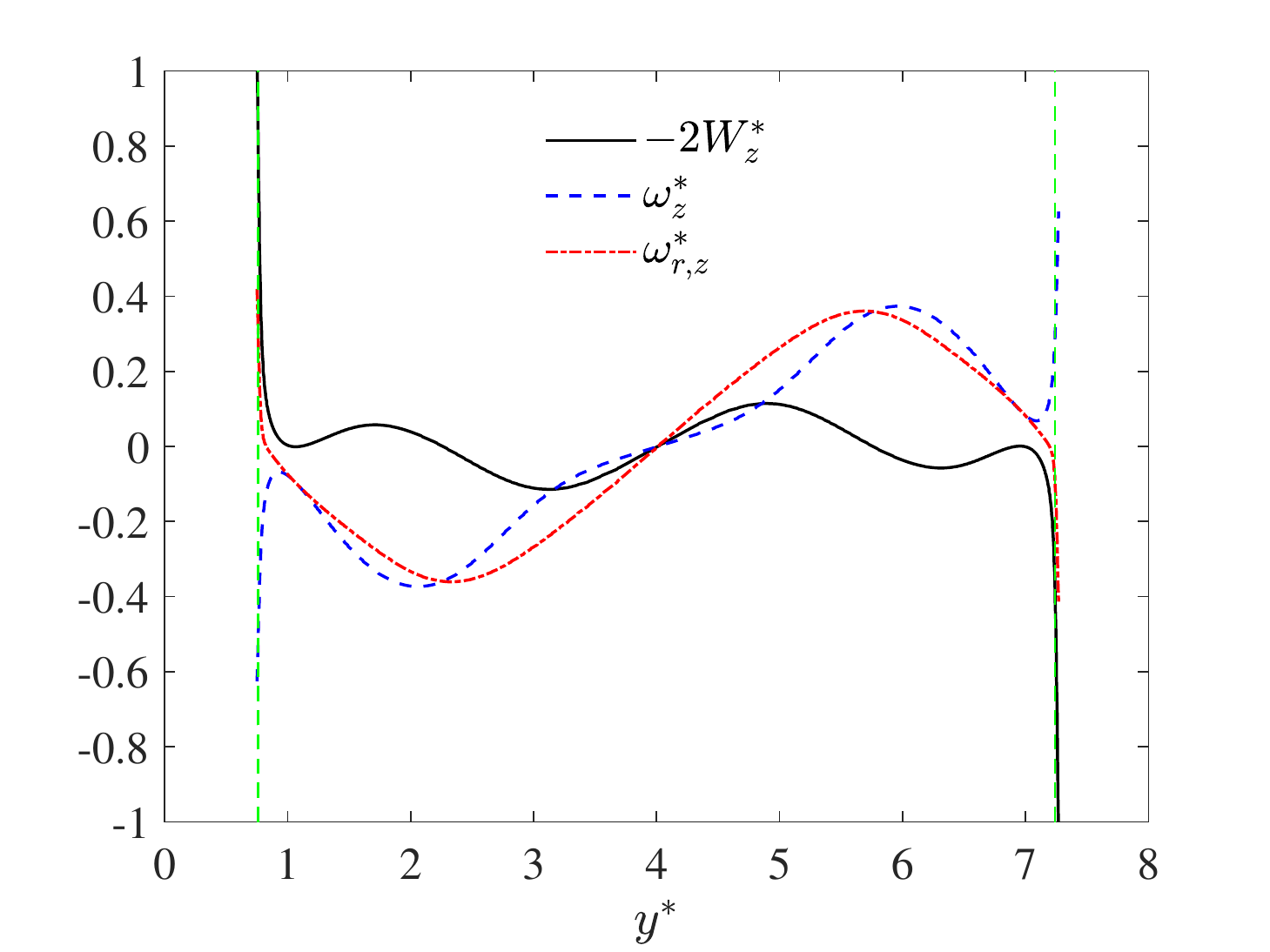}
			\label{tf1}
		\end{minipage}%
	}%
	\subfigure[]{
		\begin{minipage}[t]{0.5\linewidth}
			\centering
			\includegraphics[width=1.0\columnwidth,trim={0.1cm 0.1cm 0.1cm 0.3cm},clip]{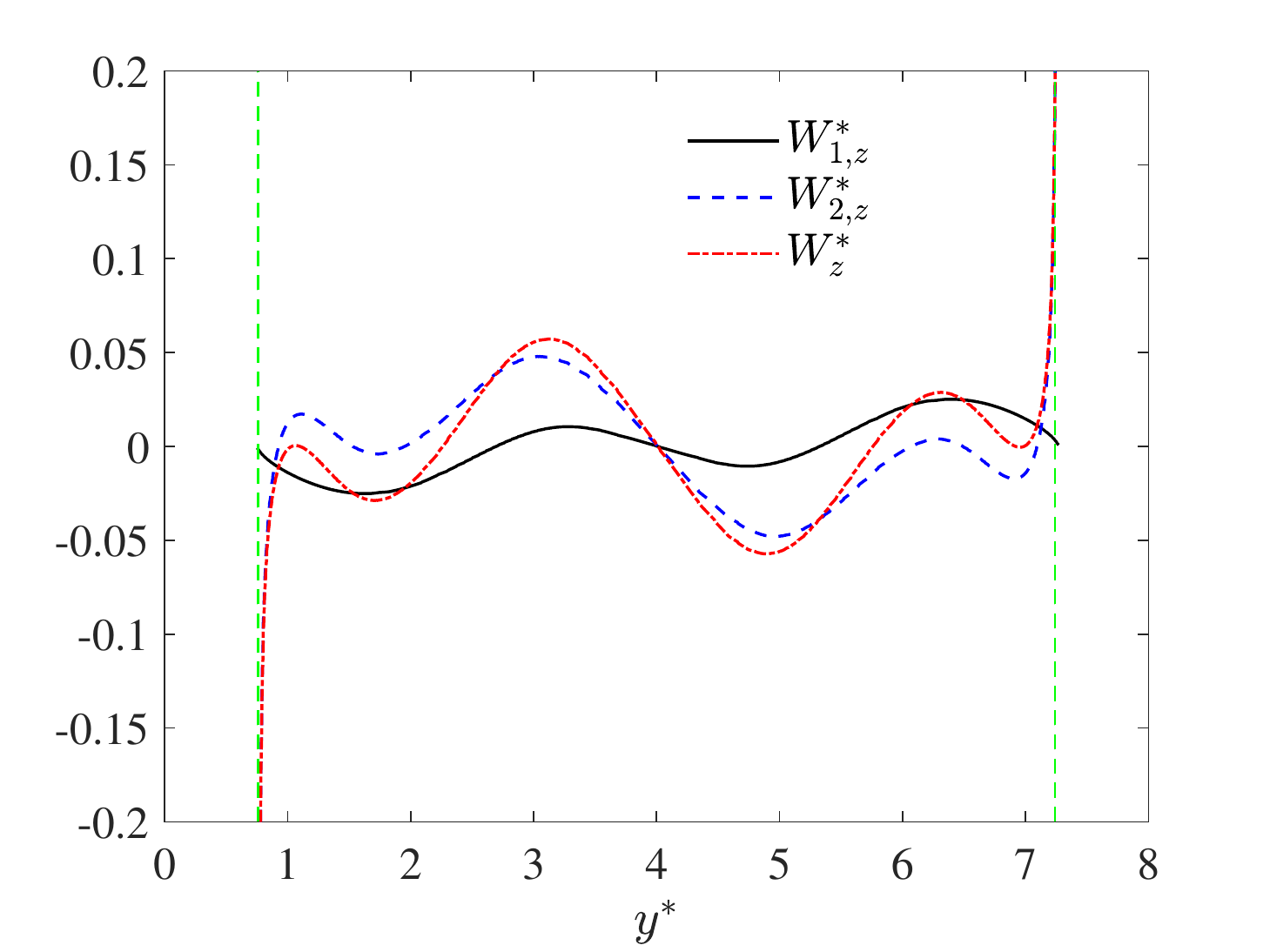}
			\label{tf2}
		\end{minipage}%
	}%
	
	\subfigure[]{
		\begin{minipage}[t]{0.5\linewidth}
			\centering
			\includegraphics[width=1.0\columnwidth,trim={0.1cm 0.1cm 0.1cm 0.3cm},clip]{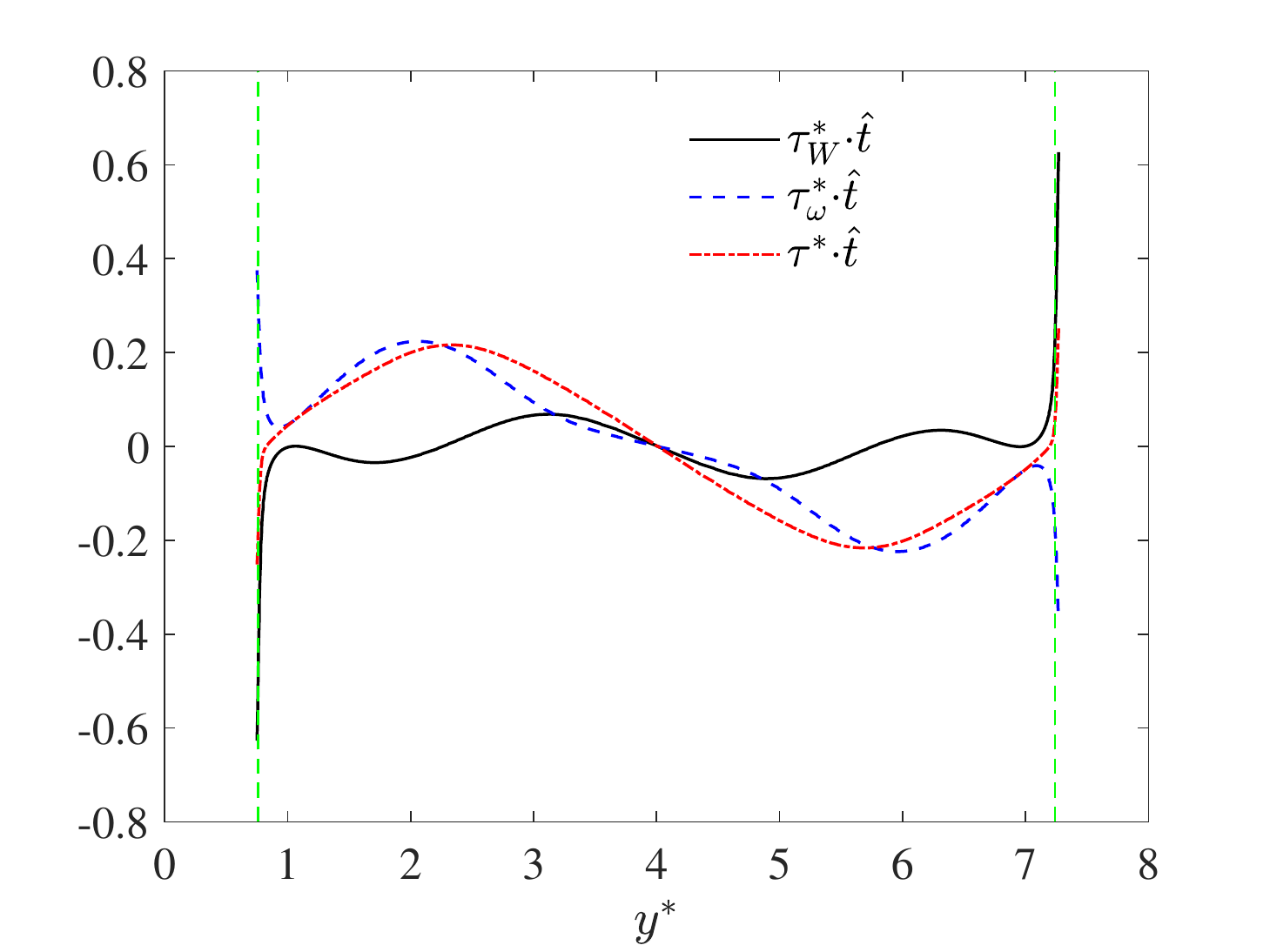}
			\label{tf3}
		\end{minipage}%
	}%
	\subfigure[]{
		\begin{minipage}[t]{0.5\linewidth}
			\centering
			\includegraphics[width=1.0\columnwidth,trim={0.1cm 0.1cm 0.1cm 0.3cm},clip]{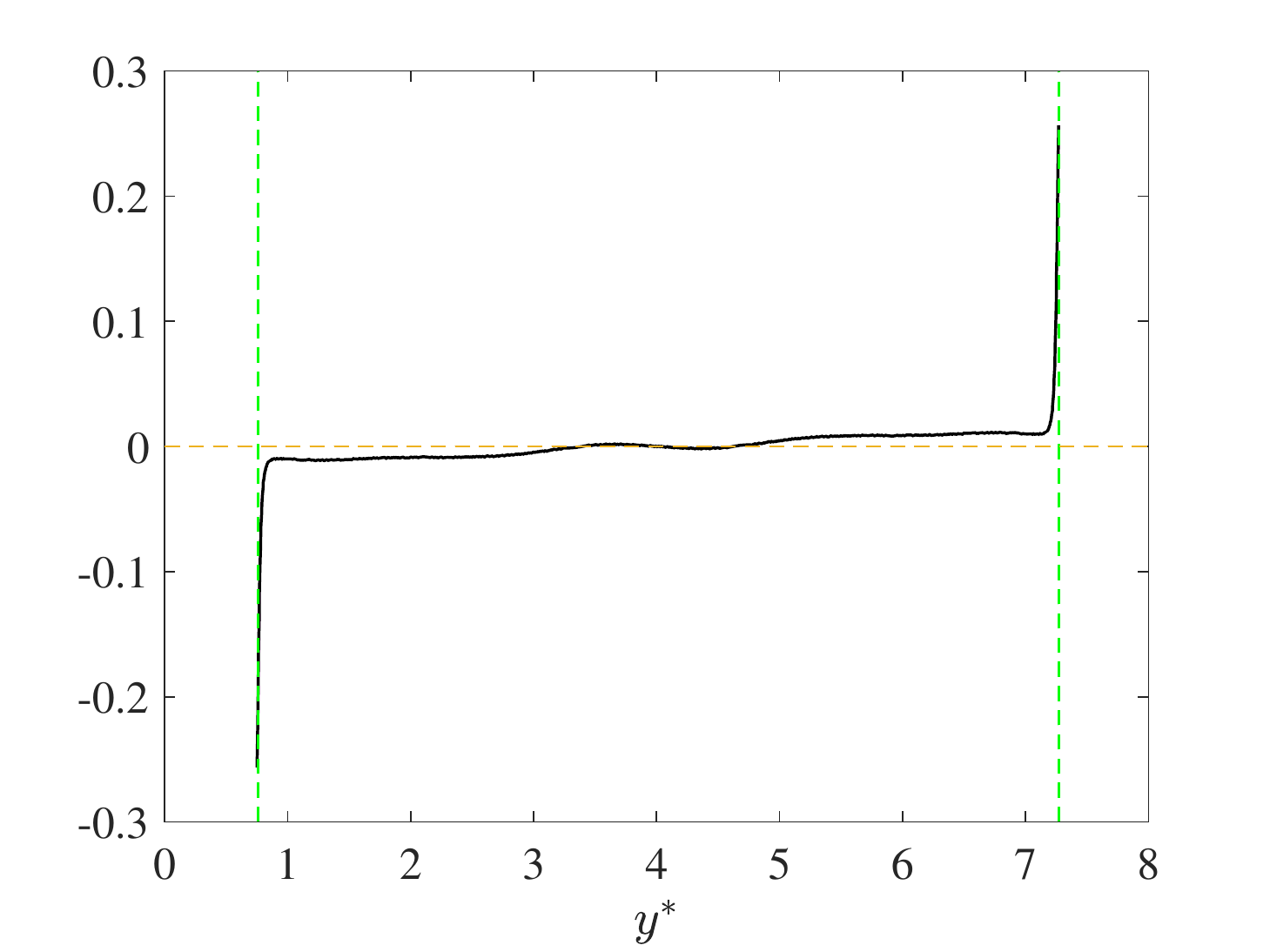}
			\label{tf4}
		\end{minipage}%
	}%
	\caption{Distributions of normalized physical quantities along the liquid-vapor interface $S$. (a) Additional vorticity due to the angular velocity $-2W_z^{*}$, vorticity $\omega_{z}^{*}$ and relative vorticity $\omega_{r,z}^{*}=\omega_{z}^{*}-2W_{z}^{*}$; (b) $W_{1,z}^{*}$, $W_{2,z}^{*}$ and $W_{z}^{*}=W_{1,z}^{*}+W_{2,z}^{*}$; (c) surface deformation stress $\bm{\tau}_{\bm{W}}^{*}\bm{\cdot}\hat{\bm{t}}$, vorticity-induced surface shear stress $\bm{\tau}_{\bm{\omega}}^{*}\bm{\cdot}\hat{\bm{t}}$ and total surface shear stress $\bm{\tau}^{*}\bm{\cdot}\hat{\bm{t}}$; (d) surface pressure gradient $[\partial\bar{p}_{h}^{*}/\partial{s}^{*}]_{S}$. The quantities are normalized by (a) $U_{0}/D_{l}$, (b,~c) $\mu_{l}U_{0}/D_{l}$ and (d) $\rho_{l}U_{0}^2/D_{l}$, respectively. Two green dash lines indicate the locations of the contact points, namely, $(x^{*},y^{*})=(0,0.76)$ and $(0,7.24)$.} 
	\label{tf1234}
\end{figure}
The objective of this subsection is to give an example on analyzing IEF across the interface $S$ based on Eqs.~\eqref{pp0} and~\eqref{pp}. To this end, associated surface physical quantities are extracted from the DNS data using
a self-developed postprocessing code based on the Matlab platform, which is validated and briefly introduced in Appendix~\ref{ESPQ}. Without losing generality, we still focus on the time instant $t^*=4.8$ during the droplet spreading process (see Fig.~\ref{dropimpact_V2_60dg_streamlines}).
Undoubtedly, the formation of outer primary vortex and inner separation bubble are 
accompanied by some interesting variation and coupling of surface quantities.
Since the arc length $s$ (measured from the left contact point) has one-to-one mapping with respect to the horizontal coordinate $y$, all the following figures are presented using $y$.

First, Fig.~\ref{tf1} shows extracted surface angular velocity $\bm{W}=W_{z}\bm{k}$, surface vorticity $\bm{\omega}=\omega_{z}\bm{k}$ and relative vorticity $\bm{\omega}_{r}=\omega_{r,z}\bm{k}$ on the interface $S$, where $\bm{k}=\bm{i}\times\bm{j}$ is the unit base vector perpendicular to the paper, $\bm{i}$ and $\bm{j}$ are the unit base vectors along $x$ and $y$ directions, respectively. 
For the interfacial region away from the contact points, the relative vorticity $\omega_{r,z}$ is dominated by the surface vorticity $\omega_{z}$, and is slightly modified by the additional vorticity $-2W_{z}$ due to the surface angular vorticity. 
Interestingly, the case becomes different when approaching the contact point along the interface $S$. In a small vicinity of contact point, $-2W_{z}$ approaches a finite positive value with relatively high magnitude, while $\omega_{z}$ goes to a finite negative value. The superposition of these two contributions gives a positive relative vorticity $\omega_{r,z}$ in this small vicinity.
\begin{figure}[h!]
	\centering
	\subfigure[]{
		\begin{minipage}[t]{0.5\linewidth}
			\centering
			\includegraphics[width=1.2\columnwidth,trim={2.0cm 0.0cm 0.1cm 0.6cm},clip]{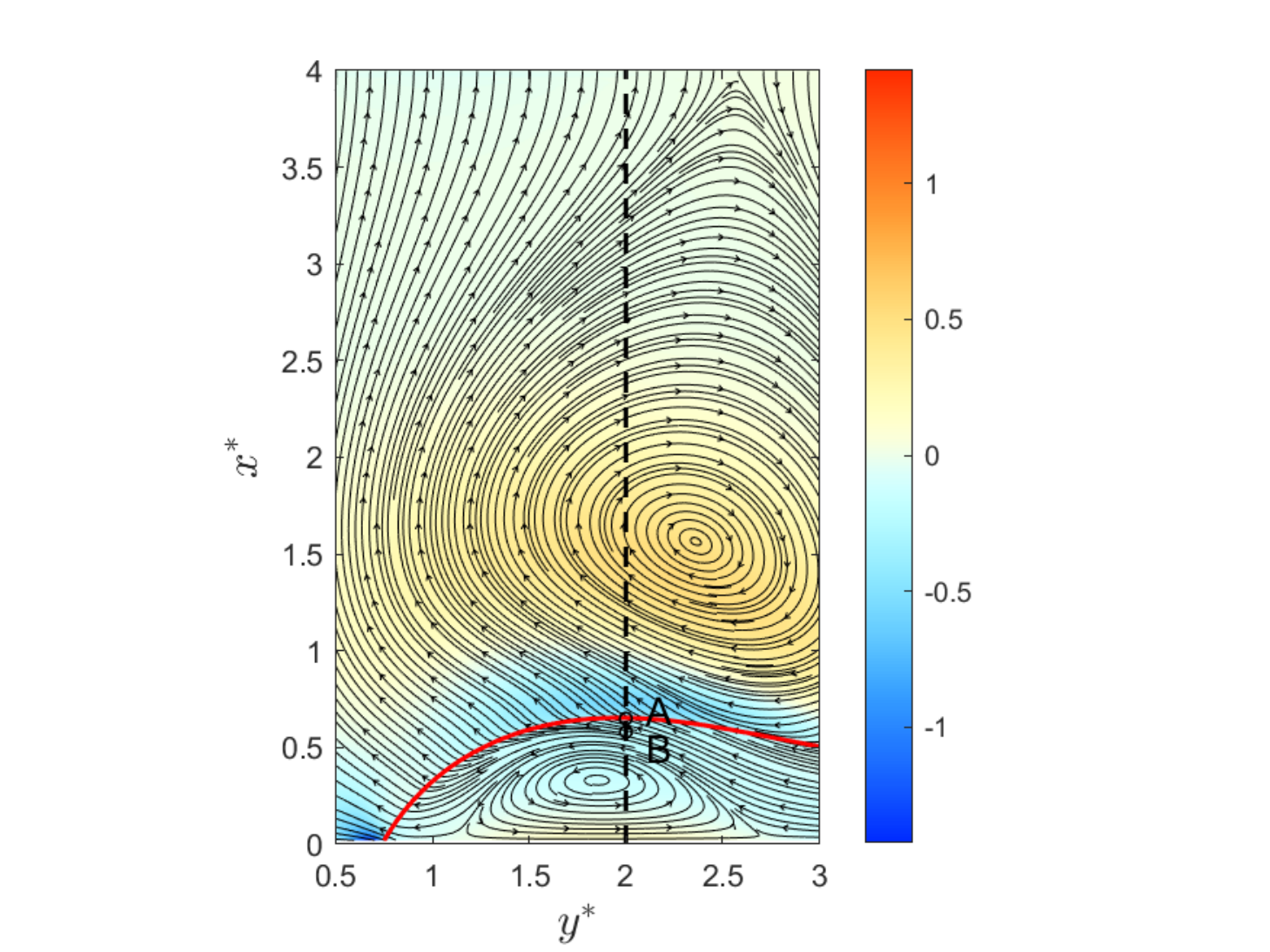}
			\label{yy1a}
		\end{minipage}%
	}%
	\subfigure[]{
		\begin{minipage}[t]{0.26\linewidth}
			\centering
			\includegraphics[width=1.25\columnwidth,trim={1.0cm 0.0cm 0.1cm 0.7cm},clip]{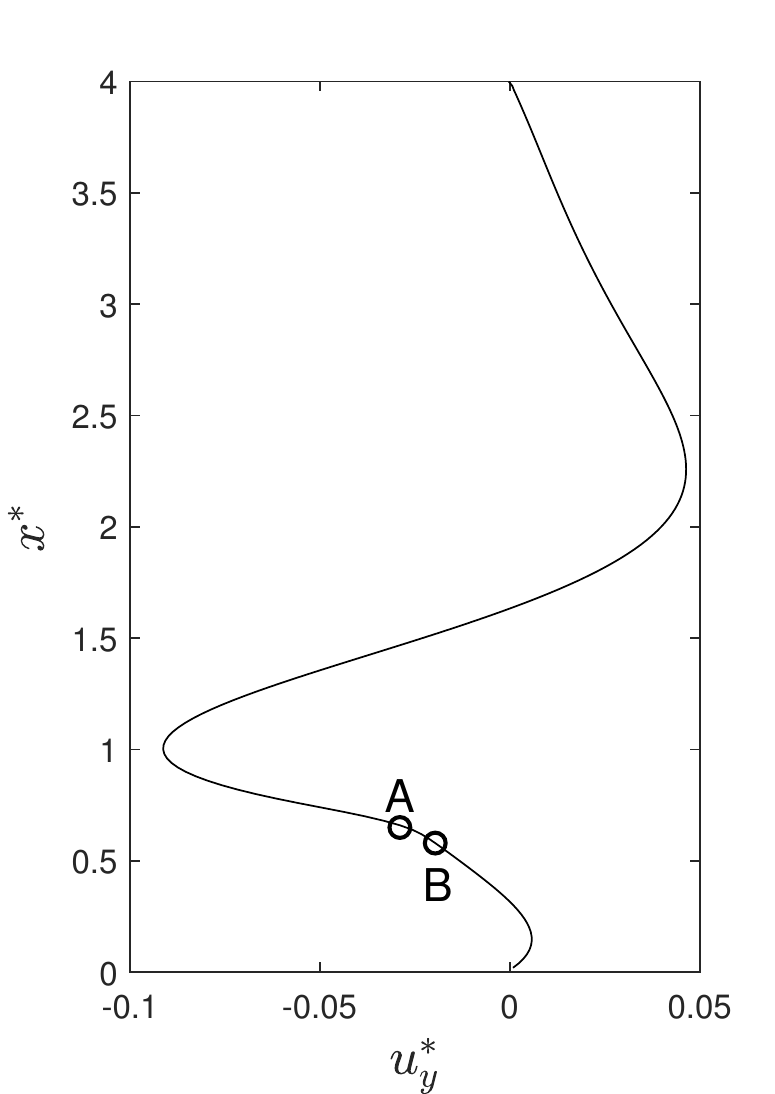}
			\label{yy1b}
		\end{minipage}%
	}%
	\caption{(a) Streamlines superposed on the contour map of the normalized vorticity $\omega_{z}^{*}\equiv\omega_{z}/(U_{0}/D_{l})$. The red solid line represents the interface $S$. The black dash line passes through the separation bubble along the $x$-direction, where $A$ denotes an intersection point on the interface and $B$ is a point very close to $A$. (b) Normalized horizontal velocity component $u_{y}^{*}=u_{y}/U_{0}$ along the black dash line marked in (a).} 
	\label{yy1}
\end{figure}

Next, using Eqs.~\eqref{Wpi} and~\eqref{Wn}, the surface angular velocity can be decomposed as $W_z=W_{1,z}+W_{2,z}$, where  $W_{1,z}=\kappa[u_{\pi}]_{S}$, $W_{2,z}=\partial[u_{n}]_{S}/\partial s$ and $\kappa$ is the curvature of the curve. 
$W_{1,z}$ is interpreted as the angular velocity of the circular motion with the radius $R=\kappa^{-1}$ (when $\kappa\neq{0}$) and the tangential interfacial velocity $[u_{\pi}]_{S}$. 
$W_{2,z}$ is determined by the rate of change of the surface normal velocity $[u_n]_{S}$ with respect to the arc length parameter $s$. As shown in Fig.~\ref{tf2}, $W_z$ is mainly contributed by $W_{2,z}$, whose magnitude is modified by $W_{1,z}$ especially in the peak region.
Fig.~\ref{tf3} shows the surface shear stresses projected on the unit tangential vector $\hat{\bm{t}}$ of $S$. The distribution of surface shear stresses show similar trends as those of surface vorticities in Fig.~\ref{tf1}, because the former is directly determined by the latter on $S$ according to Eq.~\eqref{Skin_friction}.
The total surface shear stress $\bm{\tau}\bm{\cdot}\hat{\bm{t}}$ is mainly contributed by the vorticity-induced surface shear stress $\bm{\tau}_{\bm{\omega}}\bm{\cdot}\hat{\bm{t}}$ in the regions away from the contact point and by the surface deformation stress $\bm{\tau}_{\bm{W}}\bm{\cdot}\hat{\bm{t}}$ in the vicinity of the contact points. In addition, we note that the total shear stress $\bm{\tau}\bm{\cdot}\hat{\bm{t}}$ is positive (negative) in the region $y^*\in[1.5, 4]$ ($y^*\in[4, 6.5]$). 

The sign of $\bm{\tau}\bm{\cdot}\hat{\bm{t}}$ is physically reasonable, which can be further validated from Fig.~\ref{yy1}.
Fig.~\ref{yy1a} shows a zoom-in view of the separation bubble close to the left contact point, where a vertical black dash line is selected for representative analysis. We use $A$ to denote an intersection point on the interface $S$ and $B$ is a point approaching $A$. Fig.~\ref{yy1b} shows the distribution of horizontal velocity component $u_y$ along the black dash line. It is clear that $u_y$ is negative at both $A$ and $B$ while $A$ moves faster than $B$. Therefore, the fluid element at $B$ should impose positive surface shear stress on $A$, which is consistent with the variation of the total shear stress $\bm{\tau}\bm{\cdot}\hat{\bm{t}}$ shown in Fig.~\ref{tf3}. Besides, it is noted that extremely high magnitude of surface pressure gradient $[\partial\bar{p}_{h}/\partial{s}]_{S}$ also exists in a small vicinity of each contact point, as illustrated in Fig.~\ref{tf4}.
\begin{figure}[h!]
	\centering
	\subfigure[]{
	\begin{minipage}[t]{0.5\linewidth}
		\centering
		\includegraphics[width=1.0\columnwidth,trim={0.1cm 0.1cm 0.1cm 0.6cm},clip]{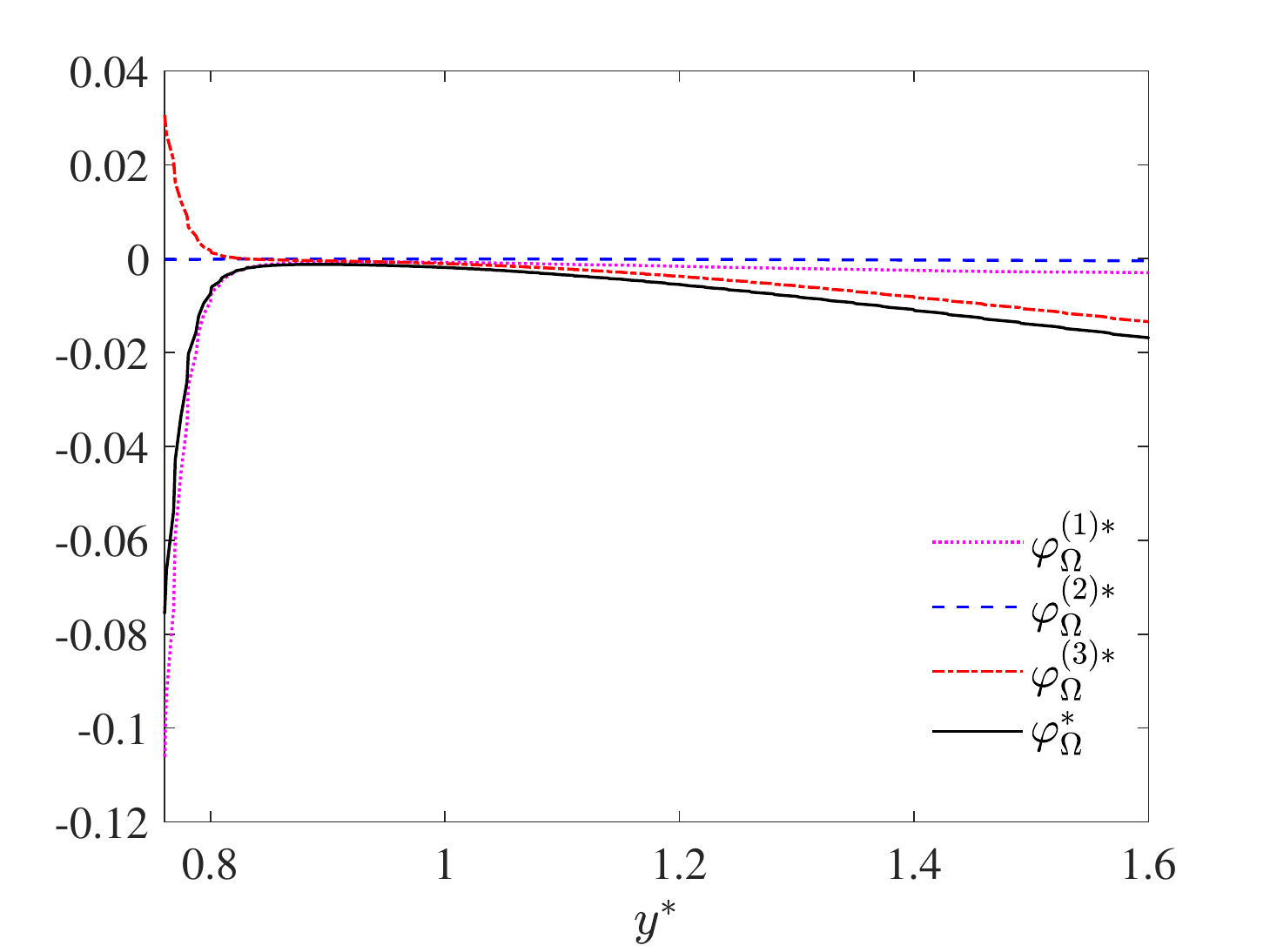}
		\label{yy2a}
	\end{minipage}%
}%
	\subfigure[]{
	\begin{minipage}[t]{0.5\linewidth}
		\centering
		\includegraphics[width=1.0\columnwidth,trim={0.1cm 0.1cm 0.1cm 0.6cm},clip]{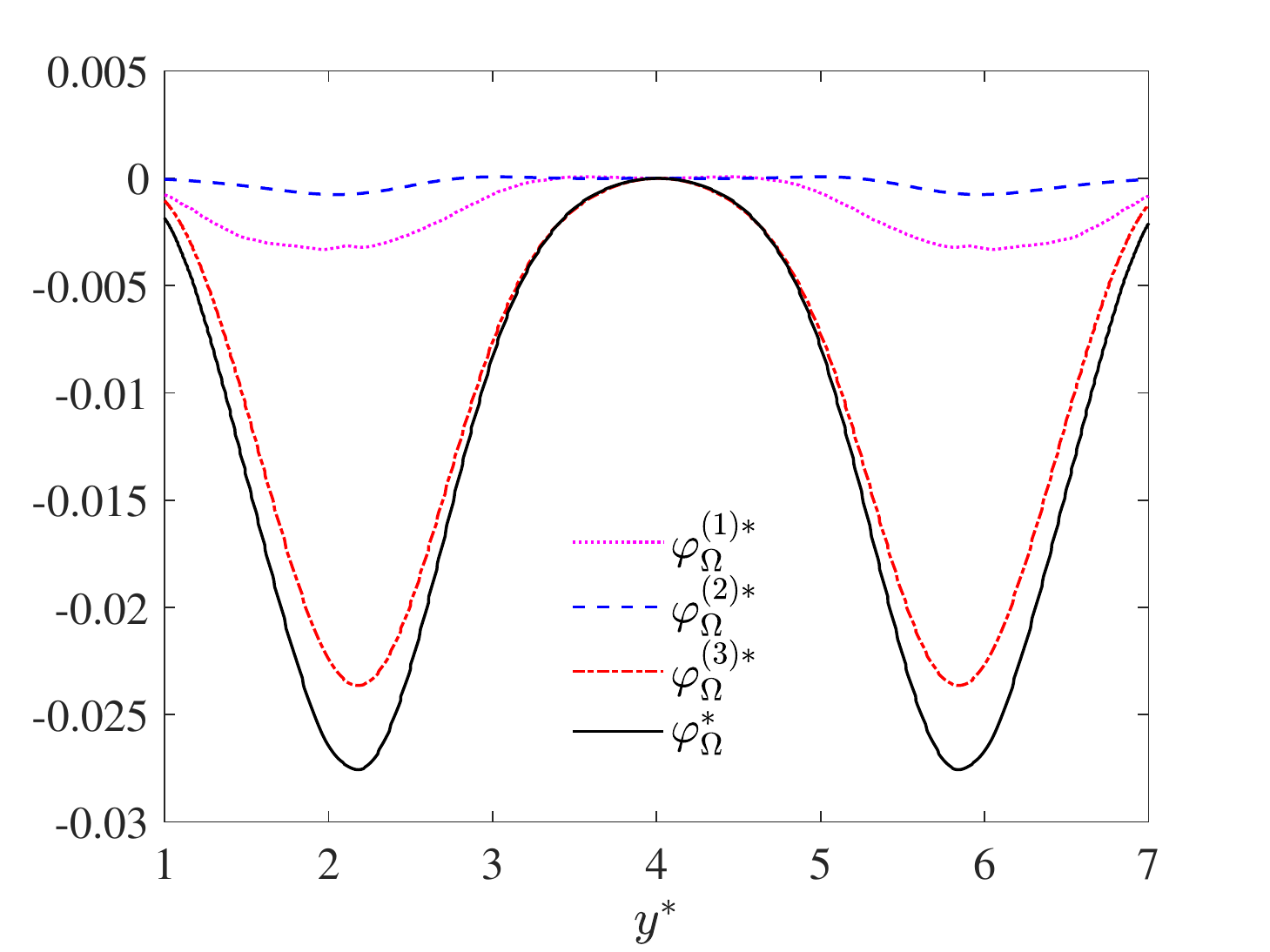}
		\label{yy2b}
	\end{minipage}%
}%
	\caption{Comparison of different contributions to the normalized total IEF $\varphi_{\Omega}^{*}\equiv\varphi_{\Omega}/(\rho_{l}U_{0}^3/D_{l}^2)$ along the liquid-vapor interface $S$ at $t^*=4.8$. (a) Near the contact point $(x^*,y^*)=(0,0.76)$; (b) Near the center region.} 
	\label{yy2}
\end{figure}

Finally, different contributions to the IEF $\varphi_{\Omega}$ are compared in Fig.~\eqref{yy2}. It is clearly seen that the contribution from $\varphi_{\Omega}^{(2)}$ due to the surface acceleration is negligibly small compared to $\varphi_{\Omega}^{(1)}$ and $\varphi_{\Omega}^{(3)}$. 
Most importantly, we note that $\varphi_{\Omega}$ remains negative along the interface $S$, which implies that the enstrophy is diffusing from the region outside the droplet to the droplet interior.
The increased enstrophy inside the droplet through the IEF will make up the viscous dissipation and resist the energy conversion from kinetic energy to surface tension energy, which could be important to the maintenance of separation bubbles during the droplet spreading process. 

Nevertheless, the dominant physical mechanism for this negative IEF is different for the regions close to and away from the contact point, which is analyzed in detail as follows.
As displayed in Fig.~\ref{yy2a}, in a small vicinity of the contact point ($y^*\in[0.76,0.8]$), both $\varphi_{\Omega}^{(1)}$ and $\varphi_{\Omega}^{(3)}$ achieve their highest magnitudes at the contact point, although with their signs being different ($\varphi_{\Omega}^{(1)}<0$ and $\varphi_{\Omega}^{(3)}>0$). Since $\lvert\varphi_{\Omega}^{(1)}\rvert$ is notably higher than $\lvert\varphi_{\Omega}^{(3)}\rvert$, $\varphi_{\Omega}$ remains negative in this small vicinity with its highest magnitude appearing at the contact point, which is dominated by the coupling between the vorticity-induced surface shear stress ($\bm{\tau}_{\bm{\omega}}\bm{\cdot}\hat{\bm{t}}>0$) and the surface pressure gradient ($[\partial\bar{p}_{h}/\partial{s}]_{S}<0$), according to Eq.~\eqref{phia}.

In constrast, $\varphi_{\Omega}$ and different contributions show relatively low magnitudes away from the contact points. As illustrated in Fig.~\ref{yy2b}, different from their distributions shown in Fig.~\eqref{yy2a}, both $\varphi_{\Omega}^{(1)}$ and $\varphi_{\Omega}^{(3)}$ are negative, while $\lvert\varphi_{\Omega}^{(3)}\rvert$ is much higher than $\lvert\varphi_{\Omega}^{(1)}\rvert$. Two distinct negative peak regions are observed for both $\varphi_{\Omega}^{(1)}$ and $\varphi_{\Omega}^{(3)}$, particularly for $\varphi_{\Omega}^{(3)}$. 
This observation can be explained from two aspects.
On one hand, each peak region just corresponds to the interfacial portion between an outer primary vortex and a separation bubble inside the droplet, where the shearing effect is much stronger compared to the centeral portion of the interface $S$. On the other hand, according to Eq.~\eqref{phic2}, we have $\varphi_{\Omega}^{(3)}\propto-\lVert\bm{\nabla}\rho\rVert_{S}\bm{\tau}_{\bm{\omega}}\bm{\cdot}\bm{\tau}_{S}$. Since $\bm{\tau}_{\omega}\bm{\cdot}\hat{\bm{t}}$ and $\bm{\tau}\bm{\cdot}\hat{\bm{t}}$ basically have same signs away from the contact points (see Fig.~\ref{tf3}), it follows that $\varphi_{\Omega}^{(3)}$ is negative in these two peak regions.
\section{Conclusions and discussions}\label{Conclusions and discussions}
In this paper, a theoretical study is performed to derive the exact decompositions of the boundary enstrophy flux (BEF) $f_{\Omega}$ and the interfacial enstrophy flux (IEF) $\varphi_{\Omega}$ for two-phase viscous flow with diffuse interface.
All the physical mechanisms causing the BEF and IEF are clearly elucidated using the obtained exact relations in the boundary vorticity dynamics.
In order to demonstrate the application of these relations, we simulate a droplet impact on a solid wall
by using a recently developed well-balanced discrete unified gas kinetic scheme (WB-DUGKS). Based on the analysis of the simulation data in the framework of the boundary vorticity dynamics, the new findings on the impinging droplet are summarized as follows.   

First, on the bottom solid wall, the distributions of skin friction, surface pressure and BEF are effective indicators for local flow separation and attachment in the droplet, revealing the complexity of droplet-wall interaction. 
It is found that the BEF $f_{\Omega}$ has extremely higher negative value in the regions near the moving contact points than that in other regions, indicating that the regions near the contact points are the main vorticity source on the wall.

Second, on the liquid-vapor interface, the simulation shows that the vorticity-induced surface shear stress $\bm{\tau}_{\bm{\omega}}$ and the surface deformation shear stress $\bm{\tau}_{_{\bm{W}}}$ play opposite roles in the vicinity of the contact point.
Both of them and the surface pressure gradient $\bm{\nabla}_{S}[\bar{p}_{h}]_{S}$ exhibit higher magnitudes there compared to the region away from the contact point. The IEF $\varphi_{\Omega}$ is negative on the interface. 
The IEF is dominated by the coupling between 
$\bm{\tau}_{\bm{\omega}}$ and $\bm{\nabla}_{S}[\bar{p}_{h}]_{S}$ in a small vicinity of each contact point.
There are also two negative peaks with lower magnitudes in the IEF around the central region, which are controlled by the dot product of
$\bm{\tau}_{\bm{\omega}}$ and the total surface shear stress $\bm{\tau}$, and the magnitude of density gradient $\lVert\bm{\nabla}\rho\rVert_{S}$.

The total enstrophy flux is negative on the surface of the closed droplet volume bounded by the interface and the wall, which implies the increase of
the enstrophy inside the impinging droplet.
From the perspective of vorticity dynamics, this negative enstrophy flux will maintain separation bubbles (vortices) by balancing the viscous dissipation and the surface tension work during the droplet spreading process.

The present study restricts to 2D case due to the limitation of computational resource, where both the contributions caused by the surface-normal vorticity component and the surface curvature in Eqs.~\eqref{phie} and~\eqref{fomega4} vanish. Nevertheless, they will certainly contribute to $f_{\Omega}$ and $\varphi_{\Omega}$ in 3D case and greatly enrich the flow physics, such as the experimental and numerical studies about droplet impact on a solid surface reviewed by Josserand and Thoroddsen.~\cite{Josserand2015} 
Further exploration can be performed by applying the boundary vorticity dynamics to more complex systems involving a number of 3D droplets and vapour bubbles, as reported in Saddle et al.~\cite{Saade2021}, Prosperetti~\cite{Prosperetti2017} and Elghobashi.~\cite{Elghobashi2019}
Extended studies to physical problems involving multiple interfaces (bubbly
clouds) and bubble trapping close to a free surface are of great interest, as well as
problems involving cavitation, with coalescence and collapse phenomena.
The evaporation and condensation phenomena can also be investigated by incorporating the energy equation to the present framework. 

%==========================================
\begin{acknowledgments}
T. Liu is partially supported by the John O. Hallquist Endowed Professorship and the Presidential Innovation Professorship. The authors thank the reviewers for their good suggestions on improving the quality of the paper.
\end{acknowledgments}

\section*{Conflict of Interest}
The authors have no conflicts to disclose.

\section*{DATA AVAILABILITY}
The data that support the findings of this study are available
from the corresponding author upon reasonable request.

\appendix

\section{Proof of Eq.~\eqref{xx2}}\label{ProofLeeLin}
Eq.~\eqref{density_profile} is the analytical solution for a flat surface at equilibrium, which satisfies the condition
\begin{eqnarray}\label{r1}
\psi(\rho)=\frac{1}{2}\kappa\frac{d^2\rho}{d\zeta^2}=\frac{1}{2}\kappa\lVert\bm{\nabla}\rho\rVert^2.
\end{eqnarray}
In most cases when the interfacial thickness is sufficiently small compared to the characteristic length scale, Eq.~\eqref{r1} actually represents the 
leading-order approximation in the presence of a curved surface. Therefore, it can be viewed as an useful equivalent substitution for the double-well bulk free energy density $\psi$ in theoretical analysis.

For the hydrodynamic pressure $p_{h}$, by using Eqs.~\eqref{p0},~\eqref{murho},~\eqref{mu0} and~\eqref{r1}, we have
\begin{eqnarray}\label{r2}
p_{h}&\equiv&\rho\mu_{\rho}=\rho\mu_{0}-\kappa\rho\bm{\nabla}^2\rho\nonumber\\
&=&p_0+\psi-\kappa\rho\nabla^2\rho\nonumber\\
&\approx&p_0-\kappa\rho\nabla^2\rho+\frac{1}{2}\kappa\lVert\bm{\nabla}\rho\rVert^2\nonumber\\
&=&p_{h}^{L-L}.
\end{eqnarray}

Similarly, the density gradient force $\rho\bm{F}_{\nabla}$ can be evaluated as
\begin{eqnarray}\label{Fnabla}
\rho\bm{F}_{\nabla}
&\equiv&\mu_{\rho}\bm{\nabla
}\rho=\left(\frac{\partial\psi}{\partial\rho}-\kappa\nabla^2\rho\right)\bm{\nabla}\rho\nonumber\\
&=&\bm{\nabla}\psi-\kappa\nabla^2\rho\bm{\nabla}\rho\nonumber\\
&=&\bm{\nabla}\psi+\kappa\rho\bm{\nabla}\nabla^2\rho-\kappa\bm{\nabla}(\rho\nabla^2\rho).
\end{eqnarray}
By using Eq.~\eqref{id1} to replace the second and the third terms in the right hand side of Eq.~\eqref{Fnabla}, we obtain
\begin{eqnarray}\label{Fnabla2}
\rho\bm{F}_{\nabla}=\bm{\nabla}\psi+\bm{\nabla}\left(\frac{1}{2}\kappa\lVert\bm{\nabla}\rho\rVert^2\right)-\kappa\bm{\nabla}\bm{\cdot}
\left(\bm{\nabla}\rho\bm{\nabla}\rho\right).
\end{eqnarray}
Substituting $\psi$ in Eq.~\eqref{Fnabla2} with Eq.~\eqref{r1}, we have
\begin{eqnarray}\label{r3}
\rho\bm{F}_{\nabla}
=\kappa\bm{\nabla}\bm{\cdot}\left(\lVert\bm{\nabla}\rho\rVert^2\bm{I}-\bm{\nabla}\rho\bm{\nabla}\rho\right)\approx\bm{\nabla}\bm{\cdot}\bm{\Phi}.
\end{eqnarray}
From Eqs.~\eqref{r2} and~\eqref{r3}, Eq.~\eqref{xx2} is proved.

\section{Interface mean curvature and its physical interpretation}\label{curvature}
On one hand, we note that the Laplacian of the density field on the surface $\left[\nabla^2\mathcal{\rho}\right]_{S}$ can be evaluated as
\begin{eqnarray}\label{wallnormalF2}
	\left[\nabla^2\rho\right]_{S}=\nabla_{S}^{2}\rho_{S}-tr(\bm{K})\left[\frac{\partial\rho}{\partial n}\right]_{S}+\left[\frac{\partial^2\rho}{\partial n^2}\right]_{S}.
\end{eqnarray}
Since $\rho_{S}$ takes a constant value and $\left[\partial\rho/\partial{n}\right]_{S}=\lVert\bm{\nabla}\rho\rVert_{S}$, from Eq.~\eqref{wallnormalF2}, we obtain
\begin{eqnarray}\label{xx3}
	tr(\bm{K})
	=-\frac{1}{\lVert\bm{\nabla}\rho\rVert}_{S}\left\{
	\left[\nabla^2\rho\right]_{S}
	-\left[\frac{\partial^2\rho}{\partial n^2}\right]_{S}\right\}.
\end{eqnarray}
Then, by using the leading-order approximation (Eq.~\eqref{density_profile}) in the presence of a curved surface, Eq.~\eqref{xx3} can be rewritten as
\begin{eqnarray}\label{mean_curvature}
	tr(\bm{K})\approx\frac{[\mu_{\rho}]_{S}}{\kappa\lVert\bm{\nabla}\rho\rVert_{S}},
\end{eqnarray}
which suggests that the mean curvature of $S$ is approximately proportional to the chemical potential on $S$, and is inversely proportional to the magnitude of the density gradient at the same location.
On the other hand, we claim that the mean curvature of $S$ is only determined the distribution of the unit normal vector field $\bm{n}_{S}$, without requiring its definition off $S$.
Considering that a small vicinity of the interface $S$ is filled by a group of density isosurfaces $\rho=constant$ (with different constants), an extended unit vector field $\hat{\bm{n}}$ can be well defined in this small vicinity, namely, $\hat{\bm{n}}\equiv\bm{\nabla}\rho/\lVert\bm{\nabla}\rho\rVert$. Obviously, the restriction of  $\hat{\bm{n}}$ on $S$ gives the unit normal vector field $\bm{n}_{S}$, namely, $\left[\hat{\bm{n}}\right]_{S}=\bm{n}_{S}$. As a result, the mean curvature of $S$ can be equivalently evaluated using $\hat{\bm{n}}$, 
\begin{eqnarray}\label{w1}
	tr(\bm{K})=-\left[\bm{\nabla}\bm{\cdot}\hat{\bm{n}}\right]_{S}
	=-\frac{1}{\lVert\bm{\nabla}\rho\rVert_{S}}\left[\nabla^2\rho-\frac{(\bm{\nabla}\rho\bm{\cdot}\bm{\nabla})\lVert\bm{\nabla}\rho\rVert}{\lVert\bm{\nabla}\rho\rVert}\right]_{S}.
\end{eqnarray}
Eq.~\eqref{w1} is already presented in the framework of phase-field by Sun and Beckermann.~\cite{Sun2007} The method of calculating the mean curvature can be found in Echebarria {\it et al.}~\cite{Echebarria2004}

\section{Code validation for prescribed contact angles}\label{Code validation}
\begin{figure}[h!]
	\centering
	\subfigure[]{
		\begin{minipage}[t]{0.5\linewidth}
			\centering
			\includegraphics[width=0.8\columnwidth,trim={0.1cm 1.5cm 0.1cm 2.2cm},clip]{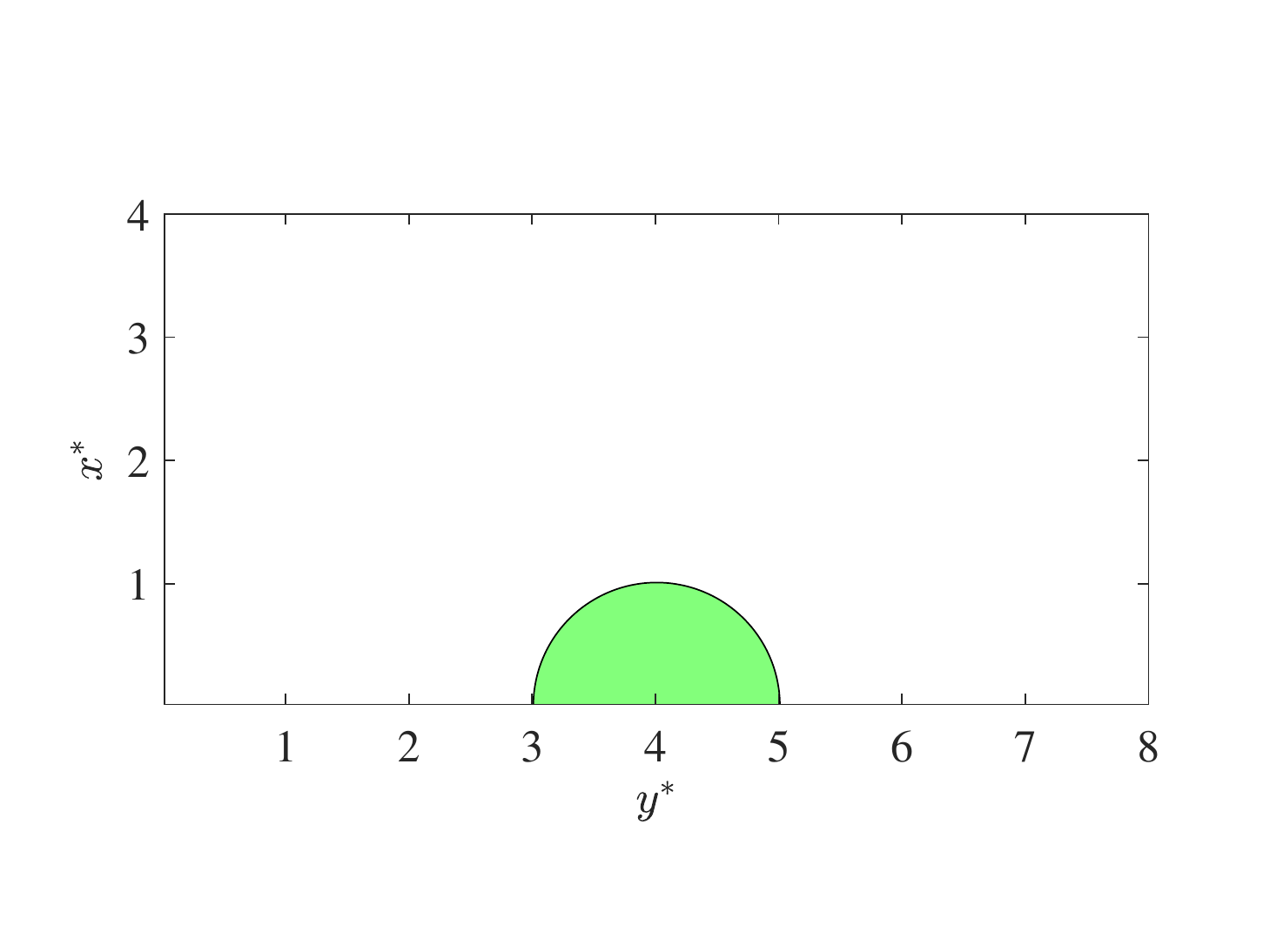}
			\label{static_initial}
		\end{minipage}%
	}%
	\subfigure[]{
		\begin{minipage}[t]{0.5\linewidth}
			\centering
			\includegraphics[width=0.8\columnwidth,trim={0.1cm 1.5cm 0.1cm 2.2cm},clip]{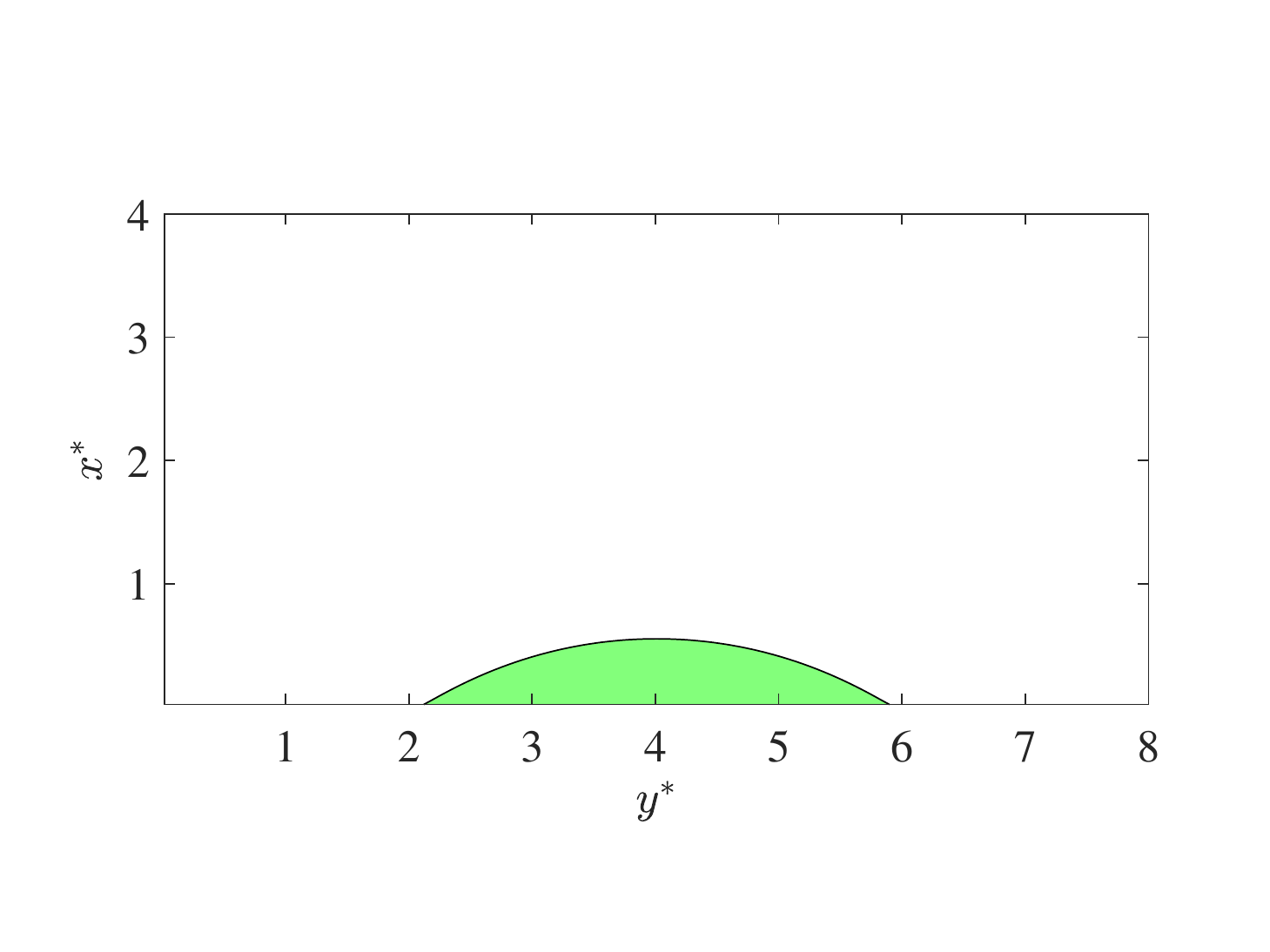}
			\label{static_deg_30}
		\end{minipage}%
	}%
	
	\subfigure[]{
		\begin{minipage}[t]{0.5\linewidth}
			\centering
			\includegraphics[width=0.8\columnwidth,trim={0.1cm 1.5cm 0.1cm 2.2cm},clip]{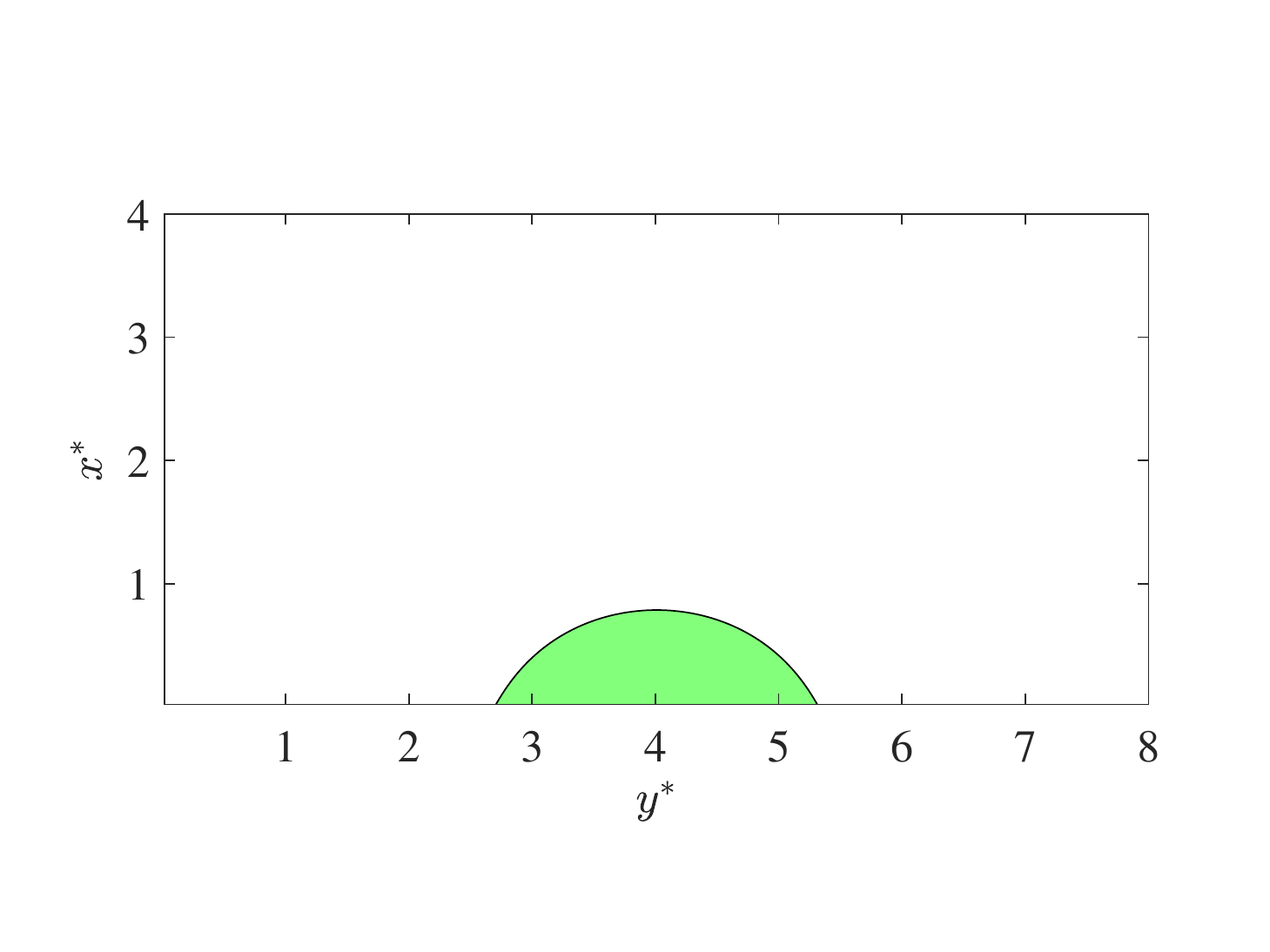}
			\label{static_deg_60}
		\end{minipage}%
	}%
	\subfigure[]{
		\begin{minipage}[t]{0.5\linewidth}
			\centering
			\includegraphics[width=0.8\columnwidth,trim={0.1cm 1.5cm 0.1cm 2.2cm},clip]{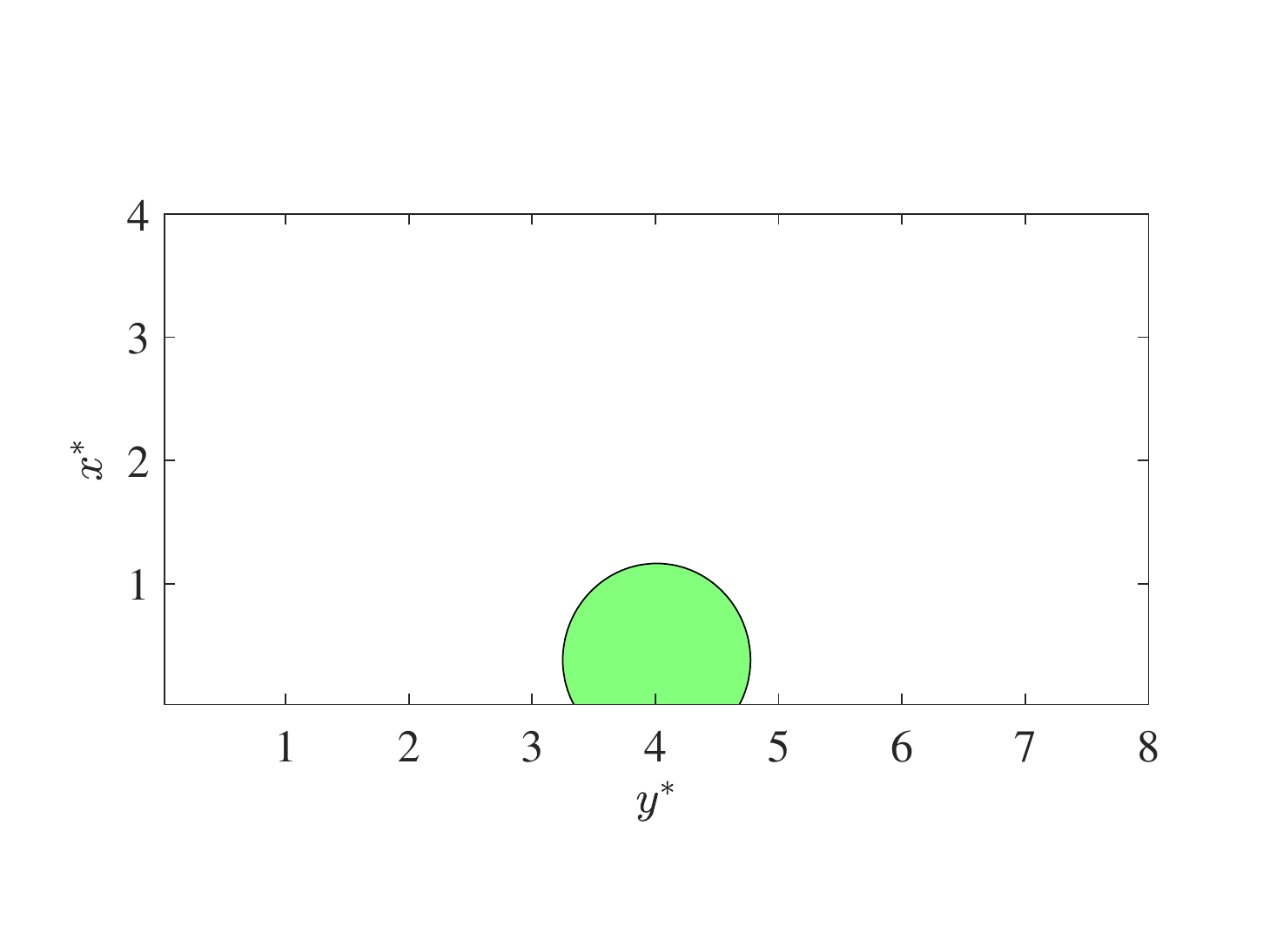}
			\label{static_deg_120}
		\end{minipage}%
	}%
	\caption{Numerical prediction of droplet equilibrium shapes for prescribed contact angles, (a) initial shape, (b) $\theta=\pi/6$, (c) $\theta=\pi/3$, (d) $\theta=2\pi/3$. The black lines indicate the droplet surface $S$ with $\rho=(\rho_{l}+\rho_{g})/2$ for all the subfigures.} 
	\label{Steady_shape_angles}
\end{figure}

In order to validate the effectiveness of the geometric boundary condition (Eq.~\eqref{geobc}) in the WB-DUGKS, a semicircular droplet on a wetting wall with different prescribed contact angles ($\theta=\pi/6$, $\pi/3$, and $2\pi/3$) is simulated. The numerical settings and boundary conditions are the same as those described in Section~\ref{Description of physical problem}.
Initially, a semicircular stationary droplet with the radius $D_{l}=100\Delta{x}$ is placed on the bottom wall, centered at $(x_{0},y_{0})=(0,2D_{l})$. As shown in Fig.~\ref{Steady_shape_angles}, the calculated steady droplet morphologies are reasonable for different prescribed contact angles.
Further, in order to provide quantitative comparison for different cases, we measure the numerical contact angle by using the geometric relation,~\cite{LiangHong2019} namely, $\theta^{*}=2\arctan(2H/L)$, where $L$ is the droplet spreading length on the substrate and $H$ is its maximum height. Note that the contact angle is measured on the droplet interface $S$ with $\rho=(\rho_{l}+\rho_{g})/2$. For $\theta=\pi/6$ (hydrophilic), $\pi/3$ (hydrophilic), and $2\pi/3$ (hydrophobic), the measured contact angles are respectively $\theta=30.38^{\circ}$, $60.30^{\circ}$ and $119.98^{\circ}$, which are in good agreement with the prescribed values.
The small discrepancies indicate the effective implementation of the geometric boundary condition in the WB-DUGKS. It should be noted that the geometric wetting boundary condition is also used in a conservative multilevel discrete unified gas kinetic scheme (MDUGKS) developed by Yang {\it et al.}~\cite{YangZeren2022} Since the present paper focuses on the discussion of physics, we will not go for more numerical details here.

\section{Extraction of surface physical quantities}\label{ESPQ}
\begin{figure}[h!]
	\centering
	\includegraphics[width=0.5\columnwidth,trim={0.0cm 0.0cm 0.0cm 0.0cm},clip]{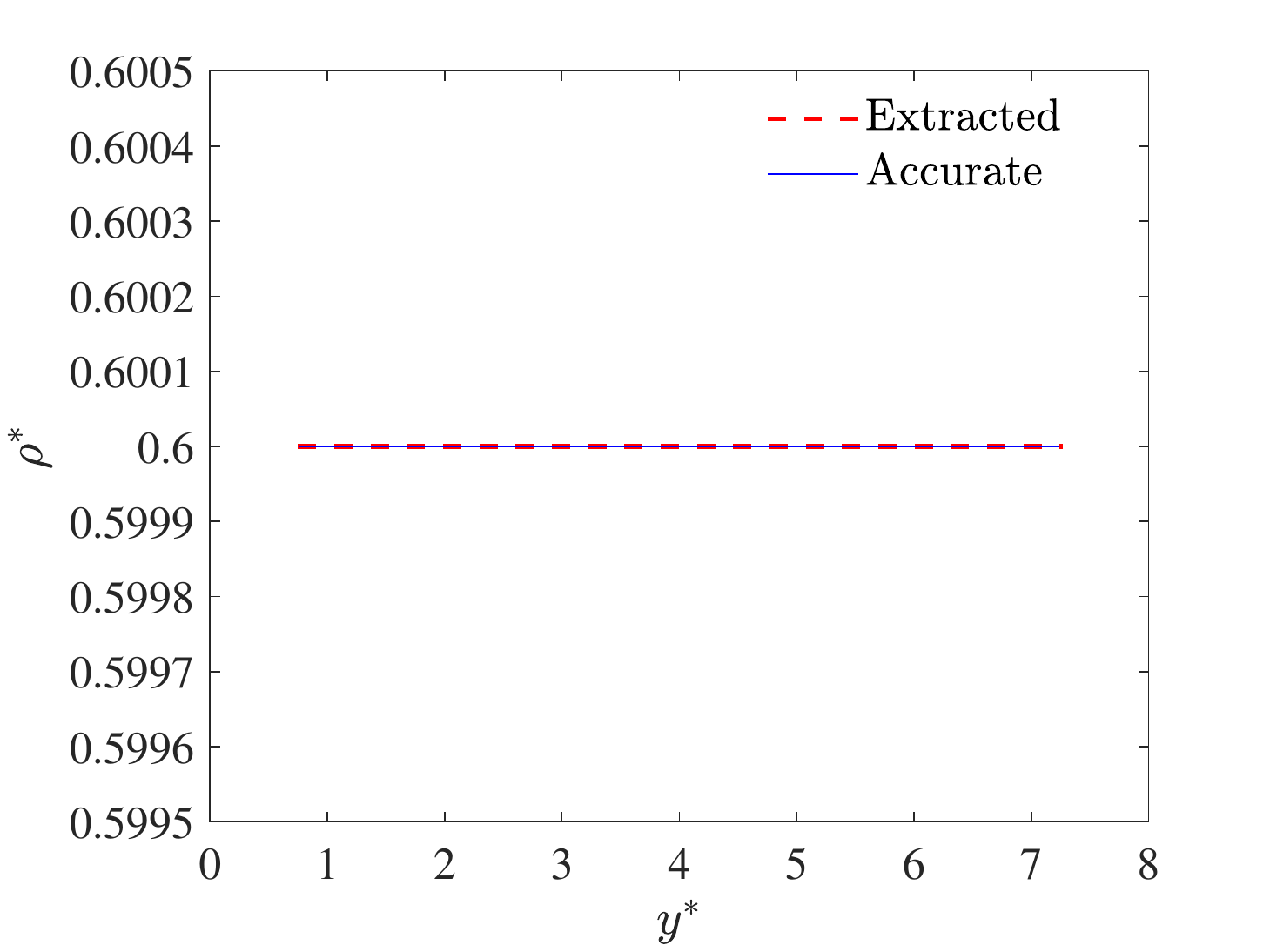}
	\caption{Comparison of the extracted and accurate normalized desnity $\rho^{*}\equiv\rho/\rho_{l}$ on the interface. } 
	\label{Extract_density}
\end{figure}
\begin{figure}[h!]
	\centering
	\subfigure[]{
		\begin{minipage}[t]{0.5\linewidth}
			\centering
			\includegraphics[width=1.0\columnwidth,trim={0.1cm 0.1cm 0.1cm 0.6cm},clip]{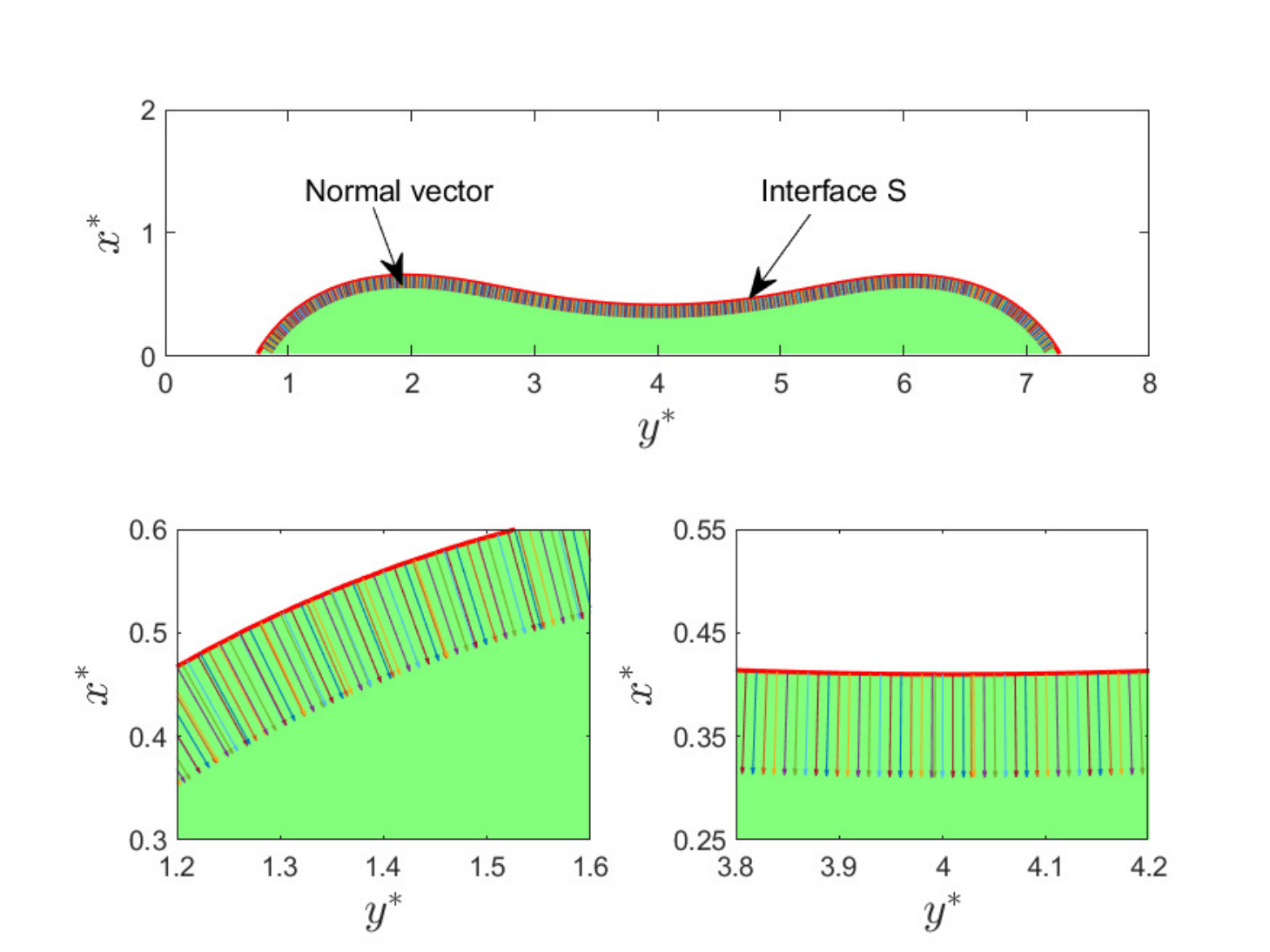}
			\label{Extract_normal_vector}
		\end{minipage}%
	}%
	\subfigure[]{
		\begin{minipage}[t]{0.5\linewidth}
			\centering
			\includegraphics[width=1.0\columnwidth,trim={0.1cm 0.1cm 0.1cm 0.6cm},clip]{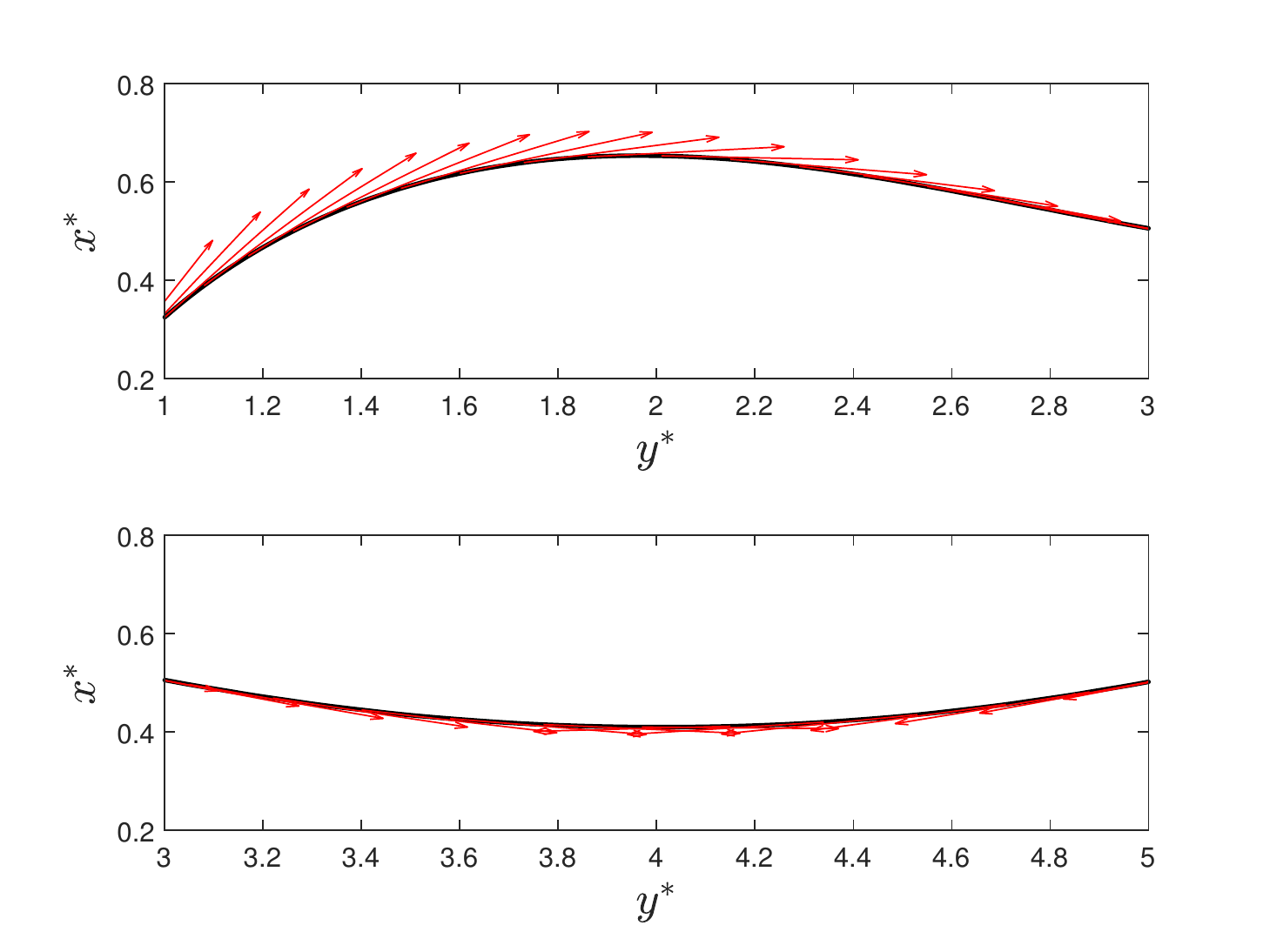}
			\label{Extract_shear_stress}
		\end{minipage}%
	}%
	\caption{(a) Extracted unit normal vector field $\bm{n}_{S}$ along the interface $S$ defined by $\rho=(\rho_{l}+\rho_{g})/2$ with two zoon-in views. (b) Extracted surface shear stress field $\bm{\tau}_{S}$ along the interface $S$.} 
	\label{f2}
\end{figure}
Our numerical simulation is performed using regular Cartesian coordinate system, which is not inherently consistent with the varying shape of the droplet interface. In order to extract surface physical quantities from the simulation data for physical analysis, a well-designed post-processing code is developed based on the Matlab platform. The post-processing code has been carefully validated for different cases. For example, we select some results to demonstrate the reliability of the code. Fig.~\ref{Extract_density} shows the extracted normalized density on the isosurface $\rho^*\equiv[\rho]_{S}/\rho_{l}=0.6$. The extracted density field is in excellent agreement with the theoretical one. Moreover, Figs.~\ref{Extract_normal_vector} and~\eqref{Extract_shear_stress} show the extracted unit normal vector field $\bm{n}_{S}$ and surface shear stress field $\bm{\tau}_{S}$ on the interface $S$, which are reasonably agreement with the surface geometry.

%\setcounter{figure}{0}
%\setcounter{table}{0}

%\nocite{*}
\bibliography{ChenPOFref}% Produces the bibliography via BibTeX.
\newpage

\end{document}